\documentclass[12pt,twoside]{article}
\usepackage[mathscr]{eucal}
\usepackage{amsmath,amsfonts,amssymb,amsthm}
\usepackage{hyperref}
\usepackage{times}
\def\mt1{Metsaev:1998it}

\def\berk{berk}

\def\ci{\cite}
\newcommand{\be}{\begin{equation}}
\newcommand{\ee}{\end{equation}}

\def \td {\tilde}

\def \bi{\bibitem}
\def \la {\label}

\def \l {\lambda}
\def\foot{\footnote}

\def \adss {$AdS_5 \times S^5$\ }
\newcommand{\rf}[1]{(\ref{#1})}
\def \ov {\over}

\def \ha{{1\ov 2}}

\def \del {\partial}

 \def \bb {\bar \begin{equation}ta}

\def \bi{\bibitem}
\def \la {\label}

\def \l {\lambda}
\def\foot{\footnote}

\def \adss {$AdS_5 \times S^5$\ }

\def \ov {\over}

\def \varpi {{\rm w}}

\def \rt {{\rm t}}

\def \adss {$AdS_5 \times S^5$\ }

\def \bb {{\bar  \begin{equation}ta}}

\def \inv {^{-1}}

  \def \te {\theta}

\def \dpp {\del_+}
\def \dmm {\del_-}
\def \k {\kappa}
\def \ci {\cite}

\def\tr{{\rm Tr}}
\def\str{{\rm STr}}

\def\gf{f}
\def\gg{g}

\newcommand{\psr}{\Psi_{{}_R}}
\newcommand{\psl}{\Psi_{{}_L}}

\renewcommand{\tilde}{\widetilde}
\renewcommand{\hat}{\widehat}

\newcommand{\bref}[1]{\textbf{\ref{#1}}}

\renewcommand{\AA}{\mathbb{A}}

\newcommand{\alga}{\Liealg{a}}
\newcommand{\algn}{\Liealg{n}}
\newcommand{\algm}{\Liealg{m}}
\newcommand{\algf}{\Liealg{f}}
\newcommand{\alghf}{\hat\algf}
\newcommand{\algg}{\Liealg{g}}
\newcommand{\algh}{\Liealg{h}}

\newcommand{\algp}{\Liealg{p}}

\newcommand{\Liealg}{\mathfrak}       %Lie alg[ebra(s)]
\newcommand{\id}{\mathbf{1}}
\renewcommand{\imath}{i}

\newcommand{\dd}{\partial}
\renewcommand{\d}{\partial}

\renewcommand{\geq}{\,{\geqslant}\,}

\newcommand{\inner}[2]{\langle #1{,}\,#2\rangle}
\newcommand{\binner}[2]{%
  {\langle}\kern-4.15pt{\langle}#1{,}\,#2{\rangle}\kern-4.15pt{\rangle}}
\newcommand{\commut}[2]{[#1{,}\,#2]}

\newcommand{\scommut}[2]{\{#1{,}\,#2\}}

\newcommand{\half}{\mathchoice{%
    \ffrac{1}{2}}{\frac{1}{2}}{\frac{1}{2}}{\frac{1}{2}}}

\newcommand{\ffrac}[2]{\raisebox{.5pt}%
  {\footnotesize$\displaystyle\frac{#1}{#2}$}\kern1pt}

\newcommand{\red}{\mathrm{red}}

\newcommand{\dl}[1]{\mathchoice{\ffrac{\dd}{\dd #1}}{\frac{\dd}{\dd
      #1}}{\ffrac{\dd}{\dd #1}}{\ffrac{\dd}{\dd #1}}}

\def\const{\mathop\mathrm{const}\nolimits}

\newcommand{\vac}{|0\rangle}

\newcommand{\fC}{\mathbb{C}}

\def\cA{\mathcal{A}}

\def\cL{\mathcal{L}}

\def \vp {\varphi}

\def \bs {\bigskip}

\numberwithin{equation}{section} \makeatletter
\@addtoreset{equation}{section}

\vspace{-1cm}

  \def \te {\theta}
  \def \vp {\varphi}
  \def \a {\alpha}
  \def \b {\beta}

\newcommand{\pp}{{\parallel}}
\newcommand{\oo}{{\perp}}
\def \A {{\cal A}}
\def \AA {{\rm A}}
\def \s {\sigma}
\def \Tr {{\rm Tr}}

\def \vt  {\vartheta}

\def \CS {{\Sigma }}

\def \P {\Phi}
\def \ns {{\rm N }}
\newcommand{\bea}{\begin{eqnarray}}
\newcommand{\eea}{\end{eqnarray}}

\def \psiL {\psi_{_L}}
\def \psiR {\psi_{_R}}
\def \vpsiL {\vpsi_{_L}}
\def \vpsiR {\vpsi_{_R}}
\def \epsL {\epsilon_{{_L}}}
\def \epsR {\epsilon_{{_R}}}
\def \PsiL {\Psi_{_L}}
\def \PsiR {\Psi_{_R}}

\def \vac {{_{\rm vac}}}

\def \vpsi{{\bf \psi}}

\def \lam {{\ell}}
\begin{document}

\vspace{ -3cm}
\rightline{Imperial-TP-AT-2007-4}

\rightline{NI07093}

\begin{center}
\vspace{0.2cm}
{\Large\bf
Pohlmeyer reduction
  \\
\vspace{0.15cm}
of   $AdS_5 \times S^5$ superstring sigma model \\
\vspace{0.2cm}
%from quantum open strings in $AdS_5 \times S^5$
%\vspace{0.15cm}
\vspace{0.4cm}
   }

 \vspace{0.5cm} {M. Grigoriev$^{a,b,c,}$\footnote{grig@lpi.ru}
 and A.A. Tseytlin$^{a,c,}$\footnote{
 %Also at
 %,
  tseytlin@imperial.ac.uk
 }}\\
 \vskip 0.13cm

{\em
$^{a}$   Blackett Laboratory, Imperial College,
London SW7 2AZ, U.K.     \\
$^{b}$  Institute for Mathematical Sciences, Imperial College,
London SW7 2PE, U.K.     \\
%$^{b}$Department of Physics, The Pennsylvania  State University,\\
%University Park, PA 16802 , USA\\
$^{c}$  Department of Theoretical Physics, Lebedev  Institute, Moscow, Russia   }

\end{center}

 \begin{abstract}
 %%%%%%%%%%%%%%%%%%%%%%%%%%%%%%%%%
 Motivated  by a desire  to find a useful 2d Lorentz-invariant
reformulation of the $AdS_5 \times S^5$ superstring world-sheet theory 
in terms of physical degrees of freedom we  construct  the  ``Pohlmeyer-reduced'' version
of the corresponding  sigma model. The Pohlmeyer reduction procedure
involves  several steps. Starting with a  coset space string sigma model in 
the conformal gauge and writing  the classical  equations in terms of currents
one  can fix the residual  conformal diffeomorphism  symmetry and kappa-symmetry
and introduce a new  set of variables  (related locally to currents but
non-locally to the original  string coordinate fields)
so that the Virasoro constraints are  automatically satisfied.
The resulting   equations can be obtained from a Lagrangian of a
non-abelian Toda  type: a gauged WZW model with an integrable potential
coupled also to a set of 2d fermionic fields. A gauge-fixed  form of the  Pohlmeyer-reduced 
theory  can be found  by integrating out the 2d gauge field of the  gauged WZW model.  
The  small-fluctuation spectrum near the trivial vacuum contains  8 bosonic  and 8 fermionic degrees
of freedom with  equal mass. We conjecture that 
the reduced  model has   world-sheet supersymmetry and is ultraviolet-finite.  
We show  that in the special case of the $AdS_2 \times S^2$ 
superstring model  the reduced theory is indeed supersymmetric: 
it is equivalent to the N=2 supersymmetric extension of the sine-Gordon model. 
\end{abstract}
\thispagestyle{empty}
\setcounter{page}{0}
\newpage

\setcounter{footnote}{0}

{\small
\tableofcontents
}

\def \eps {\epsilon}
\def \diag {{\rm diag}}
\def \TT  {{\rm T}} 
\def \rt {{\rm t}} 
\def \bs {\bigskip}
\def \km {{\mathbf{\mu}}}

\newpage
%%%%%%%%%%%%%%%%%%%%%%%%%%%%%%%%%%%%%%%%%%%%%%%%%%%%%%%%%%%%%%%%
\section{Introduction}
%%%%%%%%%%%%%%%%%%%%%%%%%%%%%%%%%%%%%%%%%%%%%%%%%%%%%%%%%%%%%%%%%%%%

String  theory in $AdS_5 \times S^5$ is  represented by a
Green-Schwarz-type \ci{gss}   action on a  supercoset 
$\frac{PSU(2,2|4)}{SO(1,4) \times SO(5)}$
\cite{mt1}. It is classically integrable \cite{bpr} and has an involved
solitonic spectrum (see, e.g., \ci{msw,afrt}).
To quantize it one may attempt  to eliminate first   unphysical degrees of freedom
 by  choosing
a kind of  light-cone gauge, i.e. 
an analog of $x^+=p^+ \tau,  \ \Gamma^+ \theta=0$. One  natural option is to 
expand near
the  null geodesic parallel to the boundary in the Poincare patch; the resulting
 gauge-fixed action  is  then  quartic in  fermions  \cite{mtt}.
 An alternative  is to use  the null  geodesic
  wrapping $S^5$ \cite{mt2}; the resulting action
 \cite{call,arut,fpz}   has  a 
rather
 complicated  structure  with many  non-linear interaction terms.
  
An apparent  disadvantage of the light-cone gauge choices is
that the  gauge-fixed action lacks manifest 2d Lorentz invariance
(beyond the quadratic  level in the fields).
This makes it hard to apply  familiar   methods of integrable
quantum field theories; in particular, the  S-matrix
for the  elementary excitations  has apparently 
less restricted form  \cite{bei,kzrm}
than in  a Lorentz-invariant case (cf. \ci{zz}).

An alternative approach which we shall explore here 
is   to impose  the  conformal
gauge condition and to perform 
 a non-local transformation of variables (from coodinates to currents)
   that solves
the Virasoro constraints
  at the classical level  while  preserving  the integrable structure.
  This   generalizes the Pohlmeyer ``reduction'' (or better   ``reformulation'')
    relating    the classical  $S^2$ sigma model 
  to the sine-Gordon model 
  \cite{pol} (see also  \ci{mz,pole,polr,ei,add}).
   A related  work  in this direction  appeared in
   \cite{mi12,mi3}.
  One is then left with the right number of  physical 
  (``transverse'') degrees of
  freedom. 
   In a certain sense, this reduction  approach may be viewed as 
   a  ``covariant analog''  of a light-cone gauge fixing.

   The resulting  ``reduced''  model should have  closely related solitonic
    spectrum to  the
original one, and one may then  raise the question if  the classical
 correspondence  between the two models
 may  extend to the quantum level. This  is   not  what happens 
 in the case of the $S^2$ sigma model  and  the sine-Gordon  model
 (one reason is  that in 
  the  reduction procedure  one uses  conformal symmetry of the 
  $SO(3)/SO(2)$ model  which does
  not survive beyond the classical level)
   but
 we  may  conjecture 
  that the relation  may still hold 
    in the very special  case  of the full \adss superstring model
 %(i.e.  including  2d fermions)
  which  should be 
  conformal at the quantum level.

\bs 

 Below  we shall first  discuss the  Pohlmeyer-type  reduction 
 for the bosonic part of the classical 
 \adss sigma model  and  then consider the full supercoset  superstring 
  theory.
As we shall see,  the application  of this  procedure
to  the bosonic part of the \adss string action
 leads to a
 {\it 2d relativistically invariant}  
  ``reduced'' theory
represented   by  a  sigma model with a potential term  which 
has  an equivalent integrable structure. It generalizes
 the sine-Gordon \cite{pol} and the
 complex sine-Gordon \cite{pol,lund} models to the case of
 the  4+4  dimensional target space.
 
%One  of our aims  below will be  to 
We shall explain how to obtain 
a  local  Lorentz-invariant 
 action  for  this   reduced  theory in terms of  ``physical'' (gauge-fixed) 
 degrees of freedom.\foot{This  was not  done explicitly  in the past for 
the  $S^n$ models with $n>3$. The existence of a local Lagrangian is an 
important issue. At  the level of equations for the 
currents or the  Lax pair
equations  there is a large freedom \ci{mz}  in how one can  choose a local
field representation  -- many classically equivalent  models  have
same-looking  Lax equations and yet  very different local field
representations (and thus inequivalent quantum structure).  When one
addresses the issue of existence of a local action the choice of the 
fundamental  fields becomes relevant.}
%This will map the original bosonic string  sigma model on $AdS_5 \times S^5$
%into a 2d Lorentz invariant   local field theory
%with an equivalent integrable structure.
We shall  follow
  the  approach of \cite{bak1,bak} (see also \ci{fernn}), in which the reduced theory
 is  interpreted as a
gauge-fixed version of a
gauged WZW theory with  a potential
 representing an integrable deformation,\foot{Viewed as a CFT deformation
  it is relevant in  compact (e.g. $S^n$) case and irrelevant in non-compact
  (e.g. $AdS_n$) case.}
  i.e.  as a special case 
 of a non-abelian Toda theory \cite{lez}.

 %%%%%%%%%%%%%%%%%%%%%%%%%%%%%%%%%%%%%%%%%%%%%%%%%%%%%%%%%%%%%%%%
 
 \bs

The  reduced model for the full \adss superstring   
(found after an appropriate kappa-symmetry gauge fixing) turns out to be a 
   2d Lorentz-covariant fermionic   generalisation of a non-abelian Toda theory
   for ${G\ov H} = {Sp(2,2)\ov SU(2)\times SU(2)} \times{Sp(4)\ov SU(2)\times  SU(2) } $
with $4+4$ dimensional bosonic target space.
Its simple structure (and the matching of the numbers of the 
 bosonic and the fermionic degrees of freedom)
suggests that it may possess 2d supersymmetry.
 Indeed, the  existence of the supersymmetry 
 can be seen directly in the special case 
 of the $AdS_2 \times S^2$ superstring theory  
 for which the reduced model   happens to be the same as the $\ns=2$ supersymmetric 
 sine-Gordon theory.

Though the relation of the reduced  model to the original conformal superstring model 
   involves a 
 non-local  transformation,  we may still expect that  
  it  should define 
  a UV finite 2d theory.
 Its  conformal invariance  is  then only  
 ``spontaneously'' broken by a scale $\mu$ (entering  the potential term
 and its fermionic counterpart) 
  that appears after fixing the residual 
 conformal diffeomorphism  freedom in the conformal gauge
 (the same  happens  in the  plane-wave  light-cone gauge case \ci{mt2}).
 If this   is indeed the case, 
 the reduced model  may  serve as
a  starting point for
understanding   the corresponding quantum \adss  superstring 
theory.\footnote{While the transformation  used to
arrive at  the reduced model is non-local
one may hope that  in an integrable  finite field
 theory the solitonic spectrum  should be  determined essentially
 by the  semiclassical  approximation \ci{dhn}
  and  it may then  be the same  in a pair of  theories with classically  equivalent
integrable structures. 
The Poisson structures
of the original and reduced models 
 are different \cite{mi12,mi3}, 
but as was shown in \ci{mi3} 
 in the light cone
formalism,  they  are actually
compatible (the sum of the two Poisson brackets
is again a Poisson bracket, i.e. it satisfies the Jacobi identity). 
%suggesting that   there is some  relation between them.
} 
Its  small-fluctuation spectrum near a natural vacuum state
 contains 8 bosonic and 8 fermionic  
dynamical degrees of
freedom  of equal mass $\mu$, and  the corresponding
relativistic (and 2d supersymmetric)   
S-matrix should have the $[SU(2)]^4$  global symmetry.\foot{Having obtained 
 the reduced model via the classical
 procedure  and  using it as a starting point for quantization
one would still need to understand how to compute
 the ``observables'' of the original
theory in terms of the  quantum reduced theory  (at the classical level
 one  can do this by
solving  the linear Lax system). In particular, one would need to compute
 the global charges of the $PSU(2,2|4)$ symmetry  group  
 as these are relevant for
  comparison with the gauge theory  side.}

\bs  

Let us  now describe   the contents   of the  paper.
We shall start in section \bref{sec:2} with a review of the Pohlmeyer 
reduction in the case of the bosonic string  models on $R_t \times S^2$ 
and $R_t \times S^3$ with
sine-Gordon  and complex sine-Gordon models as the corresponding reduced theories. 

To systematically construct the Lagrangians of  reduced models 
for higher-dimensional bosonic $SO(n,m)/SO(n-1,m)$ examples we shall first 
 explain  the relation  between 
the equations of motion  of geometrical  (``right'')  $F/G$ coset model written 
 in terms of currents  and the $G/H$  (``left-right'') gauged WZW   model (gWZW)  
  with an 
 integrable potential.
As a preparation,  we shall 
  review the classical equations of the $F/G$ symmetric-space
sigma model (sect. \bref{sec:FGcos})   and  the equations of the $G/H$ gWZW model
with a potential, i.e. of  a special case of 
the non-abelian Toda theory (sect.~\bref{sec:gWZW}).
The potential is 
 determined by a choice of an  element $T_+=T_-=T$ 
 %(which we usually set to be equal) 
 in the abelian
 subspace in the complement of the algebra $\algg$ of $G$ in the 
 algebra $\algf$ of $F$, and $H$ is   such that its algebra $\algh$ is 
 a  centralizer of $T$ in $\algg$.

In sect.~\bref{sec:red-bos} we shall show  how to relate the equations of motion of
 the $F/G$ coset model 
to those of the $G/H$ gWZW model by \ \  (i)   imposing  the so called reduction
 gauge  in the equations of  the $F/G$  model  
 written in terms of the current components, 
  and by \ (ii)   making  use of the residual  2d  conformal diffeomorphism symmetry
to eliminate an additional degree of freedom
(setting components of the stress tensor to be constant and thus satisfying the 
conformal gauge constraints of the string theory on $R_t \times F/G$). 
This will allow us
 to  solve part of the gauge-fixed
equations of motion  explicitly in terms
of a new field $g$ taking values in $G$ and the
$\algh$-valued gauge field $A_\pm$ (sect.~\bref{sec:42}). 
The resulting  system  will
turn out to be invariant under the  both  left and right
$H$ gauge symmetries. After imposing a 
special gauge condition under which  the gauge symmetry reduces
to that of the $G/H$ gWZW model  these  equations of motion
 %and the gauge-fixing conditions
 become equivalent to the ones 
 following  from the gWZW action with  a special integrable potential
   described   in sect.~\bref{sec:gWZW}.
That  the reduced equations of motion of the $F/G$
coset model can be related to those of  the  gWZW model with  an  integrable
potential was first suggested   (and checked on several examples)
in \cite{bak,fernn}. Here   we shall  explain  why this correspondence  should work in general
and  specify  the necessary conditions on the groups and the algebras involved.
We shall also note   that the potential term is equal to the  original $F/G$ coset 
Lagrangian in the reduction  gauge.

In sect.~\bref{sec:43} we shall mention the equivalence of the Lax representations 
for the $F/G$ coset and the $G/H$ gWZW models  and in sect.~\bref{sec:44}  we shall consider 
 the reduced  equations  for the $S^n= SO(n+1)/SO(n)$ coset model 
 in the $A_\pm=0$ \ci{bak}   $H$-gauge. These equations, are,  however,  non-Lagrangean
 on  physical subspace.\foot{The original observation of \ci{bak} 
 that the gWZW model with an integrable potential 
 provides a Lagrangean formulation of the reduced equations of motion of the 
 $F/G$ coset model  applied  on the extended  configuration space involving the 
 ``auxiliary'' $A_\pm$ fields. Similar construction
  was discussed in a string context in \ci{basf}.
 }

 As we shall discuss in  sect.~\bref{sec:lag}, to  get the Lagrangean equations 
  for the independent 
 $n$--1 degrees of freedom of the reduced counterpart of the $S^n$ model 
 (that generalizes the sine-Gordon and the complex sine-Gordon cases) 
   one should start  with the  gWZW action,  impose the $H$-gauge on the 
   group element $g \in G$ and  integrate out the gauge field components $A_\pm$.
 The resulting reduced  action is   that of a sigma model  with a  curved target space 
 metric (but no   antisymmetric tensor coupling) combined    with  a relevant integrable 
 potential term given universally by a cosine of  one of the  $n$--1 angles. 
 We describe few   explicit examples of reduced models for strings 
 on $R_t \times S^4$  and $R_t \times S^5$  in sect.~\bref{sec:52}. 
 The generalisation to $AdS_n \times S^n$  models is then 
 straightforward (sect.~\bref{sec:53}). 
% We shall mention  that
  %while  
  %the total potential appears to be  a function of only two of eight coordinates, 
% the {\it full} small-fluctuation spectrum near the trivial vacuum 
% is 
%  massive, containing 4+4 bosonic modes of  
% equal mass. 

In sect.~\bref{sec:spohl} we shall  turn to the \adss  superstring 
starting with   the  equations of motion 
for the ${\hat F \ov G}={PSU(2,2|4) \ov   Sp(2,2) \times Sp(4)}$ supercoset model 
(with the bosonic part $ {F \ov G}= AdS_5 \times S^5= {SU(2,2) \ov   Sp(2,2)} 
 \times { SU(4) \ov  Sp(4)}$). 
We choose  conformal gauge and   write them 
in terms of the components of the left-invariant  current of $PSU(2,2|4)$.
%\foot{The original $PSU(2,2|4)$ symmetry
% is thus ``hidden'' 
%in the reduced model formulation.} 
We use  the formulation   based on  
 $Z_4$  grading property   \ci{berk,bpr}   of
 the  superalgebra $psu(2,2|4)$.
 Fixing  a particular kappa-symmetry gauge we perform the analog of
  the Pohlmeyer reduction 
 discussed earlier for the similar  bosonic  cosets.
 An important  ingredient 
  is a generalization to the $psu(2,2|4)$ superalgebra  case 
of the Lie algebra decomposition originally used in~\cite{ei}
 in the bosonic  coset case.
 
 Introducing the  new  set of  fermionic variables   directly related 
 to the  odd components of the supercoset current
 we  show   in sect.~\bref{Lorentz}  that the reduced  system of equations 
  follows from  a 2d Lorentz-invariant Lagrangian  \rf{L-tot}. 
   Its bosonic part  is that of ${G\ov H}
   = {Sp(2,2) \ov SU(2) \times SU(2) } \times 
  { Sp(4)\ov SU(2) \times SU(2)}$ 
   gWZW model with an integrable potential 
   determined by a special diagonal matrix $T=T_\pm$ in the  even part of the 
 $psu(2,2|4)$  superalgebra. 
   In addition,  the Lagrangian  contains  a quadratic  fermionic part 
     with a  standard
  first-derivative kinetic term. The fermions interact
      ``minimally'' with  the  $H$ gauge field $A_\pm$ and  are  also coupled
     (by a ``Yukawa-type''  term)  to the bosonic field  $g \in G$. 
     We mention that as in the bosonic case, the sum of the  $\mu$-dependent 
     potential  and  ``Yukawa'' interaction terms 
     in the  reduced Lagrangian is equal to the  original superstring Lagrangian 
     written in terms of currents.

   %  An    explicit   gauge-fixed form of the reduced action in terms 
   %  of the dynamical fields only 
    %can be  found by integrating out  $A_\pm$,  and thus  contains  also a 
   %  quartic fermionic term.
    The   vacua of the theory are  described  by constant 
     $g$ taking  values in $H$; in the $A_\pm=0$ gauge 
      the small-fluctuation spectrum  near the trivial vacuum 
     consists  of 
      8 bosonic and 8   fermionic dynamical modes  of  the same  mass $\mu$.
      We comment on the interpretation of the parameter $\mu$ and 
      mention that the corresponding scattering matrix should have a global 
      $H=[SU(2)]^4$ symmetry.

 The structure of the reduced action suggests the presence  of a  2d  supersymmetry. 
 Its existence  is indeed  confirmed in sect.~\bref{sec:ads2s2} on the example 
 of a similar  $AdS_2 \times S^2$   superstring model based on the  $psu(1,1|2)$  
 superalgebra. The corresponding reduced Lagrangian is found to be the same as that of 
 the $\ns=2$ supersymmetric extension of the sine-Gordon model.
 
 There are also several  Appendices  containing  some technical details and definitions.

\iffalse

\fi

%%%%%%%%%%%%%%%%%%%%%%%%%%%%%%%%%%%%%%%%%%%%%%%%%%%%%%%%%%%%%%%%
\section{Examples of reduced  models:
 strings in $R_t \times S^2$ and $R_t \times S^3$}\label{sec:2}
%%%%%%%%%%%%%%%%%%%%%%%%%%%%%%%%%%%%%%%%%%%%%%%%%%%%%%%%%%%%%%%%%%%%

Let  us begin  with a review  of the
prototypical example:
reduction  of the
$S^2$ sigma model to   the sine-Gordon model \cite{pol}.
Starting with the action of the    sigma model on the sphere
written  in terms of the  embedding coordinates
$S=  \frac{1}{ 4 \pi \a'} \int d^2 \s \ L $  where  ($\d_\pm = \d_0 \pm \d_1$)
\begin{equation}
 \label{lj}
  L= \d_+ X^m \d_- X^m - \Lambda ( X^m X^m -1 ) \ , \ \ \ \ \ \ \  m= 1, 2,3 \ ,
\end{equation}
we get for the classical equations of motion 
\begin{equation}\label{jp}
  \dpp \dmm X^m + \Lambda X^m =0  \,, \qquad \Lambda = \dpp X^m \dmm X^m\,, \qquad
 X^m X^m =1 \,.
\end{equation}
Then the stress tensor satisfies
\begin{equation}
 \TT_{+ -}=0 \,, \quad \dpp \TT_{--}=0\,, \quad \dmm \TT_{++} =0 \,, \quad 
 \TT_{\pm \pm } = \d_\pm X^m \d_\pm X^m,
\end{equation}
so that  $\TT_{++} = f(\sigma_+), \ \ \TT_{--} = h(\sigma_-)$.
 Since the theory is classically conformally invariant
one can  apply  conformal transformations  to  put $\TT_{\pm \pm }$ into the  special
constant  form
\begin{equation}\label{coni}
\dpp X^m \dpp X^m = \km^2\,, \ \ \ \ \   \quad \dmm X^m \dmm X^m = \km^2 \,,\ \  \quad \km=\const\,.
\end{equation}
This effectively fixes one of the two  fields of $S^2$
 leaving us  with a one-dimensional ``reduced''  theory.
Indeed, one can introduce   a new field variable $\varphi$
 via the following  non-local transformation $X_m \to \vp$
\begin{equation}
\label{new}
\km^ 2  \cos 2 \varphi  =  \dpp X^m \dmm X^m\,.
\end{equation}
Then the equations   for $X^m$   \rf{jp} 
and the conditions \eqref{coni}
are  solved provided $\varphi$ is subject  to the sine-Gordon (SG) equation
$ \dpp\dmm \varphi + \frac{\km^2}{2} \sin 2\varphi=0$. The latter follows
from
\begin{equation}
\label{red}
\td L = \dpp \varphi \dmm \varphi  + \frac{\km^2}{2} \cos 2\varphi\ , 
\end{equation}
which is thus  the Lagrangian of the ``reduced'' theory.
The classical solutions and integrable structure (Lax pair, etc.)
of the original sigma model and its  reduced counterpart are then directly related.

This reduction from sigma model on $S^2$  to the SG theory has also an
equivalent interpretation
as a classical  equivalence  between  the
bosonic string theory  in $R_t \times S^2$
in  a special gauge and the SG theory.
 Indeed, starting with the Polyakov string action   containing  the time direction term
 $-\d_+ t \d_- t$  in addition to the $S^2$ term  \rf{lj}  and
  choosing the {\it conformal  gauge}
  combined with
 $t = \km \tau$ (to fix the residual  conformal reparametrisation symmetry)
 we end up with the same  conditions  \rf{coni},
  now interpreted as the conformal gauge (Virasoro) constraints.
  Then the classical string equations
 on $R_t \times S^2$  become equivalent  to the SG equation for the one remaining
  ``transverse'' degree of
 freedom parametrized by $\varphi$
  (the gauge conditions eliminate 1+1 out of 1+2  string  degrees
 of freedom).

One interesting outcome  of the above reduction is that while the  conditions
\rf{coni} obviously violate the 2d Lorentz invariance of the original theory
($t= \km \tau$  ``spontaneously breaks''  the 2d 
Lorentz invariance in the string-theory version of the reduction),
the resulting SG  theory   is still Lorentz invariant.
Note also that the   $SO(3)$  global symmetry
 of the original model \rf{lj} becomes  trivial   in the  reduced model: $\varphi$
 defined in \rf{new} is $SO(3)$ invariant.
 Given a SG solution for $\varphi$  and thus a specific value of the Lagrange multiplier 
  function
 $\Lambda=\km^ 2  \cos 2\varphi = \dpp X^m \dmm X^m$  in \rf{jp}
 one can  reconstruct  the corresponding solution for $X_m$ by solving the  linear
 equation $\dpp \dmm X^m + \Lambda X^m =0$.\foot{To find periodic solutions on $R \times S^1$ 
 one would need to start with a periodic solution of SG model 
 and  also  impose periodicity
  on $X_m$ in solving the linear system.}
 For a given  solution for $X_m$  one can then  find   the
 corresponding $SO(3)$  conserved   charges.
 Thus the classical solitonic spectra  of the two models should be in
 direct  correspondence
 (see \cite{hm,dor,oka} for some  specific examples).

%%%%%%%%%%%%%%%%%%%%%%%%%%%%%%%%%%%%%%%%%%%%%%%%%

 This classical  equivalence
 relation obviously breaks down in  quantum theory  where  there
  are UV divergences  and   mass generation in
 the $S^2$ sigma model so that  the  classical conformal invariance is broken
 (invalidating, in particular,  the argument leading to \rf{coni}).
 Still, one may  hope that  an analog of this reduction may extend to the quantum level 
in the case 
%where the $R_t \times S^2$ model is embedded into a bigger
of a  theory like \adss  
superstring which remains conformally invariant upon quantisation.

 \bigskip

 The above reduction has a straightforward generalisation to the case when
 $S^2$ is replaced by $S^3$  \cite{pol,lund}. The  reduced  model  corresponding to the string
 on $R_t \times S^3$ is the  complex sine-Gordon (CSG) model
 \begin{equation}
 \label{csgi}
   \td  L =
   %{ 1 \over 2}  (
   \dpp \varphi \dmm \varphi  +   \tan^2 { \varphi }\ \dpp \theta \dmm \theta
 + \frac{\km^2}{2} \cos 2\varphi \,.
\end{equation}
The variables $\varphi$   and $\theta$  are expressed in terms of  the $SO(4)$ invariant combinations of derivatives  of the original variables  $X_m$ ($m=1,2,3,4$)
\begin{equation} \label{kpl}
 \km^ 2  \cos 2\varphi = \dpp X^m \dmm X^m \ , \qquad
\km^3 \sin^2 {\varphi }\  \del_\pm  \theta
= \mp \frac{1}{2} \epsilon_{mnkl} X^m \dpp X^n \dmm X^k \d_\pm^2 X^l
\,.
\end{equation}
Again, the integrable structures and the soliton solutions of the two models are closely
related (see \cite{dor,oka}).\foot{Let us mention that an alternative reduced theory for the 
$S^3$  sigma model  formulated in terms of currents that also solve the Virasoro
conditions for a string on  $R_t \times S^3$   was discussed by Faddeev and Reshetikhin 
 \ci{far,zaz}. However, the FR model is not  manisfestly 2d Lorentz invariant  
 and thus appears to be less useful than  the corresponding Pohlmeyer-reduced
  theory, i.e. the CSG. The precise relation between the two models is worth further study.} 
The CSG model  can be  interpreted as a special
  case of a non-abelian Toda theory \cite{lez}
-- a massive integrable perturbation  of a gauged (coset)  WZW model
(here $SO(3)\ov SO(2)$ model) \cite{park}.\foot{The corresponding quantum S-matrix 
was discussed in \ci{doho}.}

%related to massive Thirring model via 1st order form
%of classical eqs:  $\psi = (u,v)$
%$i \dp u + v - |u|^2 v=0, \ i \dm  v + u - |v|^2 u=0$,
%quantum S-matrix is known  (Dorey, Hollowood)

Reduced equations of motion  for sigma models on 
higher spheres $S^n$ ($n=4,5, ...$)
 involve field variables related to  $SO(n+1)$  invariants built 
out of $X_m$ and its  higher derivatives $ \d_\pm X_m,$ $ \d^2_\pm X_m,$ $ \d^3_\pm X_m, ... $
(with indices contracted using   $\delta_{mk}$ and  $\epsilon_{m_1... m_{n+1}}$); they 
were found in \cite{polr} (see   also \cite{pole,add}).
The resulting equations  were not, however, derivable from a local Lagrangian.

It was later shown in \cite{bak}
that they can be obtained as a  particular gauge-fixed version
of the  classical equations of the $SO(n)\ov SO(n-1)$  gauged
WZW model with an  integrable potential
term.
 This  provided a Lagrangean formulation of these equations on the {\it  extended}
field space including  the 2d gauge field  $A_\pm$ of the  gWZW  model.

This construction  gives   a   strong indication that
there should exist an 
%equivalent 
alternative 
  version  of the classical reduced equations of motion 
which is {\it manifestly} Lagrangean,
i.e.  that  can be derived from an action containing 
only physical ``reduced'' set of fields as was 
 found  in the previous cases of the SG and CSG models.

The reason  for this  expectation   
is that the classical equations  written in the Lax-pair form
admit different ``gauge-equivalent'' \cite{mz}
 versions related by (non-local) field  redefinitions.\foot{
This is a  classical gauge equivalence when
 gauge transformations at the level of Lax equations lead to equivalent
 integrable systems. The resulting  non-local relation at the level of field theory models 
  does  not,  in general,  extend to the  quantum level,
 cf. \cite{nappi,frid}.}
 This  was  already noticed in \cite{add} in the $S^3$ case  where  the field variables
 corresponding to the CSG model were related by a non-local transformation
 to the variables of the reduced model of \cite{polr}.

 \bigskip

Below we shall present an  explicit  form of the reduced  Lagrangian models
for the string on $R_t \times S^4$  and $R_t \times S^5$; 
the  $AdS_n$ versions can be found by an analytic continuation.
One is then able to write down the  reduced Lagrangian for the bosonic part of the 
\adss theory. 
The basic idea  is to follow \cite{bak} and start   with the
$SO(n)\ov SO(n-1)$  gWZW model with a relevant  integrable perturbation term
but instead of fixing the gauge field  $A_\pm =0$ as in \cite{bak}
 fix the gauge on the group element  and { integrate
 out} the gauge field $A_\pm$
as in \cite{old,cres,fl,bs1} (see also \ci{fernn,mira}). 
In the case of the $SO(3)\ov SO(2)$ (or equivalently $SU(2)\ov U(1)$) model that procedure
immediately explains the appearance of the familiar  $D=2$ target space 
metric in the CSG action \rf{csgi} as was originally observed in~\cite{pars}.

\bs

The   construction of the reduced models
  based on the conformal gauge and fixing
the remaining conformal transformations by $t= \km \tau$ condition
was applied  above to a
string on  $R_t \times S^n$.  The same  can be done  for the 
bosonic string  model on $AdS_n \times S^1$  in conformal gauge and with 
fixing  the residual conformal symmetry 
choosing the $S^1$ angle $\alpha$  equal to $ \km \tau$. 
Denoting the   embedding coordinates 
of $AdS_n$ as $Y_s$  (with  $Y^s Y_s = - Y^2_0 - Y^2_{-1} + Y^2_1 + ...+ Y^2_n= -1$)
 the $AdS_n$ Lagrangian   is then the analog of \rf{lj}
 \be \la{adsa}
 L= \del_+  Y^s \del_-  Y_s   - \td \Lambda ( Y^s Y_s +1)  \ , \ee
 with the equations of motion and conformal gauge constraints  being 
 \bea 
 &&\del_+\del_-  Y_s  +\td \Lambda  Y_s=0, \ \  \ \ \ \ \td  \Lambda= - \del_+  Y^s \del_-  Y_s \
 , \  \ \ \ \ \   Y^s   Y_s   =-1     \
 , \la{vn} \\   \la{vnm}
&& \del_+  Y_s \del_+  Y^s= - \km^2 \ , \  \ \ \ \ \ \ 
  \del_-  Y_s \del_-  Y^s= - \km^2 \ . \  \ 
\eea
 By concentrating  on the plane  formed  by the normalized  vectors $\del_+  Y^s$ 
 and $\del_-  Y^s$  (orthogonal to $Y^s$)  one can see that their scalar product 
 can be set equal to 
 \be \la{scn}
   \del_+ Y^s \del_- Y_s = - \mu^2 \cosh 2 \phi     \   , \ee
where $ \phi$  is a new variable (cf. \rf{new}). 
Then in the   $AdS_2$  case  we get $\del_+ \del_- \phi + { \km^2 \ov2} \sinh 2 \phi =0$
which follows from the reduced Lagrangian (cf. \rf{red}) 
\be
\td L= \del_+\phi\del_- \phi - {\km^2 \ov 2} \cosh 2 \phi  \ . \la{hasg} 
\ee 

\bs 
Let  us now  explain how the above  special examples 
 can be generalized to the case of the bosonic string on 
$AdS_n \times S^n$. Denoting the   embedding coordinates 
of $AdS_n$ as $Y_s$ and the coordinates of $S^n$ as $X_m$  
the conformal gauge condition means
the vanishing of the total stress tensor,
\be \la{too}
\TT_{++} (Y) + \TT_{++} (X) =0 \ , \ \ \ \ \ \  \ \ \ \
\TT_{--} (Y) + \TT_{--} (X) =0 \ . \ee
Since in the conformal gauge the equations of motion for
 $Y_s$ and $X_m$ factorize,
the corresponding  stress tensors are separately traceless and conserved.
%Then also the equations of motion for $AdS_n $ and $ S^n$ parts decouple.
Then instead of using $t= \km \tau$ or  $\alpha = \km \tau$ conditions ($t$ is
now  part of $AdS_n$  and  $\alpha$  -- part of $S^n$)
we can     fix the residual   conformal transformation  freedom
 ``implicitly''
 by following \ci{pol}  and 
 demanding as in \rf{coni}  that $\TT_{\pm \pm } (X) =\km^2=\const$.
Then \rf{too} implies  that
\be \la{soo}
\TT_{\pm \pm } (X) =\km^2 \ ,\ \ \ \ \ \ \
\ \ \ \
\TT_{\pm \pm } (Y) =-\km^2 \ . \ee
We thus get two decoupled
 $AdS_n$ and $S^n$ sigma models  with the   constraints  \rf{soo}, 
 to which we can separately apply the Pohlmeyer's 
  reduction procedure.
 That  eliminates 1+1  out of $n+n$ degrees of freedom, leaving us with
 an action for  only the $(n-1) + (n-1)$ 
 physical degrees of freedom.
 
\bs
 Later in section 6  we shall discuss a generalisation of this
 reduction  procedure to  the presence
 of the superstring  fermions when the $AdS_n$ and $S^n$ 
  parts are no longer decoupled.
% will no longer factorize.

%%%%%%%%%%%%%%%%%%%%%%%%%%%%%%%%%%%%%%%%%%%%%%%%%%%%%%%%%%%%%%%%%%%%%%%%%%%%%%%%%%%%%%
\section{Coset sigma model and the corresponding
 gauged WZW model with an integrable potential}

Let us give   a  short review of
 a  coset sigma model  (of which $S^n$ model is a special case) 
 and the associated gauged WZW model.
This will set up the notation for section \bref{sec:red-bos} where we
 are going to construct an explicit
change of variables which relates 
%an appropriately gauge-fixed
the  $F/G$
coset sigma model to  certain  
$G/H$ gauged WZW model with a
 potential,  giving  an  explicit realisation
of the relationship originally proposed  in~\cite{bak}.
%In section 5  we shall discuss how to find the Lagrangian
%for the

%%%%%%%%%%%%%%%%%%%%%%%%%%%%%%%%%%%%%%%%%%%%%%%%%%%%%%%%%%%%%%%%%%
\subsection{$F/G$ coset sigma model}\label{sec:FGcos}

Let  $G$ be a subgroup  of a Lie group  $F$    and   $M=F/G$ be 
a coset space. Let us  assume that
the Lie algebra  $\algf$  of  $F$
is equipped with a positive-definite
invariant bilinear form $\inner{\,\, }{\,\, }$; explicitly,  let $F$ be
a matrix group and $\inner{a}{b}=\tr(ab)$. In addition let $F/G$ be a symmetric
space  which  is the case when
\be \la{sym}
\algf= \algp \oplus \algg \ , \qquad
 \commut{\algg}{\algg}\subset \algg \,, \qquad
\commut{\algg}{\algp}\subset \algp \,, \qquad 
\commut{\algp}{\algp}\subset \algg \ , \qquad
\ee
 where $\algp$ denotes the orthogonal
complement of the algebra  $\algg$  of $G$ in  $\algf$.
%    with respect to $\inner{\cdot}{\cdot}$.

The action of the sigma model on $F/G$ is given by
\begin{equation}
\label{g}
S=-\half\int d^2\sigma \  \eta^{ab}\ \tr(P_a P_b)\,, \qquad
P_a= (\gf^{-1}\d_a \gf)_\algp \ ,
\end{equation}
where $(...)_\algp$ denotes the orthogonal projection to $\algp$, i.e.
\be \la{cur}
J=\gf^{-1}d\gf=\A+P \ ,   \ \ \ \ \ \ \ \ \ 
\A= J_\algg \in\algg\  ,  \ \ \ \ \ 
P=J_\algp \in\algp\ . 
\ee
The action
is invariant under the $G$ gauge transformation $\gf \to \gf \gg$ for an arbitrary
$G$ valued function $\gg$. Indeed, under this transformation
$
J=\gf^{-1}d\gf \to \gg^{-1}(\gf^{-1}d\gf)\gg+\gg^{-1}d\gg
$
so that $P$ transforms into $\gg^{-1}P \gg$ ensuring the invariance
of the Lagrangian. The current $J$ and therefore the action is also invariant under the global
$F$  symmetry $f\to f_0f$ for any constant $f_0\in F$.
 Furthermore,
 the classical coset sigma model action is invariant under the 2d conformal
transformations.

The equations of motion take the form
\begin{equation}\la{eeq}
D_a P^a=0\,,\qquad \ \ D_a=\d_a +\commut{\A_a}{     \   \ }\,, \quad\ \ \ \
\A_a=(\gf^{-1}\d_a \gf)_\algg\,.
\end{equation}
Using  the light-cone  coordinates $\sigma^+,\sigma^-$ they
 can  also be written as
\begin{equation}
\label{efg}
D_+P_-=0 \ , \ \ \ \ \ \ \ \   \quad
 %\text{or, equivalently,} \quad
D_-P_+=0\,.
\end{equation}
Indeed,  the zero curvature condition
for the  current  $J$  projected to $\algp$
implies
\begin{multline}
\left(\d_+J_- -\d_-J_+ +\commut{J_+}{J_-}\right)_\algp
=~\d_+P_- -\d_-P_+ + \commut{\A_+}{P_-}+\commut{P_+}{\A_-}=0\,,
\end{multline}
i.e. $D_+ P_-  - D_- P_+ =0$.  This together with \rf{eeq}, i.e.
$D_+ P_-  + D_- P_+ =0$,  then leads to \rf{efg}.\foot{Note that the global right $F$-symmetry is not seen at the level of
equations of motion written in terms of currents  because
all the currents are explicitly invariant.}

The nonvanishing components of the stress-tensor are
\begin{equation}\la{st}
\TT_{++} =-\half\tr({P_+P_+})\,, \qquad \TT_{--}=-\half\tr({P_-P_-})\,.
 \end{equation}
Equations of motion imply the  conservation law
$
\d_-\TT_{++} =0\,,  \ \  \d_+\TT_{--}=0\,.
$
Then making an  appropriate conformal transformations one
can  always set as in \rf{coni} \ 
$\TT_{\pm\pm}=\km^2$.

The Lax representation for the coset sigma model is found from the 
zero curvature condition $d\omega+\omega\wedge\omega=0$ for the Lax connection
\begin{equation}\la{lxx}
\omega=d\sigma^+(\A_++\lam P_+)+d\sigma^-(\A_-+  \lam^{-1}  P_-)\, ,
\end{equation}
i.e.
\begin{equation}
\label{lax}
\commut{\d_++\A_++\lam P_+}{\d_-+\A_-+\lam^{-1}   P_-}=0\,,
\end{equation}
where $\lam$ is a spectral parameter. The equations of motion
\eqref{efg} follow from \rf{lax}  as the coefficients
 of order  $\lam^{-1}$ and  $\lam$ terms.
 The coefficient of the order $\l^0$ term is the  $\algg$-component of the zero curvature condition
for the connection $J=\A+P$.

Let us recall  also  two representations of  the Lagrangian   of
 the $F/G$ sigma model. One is to introduce
an  explicit parametrisation of the coset $M=F/G$ as embedded into
 $ F$. If $x^i$ are coordinates on $M$,  let $dx^i J^*_i$ be a pullback
 of  $J$ to $M$. Then the Lagrangian in  \rf{g}   takes the
form
\begin{equation}
L=-\half \eta^{ab}\d_a x^i \d_b x^j   \ G_{ij} (x)
\ ,\ \ \ \ \ \  \ \ \ \  G_{ij} (x) =\tr(J^*_i(x)J^*_j(x)) \ ,
\end{equation}
where $G_{ij}$   is the metric on the coset space. Note that
by choosing a particular parametrisation of the coset we have fixed the $G$  gauge
symmetry.
An  alternative form of $L$   is found by  introducing  a
gauge field $\AA_a \in \algg $ which serves to implement the projection
of the $\algf$-current   on  $\algp$
 \begin{equation}
\la{jaa}
L=-\half\eta^{ab}\tr[  \gf(\d_a+\AA_a)\gf^{-1}  \ \gf(\d_b+\AA_b)\gf^{-1}]\,,
\end{equation}
or,  equivalently,
\begin{equation}\la{kio}
L=-\half\eta^{ab}\tr[(\gf^{-1}\d_a \gf-\AA_a)\ (\gf^{-1}\d_b \gf-\AA_b)]\ .
\end{equation}
 Substituting the  equation of motion for $\AA$ 
\begin{equation}\la{kkio}
\AA=\A =  (\gf^{-1}d \gf)_\algg 
\end{equation}
 into \eqref{jaa}
one returns back to the  original Lagrangian in~\eqref{g}.

%%%%%%%%%%%%%%%%%%%%%%%%%%%%%%%%%%%%%%%%%%%%%%%%%%%%%%%%%%%%%%%%%%%%%%%%%%%%%%%%%%%%%%%%%5

\subsection{$G/H$ gauged WZW model with an integrable  potential}\label{sec:gWZW}

%%%%%%%%%%%%%%%%%%%%%%%%%%%%%%%%%%%%%%%%%%%%%%%%%%%%%%%%%%%%%%%%%%%%%%%%%%%

As was suggested in  \cite{bak} (see also \ci{fernn}), 
a sigma model on a symmetric space $F/G$ can be reduced
 to a ``symmetric space sine-Gordon''   model
with a Lagrangean formulation in terms of the $G\ov H$ left-right
symmetrically  gauged WZW model
with a   gauge-invariant  integrable potential.\footnote{This is a special case of
a non-abelian Toda  theory \ci{lez}.
 Non-abelian Toda  models are 
  of the  two   basic types -- ``homogeneous  sine-Gordon''
  and
``symmetric space sine-Gordon'' \ci{fernn}.
 For the  first type the gWZW part of the Toda model   corresponds to
   $G\ov [U(1)]^r$ ($r$ is a rank of $G$).
 The   models of the second type 
 are reduced theories associated to  sigma models on
 compact symmetric spaces. They are quantum-integrable but their  S-matrix is not known,
  except for special
 cases of SG and CSG models.  A review can be found in \ci{fern}.}

 The
potential is  determined by a choice of two
 elements $T_+, T_-$ in the maximal  abelian subspace  $\alga$ in
the  complement  $\algp$ of the Lie algebra $\algg$ of $G$ in the algebra $\algf$
of $F$. The
  algebra $\algh$ of the subgroup $H$ of $ G$ should be the  centralizer
 of $T_\pm$ in $\algg$:  $[ \algh, T_\pm]=0$.
 Then  the action  is
 \be \la{pi}  S_\km (g, A)  = S_{\rm gWZW} (g, A) - \km^2 \int { d^2 \s \over 2\pi }
  \ {\rm Tr} (T_+ g^{-1} T_- g ) \ , \ee
 where $S_{\rm gWZW}$  is the
  action of the left-right symmetrically gauged WZW model  \ci{gwzw}
  (we omit an overall  level $k$  factor)
\begin{multline} \la{gaui}
S_{\rm gWZW}  = -\int\frac{d^2\sigma}{4\pi}\Tr(g^{-1}\d_+g g^{-1}\d_-g) + \int\frac{d^3\sigma}{12\pi}
\Tr(g^{-1}dg g^{-1}dg g^{-1}dg)\\
 -~\int { d^2 \s \over 2\pi } \Tr \big(  A_+\,
 \d_- g g\inv -
 A_- \,g\inv\del_+ g   - g\inv A_+ g  A_-  + A_+ A_- \big)\,.
\end{multline}
   Here $g\in  G$ and $A_{\pm}  \in  \algh$ (all the  fields are
    assumed to be matrices in a given representation of $F$ or of  its Lie algebra $\algf$).
 % Later  we shall 
 %assume that the  maximal Abelian subspace $\alga\subset\algp$ is one dimensional, i.e.  choose
%$T_+=T_-$ as its  basic element.

Note that using  Polyakov--Wiegmann identity  the action 
 \rf{gaui}  can be written also in the following form 
\begin{gather}
 \la{wzw}
    S_{\rm gWZW}  = S_{\rm WZW}(h^{-1} g h') - S_{\rm WZW}(h^{-1}  h') \,,  \\ 
    A_+ = h^{-1} \d_+  h \ ,  \qquad   A_- = h'^{-1} \d_-  h' \,.
\end{gather}
 To define the action with $T_\pm$ belonging to the algebra of $F$  it
is assumed that $g \in G$ is trivially (diagonally) embedded  into $F$.
The action is then invariant under the    vector  gauge transformations
with parameters taking values in  $H$:
\be\la{gkq}
g \to  h g h\inv\  , \ \ \ \ \ \ \
 \ A_a \to h  (A_a  + \del_a)  h\inv\ , \ \ \ \ \ \ \ \ \     h  \in  H \ , \ee
where  $A_a \in \algh$  and   $h\inv T_\pm h = T_\pm$ (since $[\alga,\algh]=0$).

The equations of motion following from \rf{gaui} are
\bea
%gin{multline}\quad
 &&\quad \d_-(g^{-1}\d_+g~+~g^{-1}A_+ g)-\d_+A_- 
  \nonumber \\
 && \ \ \ \ \ \ \ \ 
 +\ 
\commut{A_-}{g^{-1}\d_+g+g^{-1}A_+ g}+\km^2\commut{g^{-1}T_-g}{T_+}=0\,,
\label{weom} \\
%\eea
%\end{multline}
%\begin{equation}
&& \ \ \quad A_+=(g^{-1}\d_+g + g^{-1}A_+ g)_\algh
\,, \qquad A_-= (-\d_-gg^{-1}+gA_-g^{-1})_\algh
\label{lwc}\,. 
\eea
Note that  $g^{-1}T_-g \in \algp$ so that $\commut{T_+}{g^{-1}T_-g}\in \algm$,
where $\algg= \algm \oplus \algh$.
%(see \cite{\eilect} for commutation relations of $\alga,\algn,\algm$, and $\algh$).
In particular, the $\algh$-component of
 the first equation implies  that $A_a$ is flat,
\be \la{fla}
\d_+ A_- - \d_- A_+   + [A_+ , A_-] =0   \ .
\ee

\bs 

Let us  comment on the classical  integrability of the above model \rf{pi}.
It is well known that the equations of motion of the standard WZW model can be
written in the Lax form.
% ensuring the classical integrability of the system.
The same also applies to gauged WZW model with the  above potential.
More precisely, using $[A_a, T_\pm]=0$  one can show that equation \eqref{weom} 
can be written in the Lax form, i.e. it 
follows from  $\commut{\cL_+}{\cL_-}=0$ where ($\lam$ is a spectral parameter)
  \be \la{laxi}
\cL_+ =
 \d_+  + g^{-1} \d_+ g  + g^{-1}  A_+ g  + \lam \km  T_+ \ , \qquad
 \cL_-=   \d_- + A_-    +  \lam^{-1} \km g^{-1} T_- g \,,
\ee
or, equivalently,  from the zero curvature equation for the 
$\algf$-valued Lax connection
\begin{equation}
\label{lw}
\omega=
d\sigma^+(g^{-1}\d_+g+g^{-1}A_+ g+\lam \km T_+)
+d\sigma^{-}(A_-+\lam^{-1}{\km}g^{-1}T_-g)\ . 
\end{equation}
While the  remaining equations \eqref{lwc}  (constraints) 
do not follow from this condition, they may be considered as
consequences of \eqref{weom} in the sense that  given a solution to \eqref{weom} 
one can find a gauge transformation such that the transformed solution  
satisfies \eqref{lwc}. 

This is possible because eq.  \eqref{weom} has a
{\it larger  gauge symmetry}  than the original gWZW model
\rf{gaui}: it is invariant under the  $H\times H $ gauge  symmetry 
\begin{equation}
\label{gs}
g \to h^{-1}g \bar h\,, \quad \ \ \ \
A_{+}\to h^{-1}A_+h +h^{-1}\d_+ h\,, \quad\ \ \ \
 A_{-}\to {\bar h}^{-1}A_{-}{\bar h}+{\bar h}^{-1}\d_- \bar h\,,
\end{equation}
where $h$ and $\bar h$ are two arbitrary $H$-valued functions.
The symmetry of \rf{gaui} is the diagonal subgroup (with $h= \bar h$) of
 the extended ``on-shell'' gauge symmetry \eqref{gs}.
It turns out that using this extended symmetry one can fulfil the
constraints \eqref{lwc}. Further details and the 
%explicit 
proof are relegated to the Appendix~\bref{sec:l2wzw}.
We shall also use this observation 
in section \bref{sec:red-bos} below.

%\maxim{
Let us note also  that given an 
automorphism $\tau$ of the  algebra $H$ preserving the trace   one
 can fix the $H\times H$ gauge symmetry of the equations of motion
 in a more general way  so that \eqref{lwc} is replaced by 
 %by requiring that 
\begin{equation}\la{ger}
\tau(A_+)=(g^{-1}\d_+g + g^{-1}A_+ g)_\algh\,, 
\qquad A_-= (-\d_-gg^{-1}+g\tau (A_-)g^{-1})_\algh \ . 
\end{equation}
The corresponding equations \rf{weom},\rf{ger}
then follow  from 
 the Lagrangian \rf{pi},\rf{gaui}
with the replacement 
 \be\la{tut}
    A_-   \ \ \to \ \  \tau(A_-)    \ee
      in the  $A_- g^{-1} \del_+ g $ and the $g^{-1}A_+g A_-$ terms. 
The corresponding 
 gauge
symmetry  is then $g \to h^{-1}\,g\,\hat\tau(h)$ where $\hat\tau$ is 
a lift of $\tau$ from $\algh$ to $H$ (see  \ci{fernn,mira}). 
In this case the left-right symmetrically gauged WZW model is thus replaced 
by a more general asymmetrically gauged WZW model \ci{bs1,qua}.
%}

\bigskip

It was observed in \ci{bak} that since 
 the field strength of $A_a$ vanishes \rf{fla} on the equations of motion, 
 one can choose a   gauge where\foot{This gauge is thus possible
only on-shell; to  gauge away $A_a$ at the   level of the gWZW  Lagragian
 one would need some  additional local  gauge invariance.}
\be \la{az}
A_+=A_- =0 \ .  \ee
 Then   the  classical equations \rf{weom},\rf{lwc}
  reduce to 
\begin{align}
  \la{ol}
&\d_- ( g\inv \d_+ g) - \km^2 [T_+, g\inv T_- g]=0 \ ,
\\
&  ( g\inv \d_+ g)_\algh =0\ , \qquad\quad
(  \d_- g g\inv)_\algh =0\ . \la{olk} 
\end{align}
These equations happen  to be  equivalent
%(after a field redefinition,  when 
% expressed in terms of $g\inv T_- g$)
 to the equations of motion of the 
  reduced $F/G$  model found in \ci{pole,ei,add}.

Various  special cases, structure of vacua and solitonic solutions of the
equations \rf{ol},\rf{olk} were  discussed in \ci{fern,mira} and refs. there.

\bs

The set of equations \rf{ol},\rf{olk} do not  directly   follow  from a 
local Lagrangian.
As was implied   in \ci{bak},  to get a
local  Lagrangian  formulation of these  equations  one is
to go back to the action \rf{gaui}
on  a bigger configuration space involving  both  $g$ and $A_a$  
with the   gauge invariance  \rf{gkq}.

At the same time, one would like also  to have 
%This   is not, however,   totally  satisfactory,
% as one would prefer to  have 
 a reduced  action involving  only the
 independent degrees of freedom, i.e.  generalizing 
the actions  of the  SG \rf{red}  and the  CSG  \rf{csgi} models.

%\maxim{probably one should say just ``totally satisfactory'' or even weaker}

\bs

%%%%%%%%%%%%%%%%%%%%%%%%%%%%%%%%%%%%%%%%%%%%%%%%%%%%%%%%%%%%%%%%%%%%%%%%%%%%%%

Below in section \bref{sec:red-bos} we shall explain 
 why  and under which conditions
the  relation between the equations of the
 reduced theory  corresponding to the $F/G$  coset model 
  and the equations of the  $G/H$ gWZW model
  proposed  in \ci{bak} actually  works.
  Then in section \bref{sec:lag} we shall  
  suggest  how  to use this correspondence
  to find a local Lagrangian  for the physical number  of  
  degrees of freedom of the reduced model.

The  main observation  will be  that there exists an  equivalent
representation for the classical equations  following  from   \rf{pi}
(or gauge-equivalent,  
in the sense of
\ci{mz},  representation of the   Lax equations corresponding to 
 \rf{laxi}) 
in which they  admit an explicit Lagrangean formulation  without any residual
gauge invariance, thus generalizing   the 
  SG and CSG  examples. 
  Instead of the ``on-shell'' gauge $A_a=0$ used in \ci{bak} one can 
   impose an ``of-shell'' $H$-gauge   on the
group element $g$  and then solve for  the  gauge field
$A_a$. ``Integrating out''  $A_a$  then   leads to
a  sigma model  for the independent  dim($G/H$)  number of parameters in $g$
in the same way
as in the examples  of  conformal sigma models  
associated to  gWZW  models
  \ci{old,fl,bs1}.\foot{Integrating out the  gauge field
at the quantum level induces also a dilaton  \ci{old};
 there are also quantum  $\alpha' \sim 1/k$ corrections to the sigma model
background fields \ci{dvv,bs3,aat}. These will be  ignored at the classical level
we are restricted to   here.}

%%%%%%%%%%%%%%%%%%%%%%%%%%%%%%%%%%%%%%%%%%%%%%%%%%%%%%%%%%%%%%%%%%%%%%%%%%%%%%%%%%%%%%%%%%%%
\section{Reduced theory for  $F/G$ coset sigma model:{\space} \\
equations of motion}
\label{sec:red-bos}
%%%%%%%%%%%%%%%%%%%%%%%%%%%%%%%%%%%%%%%%%%%%%%%%%%%%%%%%%%%%%%%%%%%%%%%%%%%%%%%%%%%%%%%%%%%%%

The   strategy to  relate the equations of motion of the $F/G$ coset model 
to those of the $G/H$ gWZW model  will be   to impose the so called reduction
 gauge  in the equations of  the $F/G$  model \rf{efg}
 written in terms of the independent current components
  and then to make use of the 2d  conformal symmetry
to eliminate one additional degree of freedom.
% This will allow us
%  to  solve part of the gauge-fixed
% equations of motion  explicitly in terms
% of a new field $g$ taking values in $G$ and the
% $\algh$-valued gauge field $A_\pm$. The constructed system of equations 
%  will
 This will allow us
 to  solve all gauge-fixed
equations of motion but the Maurer-Cartan equation explicitly in terms
of a new field $g$ taking values in $G$ and the
$\algh$-valued gauge field $A_\pm$. The remaining system of equations
(i.e. the components of the Maurer--Cartan equation in this parametrization)
 will
turn out to be invariant under both the left and the right
$H$ gauge symmetries. We will then prove  that one can impose the 
special gauge conditions under which  the gauge symmetry reduces
to that of the $H$-gauge invariance  of the 
$G/H$ gWZW model and  the equations 
 %and the gauge-fixing conditions
 become equivalent to the ones \rf{weom},\rf{lwc}
 following  from the gWZW action with  an integrable potential
  \rf{pi}  described   in section \bref{sec:gWZW}.

%%%%%%%%%%%%%%%%%%%%%%%%%%%%%%%%%%%%%%%%%%%%%%%
\subsection{Equation of motion  in terms of currents and the 
reduction gauge}\label{sec:curr-red}
%%%%%%%%%%%%%%%%%%5

The  relation between the reduced $F/G$ model
and the  $G/H$  gWZW model  will apply under certain
special conditions on the structure of the 
Lie algebras of the groups involved.
These conditions that we will specify below  will be satisfied, 
 in particular,  in the case of the
$S^n = SO(n+1)/SO(n)$  model (and its $AdS_n$ counterpart)
which is  our  main interest  here.

 Let $\alga$ be a maximal Abelian subspace
 of the orthogonal complement  $\algp$
of the algebra $\algg$  of $G$ in the algebra $\algf$ of $F$.
Let  $\algh$ be  its centralizer in $\algg$. Following \cite{ei}
we shall assume the following conditions on the
structure of these  algebras (which represent a special case  of
 \rf{sym})
% Introducing the orthogonal complements
%one arrives at the following decomposition:
\begin{gather}
\algf=\algp\oplus\algg\,, \qquad \algp=\alga\oplus \algn\,, \qquad \algg=\algm\oplus \algh \,, \qquad  [\alga, \alga]=0  \,,\qquad  [\algh, \alga]=0\,,
  \la{ala}\\
 [\algm, \algm] \subset \algh \,,\qquad
 [\algm, \algh] \subset \algm \,,\qquad
 [\algm, \alga] \subset \algn \,,\qquad
 [\alga, \algn] \subset \algm \,.  \la{dec}
\end{gather}
%\be
 %&&\algf=\algp\oplus\algg\,, \qquad \algp=\alga\oplus \algn\,, \qquad %\algg=\algm\oplus \algh \ ,
 %\ \ \ \ \ \ [\alga, \alga]=0  \ ,\ \ \ \ [\algh, \alga]=0\ ,
 % \la{ala}\\
 %&&  [\algm, \algm] \subset \algh \ ,\ \ \ \ \
 %[\algm, \algh] \subset \algm \ , \ \ \ \ \ \
 %[\algm, \alga] \subset \algn \ , \ \ \ \ \ \
 %[\alga, \algn] \subset \algm \ , \ \ \ \ \ \
 %  %[\algh, \alga]=0, \   \ \ \ \ \ \  \ \ \   [\alga, \alga]=0
 %   \ .  \la{dec}
%\eea
Starting with  a left-invariant current $J= \gf^{-1}d\gf$ with  $f \in F$
we shall use the following notation
  for its
$\algh$, $\algm$ and $\algp$ components
\begin{equation}\la{feg}
A_a=(\gf^{-1}\d_a \gf)_{\algh}\,,\qquad
B_a=(\gf^{-1}\d_a \gf)_{\algm}\,,
\qquad P_a=(\gf^{-1}\d_a \gf)_{\algp}  \ ,
\end{equation}
i.e. $\A_a \in \algg$ in \rf{cur} is equal to $A_a + B_a$.
The equations of motion of the $F/G$ sigma model \rf{efg}
 written in terms of the {\it current components}  $A_a,B_a,P_a$
viewed as {\it  independent fields }   then
take the form
\begin{gather}
D_+P_-=0\,,\qquad \quad D_- P_+=0\,, \la{qq}\\
\d_+(A_- +B_-) - \d_-(A_+ +B_+) + \commut{A_++B_+}{A_-+B_-}=\commut{P_-}{P_+}\,, \la{eqq}
\end{gather}
where $D_\pm=\d_\pm+\commut{A_{\pm}+B_{\pm}}{\ }$.
%\bea
%&&D_+P_-=0\,,\qquad D_- P_+=0\,,  \ \ \ \ \  \ \ \ \ \ \
%  D_\pm=\d_\pm+\commut{A_{\pm}+B_{\pm}}{ \ }  \ , \la{qq}\\
%&&\d_+(A_- +B_-) - \d_-(A_+ +B_+) + \commut{A_++B_+}{A_-+B_-}=\commut{P_-}{P_+}\,  . %\la{eqq}
%\eea
%where $D_\pm=\d_\pm+\commut{A_{\pm}+B_{\pm}}{\cdot¬}$.

The choice of the
 {\it reduction gauge}~\cite{ei} is based on the ``polar decomposition'' theorem which
states that for any $k \in \algp$ there exists $\gg_0\in G$ such that
$\gg_0^{-1} k  \gg_0\in \alga$. Using the $G$ gauge freedom of the coset model
 equations of motion
one can therefore
assume that one of the components of $P_a$, e.g.,  $P_+$ is  $\alga$-valued.
Then  $D_-P_+=0$  implies
\begin{equation}
\d_- P_+=0\,,\qquad \ \ \ \ \ \ \commut{B_-}{P_+}=0\,. \la{ptt}
\end{equation}
Here we made use of the condition $ [\algm, \alga] \subset \algn$ in  \rf{dec}.
Under a  certain  regularity condition which we shall assume
(in the case when $\alga$ is one-dimensional,
e.g.,  for $F/G=SO(n+1)/SO(n)$,
 it is enough to require that $P_+\neq 0$) the
equation $\commut{B_-}{P_+}=0 $
 implies that  
 \be \la{ber}
 B_-=0  \ . \ee
To summarise, by imposing the  gauge in which $P_+ \in \alga$
  and eliminating  $B_-$  by solving $\commut{B_-}{P_+}=0 $ (i.e.
    setting  $B_-$ to zero)   one  can bring the system  of
    the $F/G$  model equations of motion \rf{qq},\rf{eqq}
    to the following form:
  %  {\bf state all assumptions clearly that allow to split  last eq in 4.3}
\bea
\label{eqb}
&&\d_-P_+=0\,, \qquad\ \ \ \ \ \  \d_+ P_- +\commut{A_+}{P_-}+\commut{B_+}{P_-}=0\,, \\
&&\d_-B_++\commut{A_-}{B_+}=\commut{P_+}{P_-}\,, \la{qb}\\
&&\d_-A_+-\d_+A_-+\commut{A_-}{A_+}=0\,, \la{b}
\eea
where \rf{qb} and \rf{b} are $\algm$ and $\algh$ projections of \rf{eqq}
(we are using the conditions \rf{ala},\rf{dec}).

In this reduction
gauge the original $G$ gauge symmetry is reduced to $H$ gauge symmetry under which the 
current component $A_{\pm}$ transforms as a gauge potential  while
$B_\pm$ and $P_\pm$ transform covariantly,
i.e. as  $ (...)   \to h^{-1}  (...)  h$. In particular,
$P_+$ is invariant because it takes values in $\alga$
 and $\commut{\alga}{\algh}=0$.

Let us note that  ~\eqref{b} implies that we can
impose the on-shell $H$  gauge  where  $A_\pm=0$. In this gauge
the equations of motion \rf{eqb},\rf{qb} take
the form:
\begin{equation}
\label{af}
\begin{gathered}
\d_-P_+=0\,, \quad \ \ \ \ \ \ \d_+ P_- =\commut{P_-}{B_+}\,, \ \ \ \ \ \ \ \ \ \ \
\d_-B_+ =\commut{P_+}{P_-}\,.
\end{gathered}
\end{equation}

%%%%%%%%%%%%%%%%%%%%%%%%%%%%%%%%%%%%%%%%%%%%%%%%%%%%%%%%%%%%%%%%%%%%%%%%%%%%%%
\subsection{Fixing conformal symmetry, field redefinition\space{} \\
 and relation to $G/H$ gauged WZW model}\label{sec:42}
%%%%%%%%%%%%%%%%%%%%%%%%%%%%%%%%%%%%%%%%%%%%%%%%%%%%%%%%%%%%%%%%%%

The first equation $\d_-P_+=0$ in~\eqref{eqb}
implies  that $P_+ = P_+ (\sigma^+)$. One can then  fix
one component of the matrix function   $P_+$  using the residual
conformal symmetry under which $P_+  d \sigma^+ = P'_+  d \sigma'^+ $.
Since in the reduction gauge $P_+$ belongs to the abelian subspace $\alga$ of $\algp$,
 then if    $\dim{}\alga=1$ (which is the case, e.g.,  for
 the $SO(n+1)/SO(n)$  coset of our interest)
 one can always assume that
$P_+ = \km T_+$ where $T_+\in\alga$ is a  constant matrix in $\algf$  which is a
basic element of $\alga$ (we may also normalize it
so that  $\tr(T_+T_+)=-2$).
This is equivalent to requiring that the corresponding component
of the stress tensor in \rf{st}  is constant, i.e.
 $\TT_{++}=\km^2$.

 Furthermore, we can use the remaining conformal symmetry
 $\s^- \to \s'^- (\s^-)$
 to fix the $\TT_{--}$  component in \rf{st} also  to be constant
 as in the original Pohlmeyer's argument.\foot{The conservation equation
  $\d_+ \TT_{--}=0 $
 can be seen directly from the second equation in \rf{af}.}
 %Furthermore, the second equation implies conservation of the $T_{--}$ component
%of the stress tensor: $\d_+ (\tr(P_-P_-)) = 0$.
% Using the remaining conformal symmetry one finally arrives at
Thus  assuming that the  
maximal Abelian subspace $\alga$ of $\algp=\algf\ominus\algg$ is
1-dimensional and using the conformal symmetry  we  arrive at
%\maxim{Here some Latex:
\begin{align}
\label{qc}
   && &P_+ =\ \km\ T_+ \,, \quad & -\half&\tr(P_- P_-)= \km^2 ,&& \\
   && &\TT_{\pm\pm}=\mu^2 \,,\quad &  &\km,\ T_+ = \const \ .&&
\end{align}
The  first condition in \rf{qc}
 fixes one independent degree of freedom  contained in $P_+$
in the case when $\dim\alga=1$
and the second condition reduces by one  the number
 of independent degrees of freedom in $P_-$.
The normalization   condition on $P_-$ can be solved  by
\be
P_-=\ \km\ g^{-1} T_- g  \ ,   \ \ \ \ \ \ \ \    \   T_- = \const \ ,   \la{gg}
\ee
where $g \in G$ is a {\it new field  variable}
  (thus non-locally related to original variable $f \in F$ in \rf{feg})
and $T_-$ is a constant matrix   which is a fixed element of $\alga$.
The existence of such $g$   follows again from the
 polar
decomposition theorem, and the requirement of  $\TT_{--}=\mu^2$ implies that
 $\tr(T_-T_-)=-2$. In the case
of  $\dim{\alga}=1$  which we are considering here 
it follows that
%from the polar
%decomposition theorem that this is a general solution,
%with
\be T_+= T_-  \equiv T \  . \la{teet}\ee
 For generality and to indicate the Lorentz index structure, 
 below we  shall often keep  the  separate notation for $T_+$  and $T_-$.

%below.

 %provided $\dim{\alga}=1$.

%In what follows we assume \eqref{c-choice} satisfied
%and use the parametrisation $P_-= \km g^{-1} T_- g$.

The  equation for $P_-$ in ~\eqref{eqb}
written  in terms of $g$ in  \rf{gg}  then becomes
\begin{equation}\la{jk}
\d_+(g^{-1}T_-g) +\commut{\A_+}{g^{-1}T_-g}=0\,,\qquad 
\A_+=A_++B_+ \ .
 \end{equation}
Considering $\A_+\in \algg$ as an unknown,
the general solution of  this equation can be written as
\begin{equation}\la{kp}
\A_+=g^{-1}\d_+g+g^{-1}A^\prime_+g \ ,
\end{equation}
where $A^\prime_+$ is an arbitrary $\algh$-valued function.
Indeed, the first term in \rf{kp}
is obviously a particular solution of \rf{jk} (since $T_-=\const$)
while the second term is a general solution of the  homogeneous equation
$\commut{\A_+}{g^{-1}T_-g}=0$   (given that  $[A^\prime_+, T_-]=0$
since $[\algh,\alga]=0$).
 Thus
\be \la{jo}
A_+ = (g^{-1}\d_+g+g^{-1} A^\prime_+g)_\algh \ , \ \ \ \ \ \ \ \  \ \ \ \ \
B_+ =  (g^{-1}\d_+g+g^{-1} A^\prime_+g)_\algm \ . \ee
% in what follows.
%{\bd this is bad logic  and needs to be changed}
In terms of the new variables $g,\  A^\prime_+,\ A_-$ the
first two equations of motion in \rf{qq} or \rf{eqb} are solved
and the remaining  equation
 \rf{eqq} (or \rf{qb},\rf{b} which are its $\algm$ and $\algh$ components)
  then takes the form
\begin{equation}
\label{1eq}
\d_-(g^{-1}\d_+g+g^{-1}A^\prime_+g)-\d_+A_-+
\commut{A_-}{g^{-1}\d_+g+g^{-1}A^\prime_+g}=\km^2\commut{T_+}{g^{-1}T_-g}\,.
\end{equation}
As  discussed in section \bref{sec:gWZW},  this equation is equivalent to 
the equations of motion of the gWZW theory \rf{weom},\rf{lwc}
in the sense that by an appropriate gauge transformation one can 
always make  the following constraints satisfied:
\begin{equation}
\label{r}
 A^\prime_+=(g^{-1}\d_+g+g^{-1} A^\prime_+g)_\algh \,, \qquad
  A_-=  (g\d_-  g^{-1}+gA_-g^{-1})_\algh \,.
\end{equation}
After renaming $A^\prime_+$ as $A_+$ these are exactly the equation of 
motion \eqref{weom} and the constraints
\eqref{lwc}.\footnote{More generally one, can conside asymmetrical gauge by introducing the appopriate $\algh$-automorphism
$\tau $. See the respective discussion in section~\bref{sec:gWZW}.}

\bs

We have  thus  shown  that the original system  of   equations of the 
$F/G$   sigma model \rf{eqb}, \rf{qb},\rf{b}  is
equivalent to the one described by the
equation~\eqref{1eq} and  the constraints \eqref{r}  with the $H$
gauge symmetry~\eqref{gs} with
$h=\bar h$.
These are the same  equations of motion \rf{weom}, 
the constraints \rf{lwc} and the gauge symmetry as corresponding
to  the action  \rf{pi} of the $G/H$ gauged WZW model \rf{gaui}
with the potential  $\sim \km^2 \tr(T_+g^{-1}T_-g)$.

 That the reduced equations of motion of the $F/G$
coset model can be related to those of  the  gWZW model with  an  integrable
potential was first suggested  in \cite{bak} (and checked on several examples
 including $SO(n+1)/SO(n)$, $SU(n+1)/U(n)$, and $SU(n)/SO(n)$
cosets). Here   we  explained 
 why this correspondence  should work in general
and  specified the necessary conditions on the groups and the algebras involved.

%%%%%%%%%%%%%%%%%%%%%%%%%%%%%%%%%%%%%%%%%%%%%%%%%%%%%%%%%%%%%%%%%%%%%%
\subsection{Gauge equivalence of  Lax representations
for  the $F/G$ coset and $G/H$ gauged WZW  models}\label{sec:43}
%%%%%%%%%%%%%%%%%%%%%%%%%%%%%%%%%%%%%%%%%%%%%%%%%%%%%%%%%%%%%%%%

%%%%%%%%%%%%%%%%%%%%%%%%%%%%%%%%%%%%%%%%%%%%%%%%%
Imposing the reduction gauge in terms of the Lax connections can be achieved in a
 directly analogous way.
 Let $\omega$
 % (depending on spectral parameter $\lam$)
  be an $\algf$-valued Lax connection defined in \rf{lxx}.
 The gauge equivalence transformation
   $\omega^\prime = f^{-1}\omega f + f^{-1}d f$
   with $f\in F$ gives a new system determined
    by a gauge-equivalent Lax connection $\omega^\prime$.
Decomposing $\omega=\omega_{\algp}+\omega_{\algg}$
one observes that  in the special case of  $f=g\in G$   the component
     $\omega_{\algp}$ transforms as
 $\omega_\algp^\prime = g^{-1}\omega_\algp g$.
  Using  the same polar decomposition
 argument as discussed above one concludes that it is
 always possible to find a $G$-valued function $g$
 such that (cf. \rf{ala})  \ \  $(\omega_\algn)_+=  (A_+ +B_+ +\lam P_+)_\algn =    0$.

Decomposing $\omega^\prime$ according to $\algf=\algp\oplus\algm\oplus\algh$
\begin{equation}
\begin{gathered}
\omega^\prime=d\sigma^+(A_+ +B_+ +\lam P_+)+d\sigma^- (A_- +B_- +
{\lam}^{-1} P_-)\,, \\
     A_\pm\in\algh\,,\quad
B_{\pm}\in\algm,\quad P_+\in\alga \ , \ \ \ \ \ \ \   P_- \in \algp \ ,
\end{gathered}
\end{equation}
one  finds as above  that the compatibility
condition implies eqs. \rf{ptt}, i.e. 
$\d_-P_+=0$ and   $\commut{P_+}{B_-}=0$; 
the latter  gives again  $B_-=0$.
 %As before under the appropriate regularity assumption
 % the second one can be algebraically solved giving $B_-=0$.
   This allows us to relate  the Lax connection to that
    with $B_-=0$, i.e.
\begin{equation}
\omega^{\prime\prime}=d\sigma^+(A_++B_++\lam P_+)+ d\sigma^-(A_-+{\lam}^{-1} P_-)\ , 
\end{equation}
whose flatness  condition implies  the last
 two equations in  ~\eqref{eqb}.\foot{
Note that this reduction is  local as $B_-=0$ is an algebraic consequence of the compatibility condition, i.e. $B_-$ is an auxiliary field.}

As for the equations  $\d_-P_+=0$ and $\d_+P_-+\commut{A_++B_+}{P_-}=0$ in
\rf{eqb}, 
%they are not algebraic.However, assuming these compatibilities
assuming they are satisfied,  one can again
use the conformal transformations to set $P_+=\km T_+$ and $\tr(P_-P_-)=-2\km^2$.
As a result, the  Lax
connection  takes the following form:
\begin{equation}
\label{ed}
\omega_{\red}=d\sigma^+(A_++B_++\lam \km T_+)+
d\sigma^-(A_-+  {\lam}^{-1}P_-)\,.
\end{equation}
Finally, using again  the parametrisation $P_-=\km g^{-1}T_- g$ and $A_++
B_+=g^{-1}\d_+g+g^{-1} A'_+g$,
one  arrives at 
 \begin{equation}
\label{par}
\omega=d\sigma^+(g^{-1}\d_+g + g^{-1} A'_+g + \lam \km T_+)+d\sigma^-(A'_-+ \lam^{-1}
 \km g^{-1}T_-g)\,,
\end{equation}
 whose compatibility condition  implies \rf{1eq}.
  It was shown in the
 previous subsection that by an appropriate gauge transformation
 one can also satisfy  the on-shell relations~\eqref{r}. We thus  find the relation  to 
 %giving therefore 
 the Lax representation of the $G/H$ gWZW model
 (cf. \rf{lwc},\rf{lw}).

%%%%%%%%%%%%%%%%%%%%%%%%%%%%%%%%%%%%%%%%%%%%%%%%%%%%%%%%%%%%%%%%%%%%%%%
\subsection{Reduced equations  of
 $S^n= SO(n+1)/SO(n)$ coset model \space{ }\\
  in the $A_\pm=0$ gauge}\label{sec:44}
%%%%%%%%%%%%%%%%%%%%%%%%%%%%%%%%%%%%%%%%%%%%%%

Let us now turn to the special case  of our interest: 
sigma model  with  a sphere  as  a   target space. 
%Here $F=SO(n+1), \ G= SO(n), \ H=SO(n-1)$
Using the standard $(n+1) \times (n+1)$
matrix representation for $F=SO(n+1)$ and its diagonally
 embedded $G=SO(n)$ subgroup
 we can choose
 $T_+=T_-$  to have only one non-zero upper $2 \times 2$ block
 so that $H=SO(n-1)$ is  also diagonally embedded into $G=SO(n)$
 (the conditions \rf{ala},\rf{dec} are then satisfied).
 In this case  we   get for  $P_\pm$ in \rf{qc},\rf{gg}
\begin{equation}
P_+= \km T_+ = \km \left(
\begin{array}{cccc}
0&1&\ldots&0\\
-1&0&\ldots&0\\
\ldots&\ldots&\ldots&\ldots\\
0&0&\ldots&0
\end{array}
\right)\,,
\qquad
P_-=
\km\left(
\begin{array}{cccc}
0&k_1&\ldots &k_n\\
-k_1&0&\ldots&0\\
\ldots&\ldots&\ldots&\ldots\\
-k_n&0&\ldots&0
\end{array}
\right)\,. \la{pil}
\end{equation}
Here  $g$ in \rf{gg}   is parametrized by $k_l$ 
and  $-\half\tr(P_+P_+)=\km^2$. Also,  
 $-\half\tr(P_-P_-)=\km^2$  is satisfied provided 
\begin{equation}
 \la{norm}
 \sum_{s=1}^n k_s k_s=1 \,.
 \end{equation}
The subalgebras $\algg= so(n)$ and $\algh=so(n-1)$
  are canonically (diagonally)   embedded 
  into $\algf= so(n+1)$. 
In addition to $B_-=0$ from \rf{ptt},\rf{ber} 
we have for $B_+= (\cA_+)_\algm$ (see \rf{jk})
\begin{equation}\la{beeq}
B_+=
\left(
\begin{array}{ccccc}
0&0&0&\ldots &0\\
0&0&b_2&\ldots&b_n\\
0&-b_2&0&\ldots&0\\
\ldots&\ldots&\ldots&\ldots&\ldots\\
0&-b_n&0&\ldots&0\\
\end{array}
\right)\,.
\end{equation}
 In this  case the equation
$\d_+P_-+\commut{A_+}{P_-}=\commut{P_-}{B_+}$
in \rf{eqb}
can be  solved algebraically for $B_+$
giving \rf{beeq}  with 
\begin{equation}
 b_l=\frac{\d_+k_l+\commut{A_+}{k}_l}{\sqrt{1
 -\sum_{m=2}^n k_mk_m}}\,,\ \ \ \  \qquad l=2,\ldots,n\,.
\end{equation}
Fixing the  $H=SO(n-1)$  on-shell gauge as
\be \la{aa}
A_+=A_-=0
\ , \ee
  the third equation in \rf{eqb}  then gives the following
reduced system of equations for the remaining
 $n-1$ unknown functions $k_2,...,k_n$ ($k_1$ is determined from  \rf{norm})
  ~\cite{add}
\begin{equation}
 \d_-\frac{\d_+k_l}{\sqrt{1-\sum_{m=2}^n k_mk_m}}=-\km^2 k_l\,,
  \qquad l=2,\ldots,n\,. \la{kkk}
\end{equation}
This is  the same  reduced system that follows both from the
 $SO(n+1)/SO(n)$ coset model \ci{pole,ei}  and the  $SO(n)/SO(n-1)$ gWZW model
 in the $A_\pm=0$ gauge \ci{bak}. 
 %The equation \rf{kkk} does not, however, 
 %follow from a local 
 %action for $k_l$ only. 

%\maxim{
The  point $g=\id$   is an obvious  vacuum for eq. \rf{weom}
in the $A_\pm=0$ gauge, i.e. a trivial solution 
of \rf{ol},\rf{olk}  with 
$T_+=T_-$.
%$T_+=T_-$
 According to \rf{gg},\rf{pil}  it  corresponds to 
\be \la{vbq} k_2=...=k_n=0   \ . \ee
 The massive 
fluctuations near this 
vacuum in the gauge \rf{aa} 
are described by the  $H=SO(n-1)$  invariant equation \rf{kkk}, i.e.
\begin{equation}
 \la{fluv}
 \del_+ \del_- k_l  + \km^2 k_l + O(k^2_l) =0 \ .
\end{equation}

\bs

It is convenient
 to rewrite the equation \eqref{kkk} in terms of the  new variables 
 $(\vp, u_m)$ defined so that  \rf{norm}  is satisfied  
\begin{equation}
 \la{vb}
 k_1= \cos2\vp\,,\qquad k_l = u_l\  {\sin2\vp}\,,
\qquad
u_l u_l =1  \,,\qquad l=2,\ldots,n \ ,
\end{equation}
getting \cite{add}
\begin{equation}
\begin{aligned}
\label{eq-add}
\d_+\d_-\vp-\half\tan 2\vp  \  \d_+u_l\d_-u_l  +\frac{\km^2}{2}\sin 2\vp&=0\,,\\
 \d_+\d_- u_l + (\d_+u_{m}\d_-u_{m})\  u_l +
\frac{2}{\sin{2\vp}}
  (\cos{2\vp}\ \d_+\vp
 \d_-u_l+
 \frac{1}{\cos{2\vp}}\ 
% \tan {2\vp}\ 
 \d_-\vp \d_+  u_l )&=0\,.
\end{aligned}
\end{equation} 
Besides the obvious $SO(n-1)$ symmetry these equations are
invariant under the following formal  transformation 
\be \la{sy}  \varphi\ \to \  \varphi+\frac{\pi}{2} \ , \ \ \ \ \ \ \ \ \ \ \ \ \ \ 
\mu^2\to-\mu^2  \ . \ee 
 In  the case of  $F/G=SO(4)/SO(3)$, i.e. CSG  as a reduced model, 
this formal transformation   relates the 
 two 2d dual reduced 
 models  with T-dual
   target  space metrics in the corresponding 
 reduced Lagrangians  \ci{bak1,pars,mira}.\foot{In this   case of
  $SO(3)/SO(2)$ gWZW model this duality 
 is also related with  the vector ($g \to h^{-1} g h$) or the axial ($g \to h g h$)
 gauging \ci{dvv,bs2}.}

\bs

Let us briefly describe  the modifications of the above construction in the case of the 
$AdS_n=SO(2,n-1)/SO(1,n-1)$
coset model. The vector-space 
signature is diag$(-1,-1,1,\ldots,1)$ and the 
subgroup $G=SO(1,n-1)$ is diagonaly embedded. In the 
standard representation of $\algf=so(2,n-1)$ the element $T_+=T_-$ can be choosen to have
the same form as in  \eqref{pil} while the condition \eqref{norm} takes the form
$k_1k_1 -\sum_{m=2}^n k_m k_m=1$. Equation \eqref{kkk} is then replaced by 
\begin{equation}
 \d_-\frac{\d_+k_l}{\sqrt{1+\sum_{m=2}^n k_m k_m}}=-\km^2 k_l\,,
  \qquad l=2,\ldots,n\,. 
\end{equation}
Finally, introducing instead of \eqref{vb} the parametrization
\begin{equation}
 k_1= \cosh2\phi\,,\qquad k_l = u_l\  {\sinh2\phi}\,,\ \ \
\qquad
u_l u_l =1  \,,\qquad l=2,\ldots,n \ ,
\end{equation}
one arrives at the system \eqref{eq-add} for $\phi,u_l$ with
the obvious replacement of $\cos\vp,\sin\vp,\tan\vp$ with 
$\cosh\phi,\sinh\phi,\tanh\phi$. The two systems of equations are thus related by 
% The obvious ``mnemonic'' rule
%is then 
the replacement $\vp=\imath \phi$,  as one would 
expect from the standard analytic continuation
argument. Remarkably, the variables $u_l$ satisfy the same normalization 
condition in the $S^n$ and the $AdS_n$ cases
and both systems are invariant under the same $H=SO(n-1)$
 symmetry. Note also that in the  $AdS_n$ case
the linearized equations \eqref{fluv} have exactly the same 
form leading to the same massive fluctuations near the vacuum $g=\id$.

\bs 

Instead of using the parametrization of  $P_-$ in terms of $k_l$ in  \rf{pil} 
  we may start with a particular choice of $g\in G$ which then determines 
  $P_-$ according to  \rf{gg}.
Parametrising $g\in G=SO(n) $  by the generalized Euler angles and  expressing $P_-$
in terms of them   one arrives at a  certain multi-field
generalisation of the sine-Gordon equation  which is just another form of \rf{eq-add}
 ($\vp$ introduced in \rf{vb}
corresponds then to the first Euler angle).
 In  the  $SO(3)/SO(2)$  case this gives
the standard sine-Gordon equation
\be
g=\left(\begin{array}{cc}
\cos 2\varphi&\sin 2\varphi\\
-\sin 2\varphi& \cos 2\varphi
\end{array}
\right)\ , \qquad
k_1=\cos 2\varphi\,,\quad k_2= \sin 2\varphi\,, \ee
\be \d_+\d_-\varphi+\frac{\km^2}{2} \sin 2\varphi=0\, . \ee
In the  $SO(4)/SO(3)$  case  we can parametrize $g\in SO(3)$ as
\be
g=g_2g_1g_2\ , \ \ \ \ \ \ g_1=\exp{(2\varphi R_1)} \ , \ \ \ \ \
g_2=\exp{(\chi R_2)} \ , \ee
\be
R_1=\left(
\begin{array}{ccc}
0&1&0\\
-1&0&0\\
0&0&0
\end{array}
\right)\,,
\qquad
R_2=\left(
\begin{array}{ccc}
0&0&0\\
0&0&1\\
0&-1&0
%0&0&1\\
%0&0&0\\
%-1&0&0
\end{array}
\right)\,.
\end{equation}
The  corresponding components of the unit vector $k_s$ in  \rf{pil},\rf{vb}
 are
\begin{equation}
 k_1= \cos 2\varphi\,, \qquad  k_2=\sin 2\varphi\ \cos \chi\,,
 \qquad k_3=\sin 2\varphi\ \sin \chi \ .
\end{equation}
The equations of motion \rf{eq-add}  take the form
% (see \cite{add} for details)
\begin{equation}
 \begin{aligned}
\d_+\d_- \varphi  -   \ha \tan2\varphi\ \d_+\chi \d_-\chi+{\km^2 \ov 2} \sin 2\varphi&=0\,,\\
\d_+\d_-\chi +  \frac{2}{\sin 2\varphi}
\left(
 \cos 2\varphi\ \d_+ \varphi\  \d_-\chi+
 \frac{1}{\cos 2\varphi} \ \d_- \varphi\ \d_+\chi\right)
 &=0\,.
\end{aligned} \la{ko}
\end{equation}
%As special case of  \rf{kkk} these  equations are non-Lagrangean, but they 
%can be brought to the standard Lagrangean 
These  equations can be brought to the standard 
complex sine-Gordon form by a (nonlocal) change of 
variables (which may be interpreted as a gauge change in \rf{weom},\rf{lwc}).
Indeed, replacing $\chi$ by $\theta$ via
\begin{equation}
 \d_+\theta=\frac{\cos^2\varphi}{\cos 2\varphi}\d_+\chi\,,
  \qquad \d_-\theta=\cos^2\varphi\ \d_-\chi \ , \la{re}
\end{equation}
we get~\cite{add}:
\begin{equation}
\label{sss}
 \begin{aligned}
 \d_+\d_-\varphi-  
 \frac{\sin\varphi}{\cos^3\varphi}\d_+\theta\d_- \theta+\frac{\km^2}{2} \sin 2\varphi=0\,,\\
\d_+\d_-\theta +      
\frac{2}{\sin 2\varphi}\left( \d_+\varphi\d_-   \theta+\d_-\varphi\d_+\theta\right)=0\,,
\end{aligned}
\end{equation}
which follow from the  local CSG  Lagrangian   \rf{csgi}. 
If we replace eq.  \rf{re}
by  the transformation 
\be \la{dui}
\d_+\tilde \theta=-\frac{\sin^2\varphi}{\cos 2\varphi}\d_+\chi\ ,\qquad
  \ \ \d_-\tilde \theta=\sin^2\varphi\ \d_-\chi\ ,  \ee
we get instead of \rf{sss}  the  equations  that follow
from the analog of \rf{csgi}   with T-dual target space metric:
$ds^2 =  d \varphi^2 + \cot^2 \varphi \ d\tilde \theta ^2$.
Both the corresponding ``dual'' Lagrangian and its equations of motion 
are related, respectively, 
 to \rf{csgi} and \eqref{sss}
by the transformation \rf{sy}.
The fields $\theta$ in \rf{re} and $\tilde \theta$   in \rf{dui}  are 
related of course by the 2d duality transformation. 
% $\vp\to\vp+\frac{\pi}{2}\,,\,\, \km^2 \to -\km^2$.
% which explains why
%these two forms are dual.

\bs 

In general, the equations \rf{kkk} found in the $A_\pm=0$ gauge
 do not follow from a local
Lagrangian for the field $k_m$  (apart from the $n=2$, i.e. the  SG   case).
 In particular,  this applies to
the system \rf{ko}: one needs a nontrivial
field  redefinition \rf{re} (which is consistent only
on the equations of motion for $\varphi$)  
to get a Lagrangean  system \rf{sss}.

 Such 
 a  non-local field redefinition may be interpreted as  corresponding to a change 
of the $H$ gauge.
A    way to get a Lagrangean system of the reduced equations is to
 fix the $H$ gauge not on $A_\pm$   (as was done in \ci{bak} 
 and above in this section) 
 but on $g$, i.e. to solve the equations for  $A_\pm$ 
 in terms of the gauge-fixed  $g$.  
 We shall discuss this procedure in the next section.

%%%%%%%%%%%%%%%%%%%%%%%%%%%%%%%%%%%%%%%%%%%%%%%%%%%%%%%%%%%%%%%%%%%%%%%%%%%%%%

%%%%%%%%%%%%%%%%%%%%%%%%%%%%%%%%%%%%%%%%%%%%%%%%%%%%%%%%%%%%%%%%%%%%%%%%%%%%%
\section{Lagrangian  of  reduced  theory:  $S^n=SO(n+1)/SO(n)$ model }\label{sec:lag}
%%%%%%%%%%%%%%%%%%%%%%%%%%%%%%%%%%%%%%%%%%%%%%%%%%%%%%%%%%%%%%%%%%%%%%%%%%%%%%

As we have seen in section \bref{sec:red-bos}, the reduced equations of motion of the
$F/G$ coset model are in general gauge-equivalent to 
the equations of motion of the $G/H$ gWZW  model with a specific integrable potential.
To get a Lagrangean formulation of the reduced theory  corresponding to the 
$F/G$  model (or, equivalently,   to the bosonic string on $R_t \times F/G$ in
 the conformal gauge)
we may then start with the associated  $G/H$  gWZW  model, fix an $H$-gauge on
$g \in G$  and solve for the auxiliary  gauge field $A_\pm$.
This will produce a classically-equivalent integrable system.
Here we shall concentrate on the example  of the  $S^n$ sigma model.

\subsection{General structure of the reduced Lagrangian}\label{sec:gen-struct}

In the case of $F/G=S^{n}$, i.e.     $G/H=SO(n)/SO(n-1)$ 
 we will end up with an integrable theory  represented  by an  
  ($n$--1)- dimensional  sigma model with a potential\foot{The absence
  of the antisymmetric
   $B_{mn}$ coupling  has to do with the
 symmetric gauging of  the  maximal diagonal subgroup.}
 \be \la{si}
 L =  G_{mk} (x)\  \del_+ x^m \del_- x^k   -    \km^2 U(x)  \ .
 % \ \ \ \ \ \ \ \ \ i,j=1,..., n-1 \ ,
  \ee
  The special cases are the   $n=2$  \rf{red}  and $n=3$ \rf{csgi} examples
  discussed above. Here $x^m$  are the $n-1$ ($= \dim G - \dim H$)  independent components of
  $g$ left over after  the $H$ gauge fixing on $g$.

  In contrast to the metric of the usual geometric (or ``right'') coset $SO(n)/SO(n-1)=S^{n-1}$
 the metric  $G_{mk}$ in \rf{si}  found from the symmetrically gauged
 $G/H={SO(n)\ov SO(n-1)}$ gWZW  model
 will generically have singularities and no non-abelian
  isometries.\foot{While the gauge
  $A_\pm=0$ preserves the explicit  $SO(n-1)$ invariance of the equations of motion,
   fixing the gauge on $g$  and integrating out $A_a$
    breaks all non-abelian symmetries (the corresponding symmetries are
    then ``hidden'').} 
 %   \maxim{Moreover, as we are going to see for $\algh$ nonabelian 
 % (i.e. for $n\geq 4$ in our case) the metric $G_{mk}$ has a singularity at
 %  the vacuum. In particular,
 % the linearized equations (and therefore the mass spectrum) are in general 
 % not equivalent to the equations~\eqref{kkk} in the gauge $A_+=A_-=0$.}

Following \ci{sp} we may call these  geometries resulting from conformal 
$SO(n)\ov SO(n-1)$ gWZW  models as
 ``conformal cosets''  or ``conformal spheres'',
 with the notation  $\CS^{n-1}$.
 %= [{SO(n)\ov SO(n-1)}] $. 
 Instead of $R_{mk} =\ c \  G_{mk}$ for a standard sphere
 their 
 metric $G_{mk}$ satisfies $R_{mk} + 2 \nabla_m \nabla_k \Phi=0$  where $\Phi$ is
  the corresponding dilaton resulting from integrating out $A_a$.
  The explicit expressions for $G_{mk}$  were worked out  for a 
  few  low-dimensional cases:
   $\CS^{2}$  \ci{old}, $\CS^{3}$  \ci{cres,fl,bs1} and $\CS^{4}$ \ci{bs2}.
 %  \foot{
% It is curious that originally  refs. \ci{nemb,fl1}
%  wanted to describe $AdS$ space  or a sphere  by
 % as  coset  WZW model;  we see that there is indeed some
 % relation at the classical level
 % -- the reduced model  is expressed in terms of 
 %the metric of  a conformal coset of one
%  lower dimension.}

The potential  (``tachyon'') term   in \rf{si}   originates directly from
the $\km^2$ term in \rf{pi}. It   is
a relevant (and integrable)
 perturbation of the gWZW model  and thus also of the  ``reduced'' geometry,
so that  it should satisfy (see also  \ci{jack})
  \be \la{lap}
  {1 \ov  \sqrt G e^{-2 \P} }
\del_m ( \sqrt G e^{-2 \P}  G^{mk}\del_k) U - M^2 U =0
  \ .    \ee
 Below we shall comment
 on details of the derivation  of the metric $G_{mk}$ and  write down  explicitly the
 reduced Lagrangian \rf{si}
 for the new non-trivial
 cases of $n=4,5$, i.e. for the  string on $R_t \times S^4 $   and
  $R_t \times S^5 $,   
  which generalize the $n=3$  CSG model \eqref{csgi}.

\bs

The  $H$ gauge  fixing on $g$  and elimination of $A_a$ from
the $SO(n)\ov SO(n-1)$ gWZW Lagrangian \rf{pi}  can be done  by generalizing the
discussion  of the $n=4$  case in \ci{fl}.
The first step is   the parametrisation of $g$ in  terms of  the generalized Euler angles.
  Let us define  the 1-parameter  subgroups corresponding to the 
   $SO(n+1)$   generators  $R_{m+1, m}$  \ ($m=0,1,...,n-1$)
 \be  \la{ge}
g_m (\theta) = e^{ \theta R_{m} }, \ \  \ \ \ \ \ \ \ \ \ \ \ 
   (R_m)^j_i = (R_{m+1, m})^j_i
 \equiv
  \delta_{m}^j  \delta_{m+1,i}   - \delta_{mi}  \delta_{m+1}^j 
      \ . 
       \ee
%\maxim{
Then $T_\pm=T$ in \rf{pi} is  equivalent to the generator $R_{0}$  corresponding to $g_0$
$$ T= R_0$$ 
and the generators of the subgroup $H=SO(n-1)$ which commutes
with   $T$   contain   $R_{m+1, m}$ with $m=2,...,n-1$.
A generic element of $G=SO(n)$  can be parametrized  as

$
g=  g_{n-1} (\te_{n-1})... g_2(\te_2) g_{1} ( \te_1) h  ,
$

where $h$  belongs to $H$. A convenient $H$  gauge choice is then   \ci{fl}
\be \la{gah}
g= 
%\hat\tau(
 g_{n-1} (\te_{n-1}) ...g_2(\te_2)
 %) 
 g_{1} ( 2\vp)  
 g_2(\te_2) ...      g_{n-1} (\te_{n-1})
     \ , \ee
so that $\vp\equiv \ha  \te_1,$ and $ \te_p$  ($p=2,...,{n-1}$) are
$n-1$  coordinates on the  coset space  $\CS^{n-1}$, with $\vp$  
playing a distinguished  role.
%For generality we introduced here the automorphism enetring the general
% form of the Lagrangian
%(see the discussion of the asymmetric gauges in section \bref{sec:gWZW}).}

With this choice of the  parametrisation it turns out that the  potential $U$  in \rf{pi},\rf{si} has a universal  form
for {\it any} dimension
  $n$: 
it  is  simply proportional   to $\cos 2\vp$
  as in the SG \rf{red}  or CSG \rf{csgi}  cases.
   Indeed, since  $\commut{T_\pm}{g_k}=0$ for $k\geq 2$, 
    one finds 
\begin{equation}\la{poy}
  \tr(T_+ g^{-1}T_-g)= \tr(T_+ g_1^{-1}T_-g_1)=
2  \cos 2\varphi   \ .
%sign changed according to \vp redef
\end{equation}
%where we have introduced
% parametrisation $g_1=exp(2\varphi R_1)$. In particular,
%in the case $SO(2)$ WZW model
%one finds $g^{-1}\d_{\pm}g=2R_1\d_{\pm}\varphi$ so
%that one indeed gets standard sine-Gordon Lagrangian:
%\begin{equation}
% L=\frac{2}{\pi}(\d_+\varphi\d_-\varphi+\frac{\km^2}{2}cos (2\varphi))\,.
%\end{equation}
The  metric and the dilaton
 resulting
from  integrating out the $H$ gauge 
field $A_a$  satisfy\foot{The  dilaton field  
 should  be taken into account 
  provided the model  is defined on a curved 2d background 
and one is interested in the  Weyl invariance conditions  
(i.e. the definition of the conformal stress
tensor)
of  the 
theory on the ``restricted'' $G/H$  part of configuration space
 obtained by eliminating the $H$ 
gauge field \ci{old,dvv,aat}. In the present context 
where we started with the string theory in 
 the conformal gauge that  would require 
a re-introduction of the 2d metric in the reduced model; 
then the dilaton would  couple to the metric in the standard way
and would  enter in the definition of the stress tensor   of the 
``restricted''  sigma model  \rf{si}.  If there is indeed a path integral transformation  that 
leads from the original (super)coset model to the reduced  model,  then the latter 
can be   considered as a usual world-sheet theory   coupled  to a 2d metric
 (that  will in general depend
on moduli in the case of higher genus surfaces, etc.).
The presence of the potential term 
that ``spontaneously'' breaks the conformal symmetry 
(which was fixed by making $\mu$ constant) is unrelated to the 
dilaton coupling issue.}
 % (up to an overall constant)
\be \la{dil}
ds^2 = G_{mk} dx^m dx^k=  d\vp^2 + {\rm g}_{pq}(\vp,\te)   d \te^p d \te^q \ , \ \ \ \ \ \ \ \ \ \ \ \
\sqrt G \ e^{-2 \P}    = (\sin 2\vp )^{n-2} \ , \ee
so that the equation  \rf{lap}  is indeed
 solved  by\foot{We fix the overall normalisation constant in  the WZW  action so that
 $\alpha' k=1$.}
\be \la{uu}
U= -{1\ov 2}  \cos 2\vp \ , \ \ \ \ \  \ \ \ \ \ \ \ \  M^2 = - 4(n-1)  \ .
\ee
Let us now   make  few   remarks. 

As was already mentioned, the  reduced model \rf{si} 
%constructed using the gauge \eqref{gah}
has no antisymmetric tensor coupling term.
 The antisymmetic tensor  contribution  could  originate either from
the WZ  term in  the WZW  action in \rf{gaui} 
 or in   the process of solving for the gauge field $A_a$.
It turns out that both   contributions vanish if the 
gauge condition~\eqref{gah} is used.
Details of the proof are given in the Appendix~\bref{sec:a-vanish}.

The obvious  ``vacuum''   configurations, i.e.  extrema 
of the potential $U$   are  $\theta_p=\const$ and $ \vp = {\pi \ov 2} n, \ n=0,1,2, ...$.
The metric ${\rm g}_{pq}(\vp,\te) $   in \rf{dil} may, however, 
 be singular  near such points,  i.e.  they may not be reachable in a given coordinate system
 and more  detailed analysis may be required.

\iffalse
The possible vacua of the theory are determined by the condition $\commut{g^{-1}Tg}{T}=0$.
In terms of the paramerization~\eqref{gah} it has two families of solutions:
\begin{equation}
 g_{vac}=h \tilde h\,, \qquad g^\prime_{vac}=h e^{\pi R_1}\tilde h\,, \quad  h=const\in H\,,
\end{equation}
where $\tilde h$ is defined as follows $\widetilde{g_{n-1}\ldots g_2}=g_2\ldots g_{n-1}$.
Indeed, in this parametrization for $g=h e^{2\vp R_1}\tilde h$ the condition $\commut{g^{-1}Tg}{T}=0$
is equivalent to $\commut{e^{-2\vp R_1}T e^{2\vp R_1}}{T}=0$ as $\commut{T}{h}=0$ for any $h\in H$.
This in turn has only two inequivalent solutions $\vp=0, \vp=\frac{\pi}{2}$. However, it follows from
the explicit form of the potential \eqref{uu} that the vacua described by the 
family of solutions with $\vp=\pi/2$ are unstable and we rule them out.
 \fi

One should keep in mind  that the  gWZW  action \rf{pi} is the most general  and universal 
definition of the theory, while special gauges and parametrizations may 
have their drawbacks and may not apply globally.  For example, 
the elimination of the gauge fields $A_{\pm}$  from \rf{weom} or the gWZW action \rf{gaui}
requires solving the constraints in \rf{lwc},  i.e. $A_+=(g^{-1}A_+g+g^{-1}\d_+g)_\algh$
and $A_-=(g A_- g^{-1}+g\d_-g^{-1})_\algh$.
The corresponding  operator $(\id-Ad_{g})_\algh$ is  singular near  some points $g$ (e.g., $g= \id$)
implying that in their vicinity one should   use a different gauge
or do not directly solve for $A_\pm$. 

For example, one  may consider an asymmetrically gauged  WZW model (see \rf{tut}) 
corresponding to a more general on-shell gauge \rf{ger}; in this case one should use 
\rf{gah}   with  the left-hand-side  factor $ g_{n-1} (\te_{n-1}) ...g_2(\te_2)$
replaced  by 
$\hat\tau(
 g_{n-1} (\te_{n-1}) ...g_2(\te_2) )$  where  $\hat \tau$ is the lift of the automorphism 
 in \rf{ger}.
 However, in the case when  $\algh$ is simple (e.g., for the  $SO(5)/SO(4)$ coset)
  such an automorphism can
  always be represented
as $\tau(A)=h_\tau^{-1}A h_\tau$ for some $h_\tau \in H$;
 therefore it can not be used to remove
the degeneracy of the  operator in the $A_+ A_-$ part of 
the action.\foot{The nonsingular 
metrics known 
to arise in the SG and CSG cases are due to the fact that $\algh=0$
in the CS case and $\algh=U(1)$ in the CSG case. As we will see below,  the nonsingular metric
in the CSG case is obtained by utilizing the automorphism $\tau(A)=-A$. This automorphism
does not, however,  apply   to the case of a non-abelian $\algh$.  
}

\iffalse
%and the similar one  for $A_-$).
Taking $g=g_{vac}$ the condition that the constraints can be solved for $A_\pm$
is equivalent to the invertibility of the operator $\id-Ad_{g_{vac}}$. However, this operator is
always degenerate. This is obvious for nonsemisimple algebras while for a semisimple one any
automorphism of the form $Ad_h$ is equivalent to the automorphism from the Weyl group
and therefore have a fixed points. Another possibilty is to consider more general action
arising from the asymmetric gauge fixing by utilizing the automorphism $\tau$ 
(see the discussion in sect. \bref{sec:gWZW}). However,
in the semisimple case we are interested in, such an automorphism can always be represented
as $\tau(A)=h_\tau^{-1}A h_\tau$ for some $h_\tau \in H$ and therefore can not remove
the degeneracy. To summarize,  for $\algh$ semisimple the reduced metric is necessarily
singular at the vacuum. The nonsingular metric known to arise in SG and CSG cases are due to $\algh=0$
in the CS case and $\algh=U(1)$ in the CSG case. As we will see in what follows the nonsingular metric
in the CSG case is obtained by utilizing the automorphism $\tau(A)=-A$. This automorphism
does not extend to the nonabelian case confirming the general statement made above.
\fi

Finally, let us  note  
that both the gauge fixing and the eliminating  of $A_\pm$
 %(provided this configuration is
%  flat)
  can be implemented at the level of  the Lax connection, 
   leading to  the Lax formulation  of the reduced model
in terms of the generalized Euler angles, i.e. ensuring the
integrability of the reduced  model \rf{si}.

 Let us now  turn to specific examples.

%%%%%%%%%%%%%%%%%%%%%%%%%%%%%%%%%%%%%%%%%%%
\subsection{Examples of reduced  Lagrangians for $S^n$  models
% theory\\ from $SO(4)\ov SO(3)$ gWZW model
}\label{sec:52}

Let us first show  how to get the  Lagrangian \rf{csgi} of the
CSG model   directly  from the   $SO(3)\ov SO(2)$ gWZW model \rf{pi}.
The equation for $A_+$  following from \rf{gaui} reads:
\begin{equation}
A_+ = (g^{-1}\d_+ g  + g^{-1}A_+g)_\algh  \,.
\end{equation}
In the $SO(3)\ov SO(2)$ gWZW  case we have  from \rf{gah}  \ 
$g= g_2 (\te) g_1 ( 2 \vp)  g_2 (\te)$
so that
\begin{equation}
 \begin{gathered}
(g^{-1}\d_+ g)_\algh=
%\Pi_\algh(g_2^{-1}g_1^{-1}R_2g_1g_2+R_2)\d_\pm\theta=
(1 + \cos 2 \vp ) R_2\d_+\theta \,,\qquad
\d_-gg^{-1}=(1-\cos 2 \vp)R_2 \d_-\theta\,,\\
A_+= \frac{1 + \cos 2 \vp }{1-\cos 2 \vp }R_2 \d_+\theta\,.
\end{gathered}
\end{equation}
One finds also
\bea
&&-\frac{1}{2}\tr(g^{-1}\d_+ gg^{-1}\d_- g)=
2 (1 +\cos 2 \vp )\d_+\theta \d_-\theta+ 4 \d_+ \varphi\d_- \varphi \,,\cr
&&\tr(A_+\d_-gg^{-1})=-  2\frac{(1+\cos 2 \vp)^2}{1-\cos 2 \vp}\d_+\theta \d_-\theta \ .
\eea
Using \rf{poy}  one finally  obtains the Lagrangian
\begin{equation}
   \td  L =
   \dpp \varphi \dmm \varphi  +   \cot^2 { \varphi }\ \dpp \theta \dmm \theta
 + \frac{\km^2}{2} \cos 2\varphi \,.
\end{equation}
This Lagrangian 
is  dual to that in \rf{csgi}, i.e. the two are related by 2d duality $\theta \to \tilde \theta$.
%According to the discussion in the previous section the
As was already mentioned above, the 
CSG Lagrangian  \rf{csgi}    is directly obtained 
if we start with  the asymmetrically (``axially'')  gauged WZW 
model with $\tau(A_-)=-A_-$.\foot{In this case the parametrization \eqref{gah} takes the form
$g=\hat\tau(g_2)g_1g_2=g_2(-\theta)g_1(2\vp)g_2(\theta)$.}
Alternatively,  the two dual 
models are related  by the formal transformation \rf{sy}.

%\subsection{$SO(4)/SO(3)$ gWZW}

\bs

The explicit form of the $\CS^{n-1}$ metric \rf{dil} with   $n=2,3,4$   as found directly from the action 
\rf{pi}  with \rf{gah} is thus 
\be
ds^2_{n=2}=  d\vp^2 \ , \ \ \ \ \ \ \ \ \ \ \ \
ds^2_{n=3} =  d\vp^2 + \cot^2 \vp \ d\te^2 \ , \ee
\be \la{fli}
ds^2_{n=4} =  d\vp^2 + \cot^2 \vp \ (d\te_2 + \tan{\theta_3} \cot{\te_2} \, d \te_3)^2 +
 \tan^2 \vp\  {d \te_3^2  \ov \sin^2 \te_2 }   \ .\ee
 %where $\te,\te'$ correspond to $\te_2,\te_3$ in \rf{gah}.
 After a change of variables  ($x= \cos \te_2 \ \cos \te_3, \ y= \sin \te_3$) we get  the metric on $\CS^{3}$
     \ci{fl}
 \be \la{mee}
(ds^2)_{n=4}  =  d\vp^2 + {\cot^2 \vp \ dx^2 +
 \tan^2 \vp  \ dy^2   \ov 1- x^2 - y^2  }   \ . \ee
 %We thus we reproduce \rf{red} and \rf{csg} as the reduced models corresponding to
 %strings on $R_t \times S^{n} $ with $n=2$ and $n=3$.
 Thus in  the case of $n=4$ (i.e. for the 
 string on $R_t \times S^{4} $)  we find from \rf{mee},\rf{uu}
 that the reduced theory is described by 
  the following Lagrangian (cf. \rf{csgi})
 \be \la{csg}
   \td L =  \d_+ \vp \d_- \vp  +
    { \cot^2 { \vp
   }\ \del_+ x\ \del_- x +  \tan^2 { \vp }\ \del_+ y\ \del_- y \ov 1 - x^2 - y^2 }
\  + {\km^2\ov 2}  \cos 2\vp\ . \ee
 An equivalent  form of the  metric of $\CS^3$ 
 \rf{mee} was found in  \ci{bs1}
 \be  \la{sg}
  (ds^2)_{n=4} =   {db^2 \ov 4(1-b^2) } -  { 1+ b  \ov 4(1-b)} { dv^2\ov   v (v-u-2) }
 +    { 1-b \ov 4(1+b) } {du^2 \ov   u (v-u-2) }  \ , \ee
 as one can see  by setting $b= \cos 2\vp$, $u= - 2 y^2, \ v=  2 x^2$.
 The  metric-dilaton background for  $\CS^4$ (i.e. $n=5$) case  
 was obtained  in similar coordinates
 $(b,u,v,w)$ in \ci{bs2}.
Setting   $b= \cos 2 \vp$,\ $w= \cos \a, \ v=  \cos \b $  we get
% \bea
% (ds^2)_{n=5}  =&&{ d \vp^2  } +  \tan^2 \vp\   { d u^2 \ov (\cos \b -u) (u- \cos \a    ) }\no \\
% && +\   \cot^2 \vp\ (\cos \b - \cos \a) \bigg[
%  { d \a^2 \ov 4( u-   \cos \a)} + { d \b^2 \ov 4(\cos \b -u) }
%    \bigg] \ . \la{isg}
% \eea
\begin{multline}\quad (ds^2)_{n=5}  ={ d \vp^2  } +  \tan^2 \vp\   { d u^2 \ov (\cos \b -u) (u- \cos \a    ) }\\
+\   \cot^2 \vp\ (\cos \b - \cos \a) \bigg[
 { d \a^2 \ov 4( u-   \cos \a)} + { d \b^2 \ov 4(\cos \b -u) }
   \bigg] \ .\quad \la{isg}
\end{multline}

Together   with the  $\cos 2 \vp$ potential \rf{uu} this metric thus  defines
the reduced model  for the string on $R_t \times S^{5} $.

%%%%%%%%%%%%%%%%%%%%%%%%%%%%%%%%%%%%%%%%%%%%%%%%%%%%%%%%%%%%%%%%%%%%%%%%%%%
\subsection{Reduced model for a  bosonic string in    $AdS_n  \times S^n$  }\label{sec:53}
%%%%%%%%%%%%%%%%%%%%%%%%%%%%%%%%%%%%%%%%%%%%%%%%%%%%%%%%%%%%%%%%%%%%%%%%%%%%%%%%%%%%

 One can  similarly   find the 
reduced Lagrangians   for the $F/G= AdS_n = SO(2,n-1)/SO(1,n-1)$
coset sigma models related to the above ones 
by  an analytic continuation.
  These   reduced models  describe  strings in 
$AdS_n \times S^1$ spaces 
in the conformal gauge with the residual conformal symmetry fixed, e.g.,  by 
choosing the $S^1$ angle $\alpha$  
equal to $ \km \tau$ (cf. \rf{soo}). 

As  was already discussed at the end of section~\bref{sec:2},  the reduced model 
for strings on  $AdS_n  \times S^n$  can  then be obtained  by simply combining 
the reduced models  for strings on  $AdS_n \times S^1$ and on $R \times S^n$.\foot{Note that 
this is  {\it not} the same as  the reduced theory 
for the coset sigma model with $F/G=AdS_n  \times S^n=
 [ SO(2,n-1)/SO(1,n-1) ]\times [SO(n+1)/SO(n) ]$:
in the latter case we would set, following \ci{pol}, the components  of the 
 {\it total}  stress tensor to be equal to a constant, while 
 for a {\it string}  in $AdS_n  \times S^n$  the total stress tensor should
  vanish.  
  The reduced theory   for coset sigma model 
   $F/G= AdS_n  \times S^n$ case is of course formally 
   equivalent to the 
  reduced theory for a string on $AdS_n  \times S^n \times S^1$.
  }

For example, in the case of a string in $AdS_2  \times S^2$  we then find 
the sum of the sine-Gordon and sinh-Gordon  Lagrangians  
(cf. \rf{red},\rf{hasg})
\be \la{sinh}
\td L = \dpp \varphi \dmm \varphi  
+ \dpp \phi \dmm \phi
+ \frac{\km^2}{2} (\cos 2\varphi  - \cosh 2\phi) \ . 
\end{equation}
For a string in  $AdS_3  \times S^3$   we   get (cf. \rf{csgi})
\begin{equation}
 \label{hgi}
   \td  L =
   \dpp \varphi \dmm \varphi  +   \tan^2 { \varphi }\ \dpp \theta \dmm \theta
   +
   \dpp \phi \dmm \phi  +   \tanh^2 { \phi }\ \dpp \chi  \dmm \chi
   + \frac{\km^2}{2} (\cos 2\varphi  - \cosh 2\phi) \ . 
\end{equation}
Similar bosonic actions   are then   found for a string in  $AdS_4  \times S^4$  
and in $AdS_5  \times S^5$: one is  to 
 ``double''  \rf{csg}  and its analog corresponding to \rf{isg}.\foot{A
``mnemonic'' rule 
to get,
e.g.,   the $AdS_n$  counterparts of $S^n$ Lagrangians in 
\rf{csgi},\rf{csg} is
to change $\varphi \to i \phi$ and to change the overall  sign of the
Lagrangian.}

%NEW
Note that while the  $\cos 2\varphi$ potential is a relevant perturbation 
of the coset CFT in 
the compact $S^n$ case,     the $\cosh 2\phi$ is an irrelevant  perturbation 
of the corresponding coset CFT in the  $AdS_n$ case  (the sign of the mass term  in \rf{uu}
is opposite). We expect that in the superstring \adss case the fermionic 
contributions will make the whole theory UV finite, i.e. 
the coefficient in the potential will not run with scale
and thus it can be considered like it is 
an exactly  marginal perurbation  (the value of $\mu$ is 
arbitrary). This is what happens in the $AdS_2 \times S^2$ 
where the reduced theory  is equivalent to the (2,2) supersymmetric sine-gordon theory.

%AAT
%\maxim{The discussion below should probably be seriously rewritten or even removed}

\bs

Expanding \rf{sinh} near $\vp=\phi=0$ we get two massive fluctuation modes. 
Doing similar expansion near the  trivial vacuum in  the case of \rf{hgi}
it may seem that only two modes ($\varphi$  and $\phi$) get masses  $\mu$,  
but,  in fact,  {\it all} 
2+2 bosonic modes  become massive. Indeed, as is clear from the form of kinetic terms
in \rf{hgi}, 
the expansion near the point where all  angles are zero 
 is singular.  This is  like expanding 
near  $r=0,\ \vp=0$ on the disc $ds^2 = dr^2 + r^2 d \vp^2$; instead,   
  one is first to do a  transformation to ``cartesian'' coordinates and then expand. 
  Since $\varphi$  and $\phi$ play the role of the ``radial''  directions in the 
  2+2  dimensional  space\foot{Recall also that 
   they are related to the Lagrange multipliers for the
  embedding coordinates discussed 
  in section~\bref{sec:2} so  we are then expanding  near a point where the two Lagrange 
  multipliers have  constant ``vacuum'' values.
  % This does not, however, imply that 
 %  we get 6+6  massive
 %  modes (cf. quantum mass generation in the usual  $O(N)$ models) 
 %  since in the reduced model we are dealing directly with the  same  physical number of degrees of
 %  freedom.
 }
  their $ \frac{\km^2}{2} (\cos 2\varphi  - \cosh 2\phi)$ potential gives mass 
  to all 4 ``cartesian'' fluctuations.
  In the CSG case 
  this is the transformation that puts  the  Lagrangian  \rf{csgi} 
  into the familiar  form
  $\td L =\ha  { \del_+ \psi \del_- \psi^* +  \del_- \psi \del_+ \psi^*  
   \ov 1 - \psi \psi^* }
    - \mu^2 \psi \psi^* $
  where $\psi= \sin \vp \ e^{i \theta}$.
  
  The analogous conclusion should be true also in  the  general $AdS_n \times S^n$ case
  with $n > 3$  though there a direct demonstration of this 
  in the  gauge where $A_\pm$ are solved for is complicated 
  by the degeneracy of the  metric ${\rm g}_{pq}$   in \rf{dil}.
  As we  have  already seen  in \rf{fluv},\rf{kkk}, 
   in the $S^n$  case all the $(n-1)$ 
  fluctuation modes near the trivial vacuum 
  get mass $\mu$  if we start with the classical equations  of the reduced
   theory in the $A_+=A_-=0$
  gauge. Since the mass spectrum should be gauge-invariant,  the same should be true 
  also in other gauges/parametrizations.

  Thus in the $AdS_5 \times S^5$ case we should get  4+4 massive bosonic modes. 
  Similar conclusion  will be reached    for the fermionic fields 
  discussed in the next section  (see \rf{erm}):  all 8 dynamical fermionic modes will also have
   mass $\mu$. The ``free'' spectrum will  thus be 
   the same  as in the ``plane-wave'' limit of \ci{mt2}. 
 % An interesting question  then is how to generalize the {\it relativisic}
  %(cf. \ci{bei})  S-matrix for the CSG model \ci{doho} to 
  %the  full reduced model for \adss. 

%Let us  now  turn to   the    superstring case. 

%\newpage
%%%%%%%%%%%%%%%%%%%%%%%%%%%%%%%%%%%%%%%%%%%%%%%%%%%%%%%%%%%%%%%%%%%%%%%%%%%%%%%%%%%%%%%%%%%%%%%%%
\section{Pohlmeyer reduction of the $AdS_5\times S^5$ superstring model}\label{sec:spohl}

 The $AdS_5\times S^5$
%%{$AdS_5\times S_5$}
 superstring can be
described in terms of the Green-Schwarz version of the 
$PSU(2,2|4) \over  SO(1,4)\times SO(5)  $
(or, equivalently, $PSU(2,2|4) \over Sp(2,2)\times Sp(4)$)
coset sigma model \cite{mt1}. In the conformal gauge 
its   bosonic part is the direct sum of the $AdS_5$ and $S^5$ 
sigma models.
Below we shall apply the idea  of the Pohlmeyer reduction
to the whole  action including the fermions. The important 
new element will be  the $\kappa$-symmetry gauge fixing,  reducing the 
number of the fermionic 
degrees of freedom  to the same 8  (or 16 real Grassmann components) 
as of the bosonic ones  after the solution of the conformal
gauge constraint.

 We shall derive the corresponding reduced Lagrangian 
that generalizes the bosonic Lagrangian  discussed in section \bref{sec:lag}  above. 
We shall find that  it is invariant under the 2d Lorentz symmetry.\foot{
% and moreover 
%in the case of $psu(1,1|2)$ has also $\ns=2$ 2d world-sheet supersymmetry. 
This is  similar to what happened 
in the  expansion near the $S^5$ geodesic to quadratic order 
(i.e. plane-wave limit)
 in the light-cone gauge \ci{mt2}, but here the action 
contains all interaction terms, i.e.  is 
no longer truncated at the  quadratic level.}
%it is possible it is also conformal 

Later in section~\bref{sec:ads2s2}  we will also   consider a simpler  
 $AdS_2\times S^2$ model 
 which  is described  by a similar action for the 
 $PSU(1,1|2) \over SO(1,1)\times SO(2)$
% $PSU(1,1|2) \over Sp(1,1)\times Sp(2)$ 
coset. In this case the reduced Lagrangian happens to be invariant under the 
   $\ns=2$ 
  (i.e. (2,2))   2d supersymmetry,  and is the same 
  as  the $\ns=2$ supersymmetric sine-Gordon  Lagrangian.

%%%%%%%%%%%%%%%%%%%%%%%%%%%%%%%%%%%%%%%%%%%
\subsection{Equations of motion  in terms of currents in conformal gauge}

Let us start with some relevant definitions and notation. 
The Lie superalgebra $psl(2m|2m;\fC)$ can be identified with the quotient
 of $sl(2m|2m;\fC)$ by the central subalgebra of elements proportional to the 
 unit matrix (which belongs to 
 $sl(2m|2m;\fC)$ since  its supertrace vanishes). We are interested in 
 its real form $psu(m,m|2m)$ which is defined by the condition 
$M^*=-M$, where $^*$ is an  appropriate antilinear anti-automorphism.
% (see 
%Appendix~\bref{sec:smatrix} for more details).
This superalgebra corresponds to the Lie supergroup $\hat{F}=PSU(m,m|2m)$.

We shall consider the   superalgebra $\alghf=psu(m,m|2m)$ with $m=2$ or $m=1$ 
which admits a  $Z_4$ grading~\cite{\berk}\footnote{It appears that 
all the steps of the reduction procedure discussed below are formally valid 
 for any  value of $m$.}
\begin{equation}
\label{decompos}
\alghf=\alghf_0\oplus \alghf_1\oplus\alghf_2\oplus\alghf_3\,,
\qquad  \qquad \commut{\alghf_i}{\alghf_j}\subset \alghf_{i+j\,\mathrm{mod}\,4}\ . 
\end{equation}
In this matrix realisation one also has
 $\imath\scommut{\alghf_l}{\alghf_m}\subset \alghf_{l+m+2\,
 \mathrm{mod}\, 4}$, where $\scommut{A}{B}=AB+BA$. \foot{Note
  that for $A,B$ representing elements 
  of $psu(m,m|2m)$ their symmetrized commutator $\imath\scommut{A}{B}$ belongs to $u(m,m|2m)$
but not necessarily to $psu(m,m|2m)$.} For details see Appendix~\bref{sec:smatrix}.

The left-invariant  current $f^{-1}\del_a f,\,\,f\in \hat{F}$ can then be decomposed as
\begin{equation}
J_a=f^{-1}\del_a f= \A_a  +Q_{1a}+P_a +Q_{2a} \,,\qquad \qquad 
\A \in \alghf_0,\quad Q_1\in \alghf_1,\quad P\in\alghf_2,\quad Q_2\in \alghf_3\ .
\end{equation}
Here $\A$  corresponds to the algebra of the 
subgroup $G$ defining  the $\hat{F}/G$ coset 
(i.e. $G=  Sp(2,2) \times Sp(4)$  isomorphic to $SO(1,4)\times SO(5)$  in the \adss case), 
$P$ is the bosonic ``coset'' component, 
and  $Q_1, Q_2$ are the fermionic  (odd) currents.  

Using this    $Z_4$ split  the  Lagrangian density   of the  $AdS_5 \times S^5$
 GS superstring \ci{mt1} can be written 
  as follows \cite{\berk,roi,bpr,polk}\foot{Here the overall 
  sign is consistent with  having physical signs for the bosonic $AdS_5$ 
  and $S^5$ Lagrangians.}
\begin{equation}\la{lak}
 L_{\rm GS}= \half \mathrm{STr}(\gamma^{ab}P_a P_b + \varepsilon^{ab}Q_{1a}Q_{2b})\,,
\end{equation}
where $\gamma^{ab}= \sqrt{ - g} g^{ab} $.
% is the 
 %Weyl invariant combination of the 2d metric. 
 Written in terms of currents this  coset  action has 
  bosonic gauge symmetry with $\alghf_0$-valued gauge parameter.
  In addition to the reparametrisations it is also 
 invariant under the local fermionic $\kappa$-symmetry \ci{mt1,arutu,ald} 
\begin{equation}
\begin{gathered}
\label{kappa}
\delta_\kappa J_a=\d_a \epsilon+\commut{J_a}{\epsilon}\,, \qquad 
(\delta_\kappa \gamma)^{ab}=\frac{1}{m}\str\left( W(\commut{\imath k^a_{1(-)}}{Q^b_{1(-)}}
+\commut{\imath k^a_{2(+)}}{Q^b_{2(+)}})\right)\,,\\
\epsilon=\epsilon_1+\epsilon_2=\scommut{P_{(+)a}}{\imath k^a_{1(-)}}+\scommut{P_{(-)a}}{\imath k^a_{2(+)}}\,,\\
\end{gathered}
\end{equation}
where\footnote{Note that the definition of $\epsilon$ in \eqref{kappa} involves
 the symmetrized commutator
so that the projection from $u(m,m|2m)$ to $psu(m,m|2m)$ is assumed.}
$k_{1(-)}$ and $k_{2(+)}$ 
take values in the degree $1$ and
 degree $3$ subspaces of $u(m,m|2m)$ respectively (it is assumed that 
$k_{1(+)}=k_{2(-)}=0$). \ 
$W={\rm diag}(1,\ldots,1,-1,\ldots,-1)$ is the parity automorphism 
(see Appendix~\bref{sec:smatrix}),
and the  $(\pm)$ components are defined as:
\begin{equation}\la{vev}
 V_{(\pm)}^a=\half(\gamma^{ab}\mp\varepsilon^{ab})V_b\,.
\end{equation} 
A detailed discussion of the $\kappa$-invariance can 
be found in the Appendix~\bref{sec:kappa}.

In what follows we shall assume the {\it conformal gauge}
 condition
$ \gamma^{ab}= \eta^{ab}$.
%and use the standard light-cone worldsheet coordinates 
%$\sigma^+,\sigma^-$.
 Then (using  the standard light-cone worldsheet coordinates 
$\sigma^+,\sigma^-$)   the only nonvanishing
 components of the metric are $\gamma^{+-}=\gamma^{-+}=1$ 
 while $\varepsilon^{+-}=-\varepsilon^{-+}=1$. For any vector $V_a$ one then has
\begin{equation}
 V_{(+)+}=V_+\,,\qquad V_{(+)-}=0\,,\qquad V_{(-)+}=0\,,\qquad V_{(-)-}=V_-\,.
\end{equation}

In the conformal gauge the Lagrangian \rf{lak}  
\begin{equation}\la{lao}
 L_{\rm GS}= 
\mathrm{STr}[P_+ P_- + \half (Q_{1+}Q_{2-}-Q_{1-}Q_{2+})]\  
%AT: no 1/2 here 
\end{equation}
leads to the following  equations of motion  \ci{bpr}
\begin{equation}
\label{eom}
\begin{aligned}
\d_+P_- + \commut{\A_+}{P_-}+\commut{Q_{2+}}{Q_{2-}}=0\,, \\
\d_-P_+ + \commut{\A_-}{P_+}+\commut{Q_{1-}}{Q_{1+}}=0\,, \\
\commut{P_+}{Q_{1-}}=0\,,\qquad \commut{P_-}{Q_{2+}}=0\,.
\end{aligned}
\end{equation}
Formulated in terms of the current components
$J_\pm =\A_\pm+P_\pm+Q_{1\pm} +Q_{2\pm}$, they should be supplemented by 
the Maurer-Cartan equation 
\begin{equation}
\label{MC}
\d_-J_+-\d_+J_-+\commut{J_-}{J_+}=0\ . 
\end{equation}
In addition, one needs to take into account the 
conformal gauge (Virasoro)  constraints 
\begin{equation}
\label{vir}
\mathrm{STr}(P_+P_+)=0\,, \qquad \qquad \mathrm{STr}(P_-P_-)=0\,.
\end{equation}

Our aim below  
is to perform the Pohlmeyer-type reduction of the 
above system \rf{eom}--\rf{vir}. The bosonic part of the model is identical 
to that of the $F/G$ sigma model where the bosonic group 
$F\subset \hat{F}$  has $\alghf_0\oplus\alghf_2$  as its 
 Lie algebra  and $G$ has  Lie 
algebra $\alghf_0$. In  the $psu(2,2|4)$ case of our interest
 $\alghf_0\oplus\alghf_2$
is isomorphic to $su(2,2)\oplus su(4)$ or  $so(2,4)\oplus so(6)$ 
while $\alghf_0$ is isomorphic to $sp(2,2)\oplus sp(4)$  or 
 $so(1,4)\oplus so(5)$    (in the   $psu(1,1|2)$ case
 $\alghf_0\oplus\alghf_2  
 = su(1,1)\oplus su(2)$   and 
  $ \alghf_0 = sp(1,1)\oplus sp(2)$). 
  Because of 
the direct sum structure of
the algebras one is allowed to use the reduction gauge  separately for 
each sector,  just like in the purely bosonic case.

Performing the  reduction, requires,  besides partially fixing the $G$-gauge symmetry,
 to fix also  the $\kappa$-symmetry  gauge. 
 As we shall discuss below, this can be  achieved in two steps.
 First, we shall  impose the partial $\kappa$-symmetry gauge condition\foot{This  choice
  was suggested  by 
 R. Roiban, see also \ci{mi3}.} 
\begin{equation}\la{kg} 
 Q_{1-}=0\,,\qquad \qquad Q_{2+}=0\,,
\end{equation}
and then  apply the same procedure  as in the case of the Pohlmeyer
reduction in the  bosonic $AdS_n\times S^n$ case. 
The resulting reduced  system will  be  still
invariant under a residual $\kappa$-symmetry which can be fixed  by an 
additional gauge condition. That  will finally  make the number of the fermionic 
degrees of freedom the 
same
 as the number of the  physical bosonic degrees of freedom 
 (as in the familiar examples of the  light-cone gauge-fixed
superstring  in the flat space or in the pp-wave space).

  It will turn out that the resulting system of reduced equations of motion (that 
  originate  in particular 
  from   the Maurer-Cartan equations and thus  are first 
  order in derivatives) will follow from a local Lagrangian containing
only {\it first} derivatives of the fermionic fields. The  bosonic part of the reduced Lagrangian 
will coincide with the gauged WZW Lagrangian with the same   potential
as in  the  bosonic model discussed in section \bref{sec:lag}. 

%AAT
The possibility to make  the gauge choice \rf{kg}  can be readily justified 
as in the flat-space case 
by   using an explicit coordinate parametrization 
of the currents, i.e. by solving first the Maurer-Cartan  equations \rf{MC}.
Here we would like to use a different logic  treating all equations 
for the currents on an equal footing. Then one   way of 
 demonstrating   that the required  $\kappa$-symmetry gauge choices 
 are  allowed will  rely  on using  the consequences of 
 the reduction gauge in the bosonic
part of the model. For that  technical reason   below  we  shall  discuss the 
 reduction and the $\kappa$-symmetry gauges in parallel.

%%%%%%%%%%%%%%%%%%%%%%%%%%%%%%%%%%%%%%%%%%%%%%%%%%%%%%%

\subsection{Reduction gauge  and $\kappa$-symmetry gauge}\label{sec:red-ferm}
%%%%%%%%%%%%%%%%%%%%%%%%%%%%%%%%%%%%%%%%%%%%%%%%%%%%555

As a first step we shall  define  a decomposition $\alghf_2=\alga\oplus\algn$ where $\alga$
is the subspace of elements of the form $a_1 T^1+a_2 T^2$ such that
 $T^1$ and $T^2$ are represented by
matrices with  nonvanishing upper left and lower right blocks only 
(i.e. $T^1$ is in $su(2,2)$ and $T^2$ is in $su(4)$ parts of  $psu(2,2|4)$). More precisely, we shall choose 
\begin{equation}\la{kkg}
 T^1=\frac{i}{2}{\rm diag}(\rt,0)\,,\qquad T^2=\frac{i}{2}{\rm diag}(0,\rt)\ , 
\end{equation}
where
\begin{equation}\la{tkg}
\text{$psu(2,2|4)$ case:}
\quad
\rt={\rm diag}(1,1,-1,-1)\,,\qquad
\text{$psu(1,1|2)$ case:}\quad \rt={\rm diag}(1,-1)\,.
\end{equation}
% in the case of $psu(2,2|4)$ and $psu(1,1|2)$ respectively.
Let us also  introduce the  matrix  
\be \la{tk} 
 T=T^1+T^2\ , 
\ee 
which will play an important role in what follows. It 
  induces the  decomposition 
\begin{gather}
 \alghf=\alghf^\pp\oplus\alghf^\oo \ , \qquad 
 \zeta^\pp\in \alghf^\pp \ , \qquad  \chi^\oo\in {\alghf}^\oo\,,\\
\la{mkm}
  P^\pp\zeta^\pp=\zeta^\pp\, ,   \qquad  P^\pp{\chi^\oo}=0\ , \qquad
  P^\pp=-\commut{T}{\commut{T}{\cdot\ }}\  . 
\end{gather}
This  decomposition can also be written with the help of the projector to 
${\alghf_1}^\oo\oplus {\alghf_3}^\oo$ given by
\begin{equation}\la{mkh}
  P^\oo\chi^\oo=\chi^\oo\,, \qquad P^\oo\zeta^\pp=0\,, 
  \qquad P^\oo=-\scommut{T}{\scommut{T}{\cdot\ }}\,.
\end{equation}
Let us  note that any $\zeta\in \alghf^\pp$  can 
be written as $\zeta=\commut{T}{\lambda}$
(and   $\chi\in \algf^\oo$  can be written as  $\chi=\scommut{T}{\nu}$). In particular,  
$\commut{T}{\scommut{T}{\zeta}}
=\scommut{T}{\commut{T}{\zeta}}=0$ for any $\zeta\in\alghf$.
Moreover, $\str(\zeta^\pp\chi^\oo)=0$ for any $\zeta^\pp\in \alghf^\pp$ and
 $\chi^\oo\in \alghf^\oo$, i.e. this is an orthogonal decomposition.

The decomposition $\alghf=\alghf^\pp\oplus\alghf^\oo$ 
generalizes the bosonic decomposition \eqref{ala}
to the  superalgebra case.
 In particular,  in the bosonic sector one can easily make 
the following 
identifications:\footnote{Let us note that one can not
 define analogous decomposition in terms
of $T_\pm$ for the $SO(n)/SO(n-1)$ coset in the standard representation used in Section~\bref{sec:red-bos}
as $T_\pm$ in this representation do not induce the decomposition (cf. the explicit form \eqref{pil}).}
\begin{equation}\la{kli}
 \alga=\alghf_2^\oo\,,\qquad \algn=\alghf^\pp_2\,,\qquad \algh=\alghf_0^\oo\ , 
  \qquad \algm=\alghf_0^\pp\,,
\end{equation}
while the commutation relations \eqref{dec} follow from the $Z_4$-grading 
and the following properties:\footnote{These can be considered as defining an 
additional $Z_2$-grading on $\alghf$
 with $\alghf^\oo$
and $\alghf^\pp$ being,  respectively,  the degree $0$ and degree $1$ subspaces.}
\begin{equation}\la{mnb}
 \commut{\alghf^\oo}{\alghf^\oo}\subset\alghf^\oo\,,\qquad 
\commut{\alghf^\pp}{\alghf^\oo}\subset\alghf^\pp\,, \qquad
\commut{\alghf^\pp}{\alghf^\pp}\subset\alghf^\oo\,.
\end{equation}
The first two properties are obvious, 
 while checking the last one requires using the following identities
\begin{equation}
\label{sa-ident}
 \scommut{A}{\commut{B}{C}}=\scommut{\commut{A}{B}}{C}+
\commut{A}{\scommut{B}{C}}\,, \ \ 
\quad 
 \scommut{A}{\scommut{B}{C}}=
\commut{\commut{A}{B}}{C}+\scommut{B}{\scommut{A}{C}}\,.
\end{equation}

Let us now turn to the gauge symmetry.
Because the gauge algebra $\alghf_0$ is a direct sum of the subalgebras represented by upper-left and lower-right nonvanishing block matrices
the gauge transformations are independent. It follows that by applying the polar
decomposition theorem in each sector independently one can partially 
fix the $\alghf_0$ gauge symmetry in order to put $P_+$ into  the form 
\begin{equation}
 \la{ec} 
  P_+=p_1 T^1 + p_2 T^2\ , 
\end{equation}
where $p_1,p_2$ are some real functions. Indeed, the components of the gauge parameter taking values
in the upper-left and lower-right diagonal blocks are independent
so that we can apply the same logic as in the bosonic case in section \bref{sec:curr-red}
to each block separately. The Virasoro constraint $\str(P_+P_+)=0$ in   \rf{vir}
then implies $p_1^2-p_2^2=0$, so that, e.g.,  
 $p_1=p_2=p_+$
%(we rule out the  other solution $p_1=-p_2$ by requiring positivity 
%in each sector\footnote{What exactly one should require?})
and thus 
 \be \la{gd}
  P_+=\ p_+\ T  \ , \ \ \ \ \ \ \ \ \ \ \ \ \ T=T^1+T^2 \ . \ee 
Applying the polar decomposition theorem to $P_-$ and using the second Virasoro 
constraint in \rf{vir}  one finds that $P_-$
can be represented as follows
\begin{equation}\la{saf}
 P_-=\ p_-\ g^{-1}Tg\,,%\qquad T=T^1+T^2\,,
\end{equation}
where $p_-$ is a real function and $g$ is a $G$-valued function  (recall 
that $G$ is the Lie subgroup corresponding to the 
 Lie subalgebra $\alghf_0\subset\alghf$, i.e. 
$Sp(2,2) \times Sp(4)$  in the $PSU(2,2|4)$ case). 
In what follows we shall assume that the functions $p_+$ and $p_-$ do not have zeroes.
\bs 

Now we are ready to argue that using the $\kappa$-symmetry \rf{kappa}
one can choose the gauge \rf{kg}, i.e. $Q_{1-}=Q_{2+}=0$, provided the fermionic equations of motion
as well as the Virasoro constraints are satisfied. This basically 
follows from the fact that in the gauge where $P_+=p_+ T$ the 
equation $\commut{P_+}{Q_{1-}}=0$
implies that $Q_{1-}$ takes values in $\alghf_1^\oo$  like the 
 parameter $\epsilon_1=\imath\scommut{P_+}{k_{1-}}$ so that this 
 gauge invariance can be used to put $Q_{1-}$
to zero; an  analogous  argument    can then be  given  for $Q_{2+}$.
A complication is that 
the $\kappa$-transformation 
\rf{kappa} does not in general
preserve both the conformal gauge and the reduction gauge  and  that makes the
 precise argument more involved. 
A detailed proof of the possibility to fix \rf{kg}
 taking all this into account is given in  Appendix~\bref{sec:kappa}.

In the gauge $Q_{1-}=Q_{2+}=0$ the equations of motion \rf{eom} 
become
\begin{equation}
\label{eom-kappa}
\d_+P_- + \commut{\A_+}{P_-}=0\,, \qquad \d_-P_+ + \commut{\A_-}{P_+}=0\,,
\end{equation}
while  the   Maurer-Cartan equation \rf{MC}  splits into 
\begin{equation}
\label{MC-eom}
\begin{aligned}
\d_+\A_--\d_-\A_++\commut{\A_+}{\A_-}+\commut{P_+}{P_-}+\commut{Q_{1+}}{Q_{2-}}&=0\,,\\
\d_-Q_{1+}+\commut{\A_-}{Q_{1+}}-\commut{P_+}{Q_{2-}}&=0\,,\\
\d_+Q_{2-}+\commut{\A_+}{Q_{2-}}-\commut{P_-}{Q_{1+}}&=0\,.
\end{aligned}
\end{equation}
In the reduction gauge where $P_+=p_+T$ and $P_-=p_-g^{-1}Tg$ the second 
equation $\d_-P_+ + \commut{\A_-}{P_+}=0$ in \rf{eom-kappa} and the fact that 
$\A_-$ is  block-diagonal imply 
%AAT
that the same is true for the upper-left block projection  $ \d_- P_+^1+ [A^1_-,P^1_+]=0$. 
The latter  
implies  $\d_-\tr_1(P_+P_+)=0$
and thus also  $\d_-\tr_2(P_+P_+)=0$, where $\tr_1$ and $\tr_2$ are, respectively,
the  traces in the upper-left
and the lower-right diagonal blocks (in this notation  $\mathrm{STr}=\tr_1-\tr_2$). 
Since  $\tr_1 T^2 \neq 0$ this leads to  $\d_-p_+=0$.
 As in the bosonic case, using an  appropriate conformal transformation
$\sigma ^+ \to {\sigma^\prime}^+(\sigma ^+)$  
one can then set $p_+$ equal to  some real constant $\km$. 
Following the bosonic construction one then observes that
the first equation in~\eqref{eom-kappa} leads to  $\d_+\tr_1(P_-P_-)=0$. The conformal 
symmetry $\sigma ^- \to {\sigma^\prime}^-(\sigma ^-)$  allows one to 
set  $p_-=\km$.  Thus finally we get 
%As a result, we get 
\begin{equation}
  \la{ret} 
P_+=\ \km \ T\ ,\qquad \qquad P_-=\ \km\  g^{-1}Tg\,,\ \ \ \ \ \ \ \ \   \km=\const \ , 
\end{equation}
which is the direct  counterpart of the reduction gauge in the bosonic case (cf.
\rf{qc},\rf{gg}).  Note that in terms of the notation used in the bosonic case  here we have 
\be  T_+=T_-=T  \ \la{jow}  \ . \ee
Let us recall that the 
variable $g$  belongs to   $G$, i.e to the 
 subgroup whose Lie algebra is $\alghf_0$. There is 
  a natural arbitrariness in the choice of $g$ since $P_-$ is invariant under 
   $g\to  hg$ if  $h$ is taking values in the subgroup of 
   elements commuting with $T$. 
   %As we will see later
This description  thus has  an  additional   gauge symmetry which we shall use later.

By analogy with the bosonic case in addition to the decomposition $\alghf_2=\alga\oplus\algn$ 
we make use of the decomposition $\alghf_0=\algm\oplus\algh$ where $\algh$ is the centralizer of $\alga$
in $\alghf_0$
%\be \la{decs}
%\alghf_0=\algm\oplus\algh\ ,
%\  \ \ \ \ \ \ 
%algh=\alghf_0^\oo \ , \ \ \ \ \  [ \algh, \alga]=0 \ , \ \ \ \ \ 
%alga=\alghf_2^\oo\ , \ee
%.e. $\algh$ is centralizer of $\alga$ in $\alghf_0$
(recall that $\alga$ is the subspace of 
elements of the form $a_1T^1+a_2T^2$).\foot{In the case of our interest, 
 i.e. $\alghf=psu(2,2|4)$, the algebra 
 $\algh$ is  $[su(2)\oplus su(2)]\oplus[su(2)\oplus su(2)]$, i.e. is 
  isomorphic to  $so(4)\oplus so(4)$.}
In the present case it is useful to identify 
$\algh=\alghf_0^\oo$ and $\algm=\alghf_0^\pp$ so that
the required decomposition of the entire superalgebra 
is induced by a single element $T$ as was observed in
\eqref{kli}. Accordingly, we split 
\be 
\A_+= (\A_+)_{\algh} + (\A_+)_ {\algm} \ , \qquad \qquad   
\A_-=A_-  +  (\A_-)_\algm\ , \ \qquad  A_- \equiv  (\A_-)_\algh \in {\algh}  \ . \ee  
The second equation in ~\eqref{eom-kappa}
then implies $(\A_-)_\algm  =0$ while the first one can be solved for $\A_+$ as follows
\begin{equation}\la{se}
 \A_+=g^{-1}\d_+g+g^{-1} A_+g\,,
\end{equation}
where $A_+$ is a new field taking values in $\algh$. 

\bs 

In this way we have constructed a new parametrisation of the system in the reduction gauge:
all the bosonic  currents are now expressed in terms of the 
  $G$-valued field $g$, $\algh$-valued field $A_{\pm}$, and in addition we have 
the 
fermionic currents $Q_{1+}$, $Q_{2-}$. The equations \rf{MC-eom}
  then take the form:
\begin{equation}\label{main+Q}
\begin{aligned}
\d_-(g^{-1}\d_+g+g^{-1}A_+g)-\d_+A_-+\commut{A_-&}{g^{-1}\d_+g+g^{-1}A_+g}\\
 =& \, -
\km^2\commut{g^{-1}Tg}{T}+  \commut{Q_{1+}}{Q_{2-}}\,,\\[-5pt]
\end{aligned}
\end{equation}
\begin{equation}\label{ferm+Q}
\begin{aligned}
\d_-Q_{1+}+\commut{A_-}{Q_{1+}}=&\km\commut{T}{Q_{2-}}\,,\\
\d_+Q_{2-}+\commut{g^{-1}\d_+g+g^{-1}A_+ g}{Q_{2-}}=&\km\commut{g^{-1}Tg}{Q_{1+}}\,.
\end{aligned}
\end{equation}
These equations
%(or equivalently,   equations \eqref{main+Q}\eqref{ferm+Q}) 
are invariant under the following  $H \times H$ gauge symmetry ($H$ is the  group whose  algebra
is $\algh$):
 \begin{gather}
\label{gs+f}
g \to h^{-1}g \bar h\,, \quad A_{+}\to h^{-1}A_+h +h^{-1}\d_+ h\,, \quad   A_{-}\to 
{\bar h}^{-1}A_{-}{\bar h}+{\bar h}^{-1}\d_- \bar h\,,\\
\label{Q-gs}
Q_{1+}\to {\bar h}^{-1} Q_{1+} \bar h \,, \qquad \qquad Q_{2-}\to {\bar h}^{-1} Q_{2-} \bar h\,. 
\end{gather}
Let us note  that this symmetry is large enough to choose the gauge  $A_+=A_-=0$.
 This can be shown by a simplified version of the argument given in Appendix~\bref{sec:ils}.
  In particular, there is also a choice of a partial gauge in which 
$A_+$ and $A_-$ are components of a flat connection, i.e. $F_{+-}=0$. 

\bs 

The  equations \eqref{main+Q},\eqref{ferm+Q} admit a Lax representation.
Moreover, they can be derived  from a local Lagrangian
provided one uses the following parametrisation of the fermionic currents 
in terms of the  new fermionic variables $q_1,\ q_2$ via  
%\begin{equation}
$Q_{1+}=g^{-1}(\d_+ q_1+\commut{A_+}{q_1})g\,,\ \  Q_{2-}=\d_- q_2+\commut{A_-}{q_2},$
%\end{equation}
and imposes the  appropriate gauge condition on $A_\pm$. This gauge condition is
 analogous to the constraints
\eqref{lwc} in the purely bosonic case. However, the resulting Lagrangean system is not completely
satisfactory, in particular, it contains second (instead of usual first)
%(i.e. higher)
 derivatives of the fermions  and thus will not be discussed below.  
% As we are going to see in the next section equation \eqref{main+Q},\eqref{ferm+Q} are still invariant
% under the residual $\km$-symmetry whose fixing allows one to arrive at the Lagrangian in terms
% the original fermionic currents (more precisely their components that do not vanish in this $\km$-gauge) 
% and therefore involving at most first order derivatives of the fermions.

\subsection{Gauge-fixing  residual $\kappa$-symmetry}
%%%%%%%%%%%%%%%%%%%%%%%%%%%%%%%%%%%%%%%%%%%%%%%%%%%%%%%%%%%%%%%%%%%%%%%%%5

Besides the gauge symmetry \eqref{gs+f},\eqref{Q-gs}, the 
equations \eqref{main+Q},\eqref{ferm+Q} are also invariant under the residual $\kappa$-symmetry
which can be used to eliminate some  parts of the fermionic currents.
 To identify this symmetry
let us first introduce the new fermionic variables $Q_{1+}, Q_{2-} \to   \Psi_1, \Psi_2$:
\begin{equation}\la{psa}
\Psi_1=Q_{1+}\,,\qquad \qquad \Psi_2=gQ_{2-}g^{-1}\ . 
 \end{equation}
The equations of motion \eqref{main+Q},\eqref{ferm+Q} then take the form 
\begin{equation}
\label{main+Psi}
\begin{aligned}
\d_-(g^{-1}\d_+g+g^{-1}A_+g)-\d_+A_-&+\commut{A_-}{g^{-1}\d_+g+g^{-1}A_+g} \\
\qquad &=~ -\km^2\commut{g^{-1}Tg}{T}-\commut{g^{-1}\Psi_2g}{\Psi_1}\,,\\[-5pt]
\end{aligned}
\end{equation}
\begin{equation}\label{ferm+Psi}
 D_- \Psi_1=\km\commut{T}{g^{-1}\Psi_2g}\,,
 \quad \qquad D_+\Psi_2=\km\commut{T}{g\Psi_1 g^{-1}}\,, \ \ \ \ \ \ \ \ \ \ 
 D_\pm = \del_\pm + [A_\pm, \  ]  \ . 
\end{equation}
%Recall, that $h,\bar h$ are gauge parameters taking values in the subgroup $H\in\hat F$ corresponding to the
%subalgebra $\algh$. 
Projecting the fermionic equations \eqref{ferm+Psi} to ${\alghf_1}^\oo\oplus {\alghf_3}^\oo$ gives 
\begin{equation}\la{pps}
 D_-(\Psi_1)^\oo=0\,, \qquad D_+(\Psi_2)^\oo=0\,.
\end{equation}
Let us  choose the gauge   where (cf. the remark made below \rf{Q-gs})
\be \la{ano} 
A_+=A_-=0 \ .\ee
%This implies  that up to an appropriate gauge transformation 
%(cf. the remark made above  that there exists a gauge where
%that sets $A_+=A_-=0$ (cf. the remark made below \rf{Q-gs}) 
Then the  solution of \rf{pps} has
the form $(\Psi_1)^\oo=\psi_1(\sigma^+)$ and $(\Psi_2)^\oo=\psi_2(\sigma^-)$.

Let us now describe the residual fermionic symmetry of the equations~\eqref{main+Psi},\eqref{ferm+Psi}.
% To this end assume that we are in the gauge where $A_+=A_-=0$. 
 Under the infinitesimal
transformation 
\be  
\Psi_1\to\Psi_1+\varepsilon_1\ , \qquad \Psi_2\to\Psi_2+\varepsilon_2\ , 
\qquad  g\to g+gh\ ,  \ee  with $\varepsilon_1\in\alghf_1$, $\varepsilon_2\in\alghf_3$, and
$h\in\alghf_0$ these equations are invariant provided
%\begin{equation}
\begin{multline}\la{bis}
\d_-\d_+h+\commut{g^{-1}\d_+g}{h}-\km^2\commut{\commut{g^{-1}Tg}{h}}{T}\\
+\commut{g^{-1}\Psi_2g}{\varepsilon_1}
+\commut{g^{-1}\varepsilon_2g}{\Psi_1}+\commut{\commut{g^{-1}\Psi_2g}{h}}{\Psi_1}=0\,,
\end{multline}
%\end{equation}
\begin{equation}
 D_- \varepsilon_1=\km\commut{T}{g^{-1}\varepsilon_2g+\commut{g^{-1}\Psi_2g}{h}}\,,\qquad
 D_+\varepsilon_2=\km\commut{T}{g\varepsilon_1 g^{-1}+g\commut{h}{\Psi_1}g^{-1}}\,.
\end{equation}
Projecting the fermionic equations on $\alghf^\oo$ one finds 
that $\d_-\varepsilon_1^\oo=0$ 
and $\d_+\varepsilon_2^\oo=0$, implying  $\varepsilon_1^\oo$=$\varepsilon_1^\oo(\sigma^+)$
and $\varepsilon_2^\oo=\varepsilon_2^\oo(\sigma^-)$.
Let us consider then the projection  of the fermionic equations on 
$\alghf^\pp_1\oplus\alghf^\pp_3$  together   with the bosonic 
 equation \rf{bis}  as a system of equations 
  on $\varepsilon_1^\pp,\varepsilon_2^\pp,h$
with $\varepsilon_1^\oo(\sigma^+)$ and $\varepsilon_2^\oo(\sigma^-)$ treated as given functions
(note that their derivatives do not enter these equations).
This system of partial differential equations is not overdetermined 
and is linear in  derivatives 
so that it has
a solution for any $\varepsilon_1^\oo(\sigma^+)$ and $\varepsilon_2^\oo(\sigma^-)$, thus  giving a symmetry 
transformation of the equations~\eqref{main+Psi},\eqref{ferm+Psi}. The symmetry parameters
  $\varepsilon_1^\oo$ and $\varepsilon_2^\oo$
can,  in fact, 
 be identified as  parameters of  the residual $\kappa$-symmetry in 
\rf{kappa} as\foot{Note that in the gauge \rf{kg}
 the residual $\kappa$ symmetry is determined
 by $k_1$, $k_2$ satisfying $\d_-k_{1-}=0$ and $\d_+k_{2+}+\commut{g^{-1}\d_+g}{k_{2+}}=0$.} 
\begin{equation}
\varepsilon^\oo_1=\d_+\scommut{\km T}{\imath k_{1-}}\,, \qquad
\qquad \varepsilon^\oo_2=\d_-\scommut{\km T}{\imath g k_{2+} g^{-1}}\,,
\end{equation}
 while the 
additional terms are needed to maintain the gauge conditions we have chosen.
Finally, using \rf{pps}, i.e.
 $\d_-\Psi_1^\oo=0$ and $\d_+\Psi_2^\oo=0$ one concludes 
that $\Psi_1^\oo,\Psi_2^\oo$ can be put to zero by
the residual $\kappa$-transformations.
 In what follows we shall thus assume the gauge where 
 \be \la{psig}
 \Psi_1^\oo=\Psi_2^\oo=0 \ . \ee

 The remaining  fermionic degrees of freedom can be parametrized as follows 
\begin{equation}\la{psis}
 \Psi_{_R}=\frac{1}{\sqrt{\km}} \Psi_1^\pp\,,\qquad\qquad \Psi_{{_L}}=\frac{1}{\sqrt{\km}}
 \Psi_2^\pp\,,
\end{equation}
taking values in $\algh_1^\pp$ and $\algh_3^\pp$ respectively
(see  \rf{mkm},\rf{mkh}). 
As we shall see below  the additional factor
 $\km^{-\half}$  in  \rf{psis}  will  simplify the structure of the 
 2d  Lorentz invariant Lagrangian description of the resulting  system 
%is needed for the Lagrangian description  and 
%is  natural as it reflects the 2d  Lorentz transformation
%properties of these variables 
(cf. \rf{ret}).
 The gauge transformations of the new fermionic 
 variables read as follows
\begin{equation}
\label{gs-opsi}
\Psi_{{}_R}\to {\bar h}^{-1} \Psi_{{}_R} \bar h \,, \qquad \qquad \Psi_{{}_L}\to h^{-1} \Psi_{{}_L} h\,.
\end{equation}
The equations of motion~\eqref{main+Psi},\eqref{ferm+Psi} written in the 
gauge \rf{psig} are 
 %(\maxim{``restoring the general values of $A_+,A_-$'' -- what does it mean?}) 
\begin{equation}
\label{main+psi}
\begin{aligned}
\d_-(g^{-1}\d_+g+g^{-1}A_+g)-\d_+A_-&+\commut{A_-}{g^{-1}\d_+g+g^{-1}A_+g} \\
\qquad &=~ -\km^2\commut{g^{-1}Tg}{T}-\km\commut{g^{-1}\psl g}{\psr}\,,\\[-8pt]
\end{aligned}
\end{equation}
\begin{equation}
\label{red-f}
 \commut{T}{D_-\Psi_{_R}}=-\km(g^{-1}\Psi_{_L} g)^\pp\,,\qquad\ \  \commut{T}{D_+\Psi_{_L}}=
-\km(g\Psi_{ _R}g^{-1})^\pp\,.
\end{equation}
These equations and the 
gauge symmetries \rf{gs+f},\eqref{gs-opsi} 
define the {\it reduced}  system of equations of motion for the 
 superstring on $AdS_5\times S^5$  (or on  $AdS_2\times S^2$).
 
  The  new dynamical field variables  $g, \Psi_{_L}, \Psi_{_R}$ and $ A_+, A_- $ are components 
 of the   currents, i.e. they 
 are non-locally related to the original \adss  sigma model fields (coordinates on the supercoset).   
 Note also that the bosonic equations are second-order while 
 the fermionic equations are first-order in derivatives, as it should be 
  for a standard 2d boson-fermion system.

\bs

Finally, let us mention that one can see explicitly that the reduced system 
\eqref{main+psi} and \eqref{red-f} is integrable. 
The corresponding  Lax pair  
 encoding the  equations~\eqref{main+psi} and \eqref{red-f} is 
\begin{equation}
\begin{gathered}
\cL_-=\d_-+A_-+ {\lam}^{-1} \sqrt{\km}g^{-1}\psl g +\lam^{-2}\km g^{-1}Tg\,,\\
\cL_+=\d_+ +g^{-1}\d_+ g+ g^{-1}A_+g+ \lam \sqrt{\km}\psr+\lam^2 \km T\,.
\end{gathered}
\end{equation}
To show that the compatibility conditions $\commut{\cL_-}{\cL_+}=0 $
imply the equations of motion   ~\eqref{main+psi} and \eqref{red-f} 
 one needs to use \rf{mkm},\rf{psis}, i.e.  that   $\commut{T}
{\commut{T}{\Psi_{{{}_{L,}}{}_R}}}=-{\Psi_{{}_{L,} {}_R}}$.

%%%%%%%%%%%%%%%%%%%%%%%%%%%%%%%%%%%%%%%%%%%%%%%

\subsection{Reduced  Lagrangian: 2d Lorentz symmetry, massive spectrum\space{ }\\
 and possible  2d supersymmetry}
\label{Lorentz}

%%%%%%%%%%%%%%%%%%%%%%%%%%%%%%%%%%%%

Remarkably, it turns out that  the  equations of motion
 \eqref{red-f} and \eqref{main+psi} 
 follow from the following local  Lagrangian:
\begin{multline}
\label{L-tot}
\qquad L_{tot}=L_{\rm gWZW}+\km^2\,\mathrm{STr}(g^{-1}Tg T )\\
+{\textstyle \frac{1}{2}}\mathrm{STr}\left(\Psi_{_L}\commut{T}{D_+\Psi_{_L}}+
\Psi_{_R}\commut{T}{D_-\Psi_{_R}}\right)
+\ \km\,
\mathrm{STr}\left(
g^{-1}\Psi_{_L}g\Psi_{_R}\right)\,,\qquad 
\end{multline}
where $L_{\rm gWZW}$ represents the $G/H$ gWZW model  \rf{gaui}  with\foot{Here  
$L_{\rm gWZW}$ is given by  \rf{gaui}  with 
$\tr$  replaced by the $-\str$. The minus sign is needed to compensate
for  the definition of the supertrace 
which includes the $S^m$ sector  with a minus sign
(the use of supertrace  in the first two bosonic terms  means of course 
 just the sum 
of the reduced models for the  $AdS_5$ and the $S^5$  parts). 
The  corresponding reduced action  
  $S_{tot} =\int \frac{d^2\sigma}{2\pi} \ L_{tot}$   is  real (as can be
  seen by applying
the conjugation $*$ defined in Appendix C to the expression under the trace).}
$${G\ov H} = { Sp(2,2) \ov SU(2) \times SU(2) } \times 
{ Sp(4) \ov SU(2) \times SU(2) }$$ 
Note 
$L_{tot}$ is explicitly  $H$ gauge-invariant under~\eqref{gs+f},\eqref{gs-opsi} 
with $h=\bar h$.\foot{As  was already mentioned above, our reduction 
procedure formally applies and leads to  the Lagrangian \rf{L-tot} 
if one   starts with  any  $psu(m,m|2m)$; in particular,  
  the  $m=1$ case  corresponds to 
$AdS_2\times S^2$ superstring model.}
The dimension of the bosonic target space here is the same as the
 dimension of the $G/H$ coset, i.e. 4+4=8. The  fermionic fields 
  contain  8+8 independent real  Grassmann components (describing 8 dynamical degrees of freedom).

The variations over $g$ and $\PsiL,\PsiR$ indeed lead to \rf{main+psi},\rf{red-f}.
Thus in 
order to show that the reduced model \rf{main+psi},\rf{red-f}
is  described by ~\eqref{L-tot}
one  is to demonstrate that the constraint equations that arise from  varying this 
 action with respect to
$A_{\pm}$ represent an admissible gauge condition for the equations of motion.\footnote{Note that in
 the $AdS_2 \times S^2$ case the
 subalgebra $\algh$ is 
empty and so this step is trivial.}
These constraints read as
\bea
\label{A-const}
&& A_+=(\hat A_+)_\algh\,,  \qquad \hat A_+ \equiv  g^{-1}\d_+g+g^{-1}A_+g 
-\half\commut{\commut{T}{\PsiR}}{\PsiR}\,,
 \\
 \label{hatA}
&&  A_-=(\hat A_-)_\algh \,, \qquad
\hat A_- \equiv  g\d_-g^{-1}+gA_-g^{-1} 
-\half\commut{\commut{T}{\PsiL}}{\PsiL}\,.
\eea
In the Appendix~\bref{sec:ils} we show that  they   can be satisfied 
by an appropriate on-shell gauge transformation. Note that once these constraints are
satisfied       the original $H \times H$  ``on-shell''  gauge symmetry
\eqref{gs+f},\eqref{gs-opsi} of the equations of motion 
having  independent  $h$ and $\bar h$ parameters reduces to the $H$  gauge symmetry 
with $h=\bar h$  which is  the ``off-shell'' gauge symmetry of the 
Lagrangian  \rf{L-tot}.\footnote{
More generally, similarly 
to the purely bosonic case, 
one can consider an  asymmetric gauge determined by an automorphism $\tau$   of  $\algh$
  preserving the supertrace.
In this case the residual gauge transformations are $g 
\to h^{-1}g \hat\tau(h)$, 
$\psr\to \hat\tau(h^{-1})\psr\hat\tau(h)$
%$\psr\to \hat\tau(h^{-1}\psr\hat\tau(h)$
 with transformations of  the remaining variables unchanged. 
The Lagrangian
of the asymmetrically gauged model is given by~\eqref{L-tot}
 with $A_-$ in   
  $A_- \,g\inv\del_+ g   - g\inv A_+ g  A_-$ terms in \rf{gaui} 
 replaced with $\tau(A_-)$.    
}

\bs  
Let  us now discuss several properties of this reduced action.

The  Lagrangian   \rf{L-tot}
is formulated in terms of the left-invariant
$\hat F$ current variables (cf. \rf{ret}, \rf{psis})
that are ``blind'' to the original
$\hat F=PSU(2,2|4)$  symmetry.  Note that  since the original coset 
$\hat F/G =PSU(2,2|4)/[ Sp(2,2) \times Sp(4)]$  has the  purely
 {\it bosonic} factor $G$, 
the reduced action \rf{L-tot}   has only the {\it bosonic} global
%\maxim{Why? It can have e.g. 2d susy which is a global symmetry..}
 and  gauge    symmetries, 
i.e. it has  no  target-space supersymmetry
(but may have 2d supersymmetry, see below). 
 
 \bs 

It is interesting to notice 
that  the Lagrangian~\eqref{L-tot} can be rewritten as 
\begin{equation}\la{rerw}
L_{tot}=\hat L_{gWZW}+L_{add}\,,\ \ \ \ 
 \qquad L_{add}= \str\left[P_+P_-  + \half(Q_{1+}Q_{2-}-Q_{1-}Q_{2+})\right]\,.
\end{equation}
Here $\hat L_{gWZW}$ is the  $G/H$ bosonic 
gWZW  Lagrangian  supplemented with 
 the 
``free'' fermionic terms ${\textstyle \frac{1}{2}}\mathrm{STr}\left(\Psi_{_L}\commut{T}{D_+\Psi_{_L}}+
\Psi_{_R}\commut{T}{D_-\Psi_{_R}}\right)$
while   $L_{add}$  stands for the sum of the remaining $\km$ dependent terms
in  \eqref{L-tot}.
Here we  restored the original notations for the 
current components, i.e. used that 
$P_+=\km T, \ P_-=\km g^{-1}Tg$ (see \rf{ret}), that 
 $Q_{1+}=Q_{2-}=0$ due 
to the $\kappa$-symmetry gauge condition~\eqref{kg}, and that 
$Q_{1+}=\psr, Q_{2-}=g^{-1}\psl g$ in \rf{psa}. 
Remarkably, 
$L_{add}= \km^2\,\mathrm{STr}(g^{-1}Tg T ) +  \km\,
\mathrm{STr}\left(
g^{-1}\Psi_{_L}g\Psi_{_R}\right)    $ 
is thus   nothing but  the original superstring  Lagrangian~\eqref{lao}
rewritten in terms of the  new variables $g,\psr,\psl$.
 At the same time, the equations  following 
 from  $\hat L_{gWZW}$
  encode the Maurer-Cartan 
equations \rf{MC-eom}  for the $\hat F$  currents.
It is then clear that 
%at
once the conformal gauge (Virasoro)
 constraints are imposed, $L_{tot}$ describes,  
at least at the level of the corresponding 
 equations of motion  and up to the various gauge transformations and fixing 
 the values of the conserved quantities in terms of  $\km$, 
  the
 same field configurations  as the original
 superstring 
  sigma-model Lagrangian \rf{lak},\rf{lao}.
   An interesting question is whether  one can  implement  a similar 
   argument `off-shell''   or  even at the quantum level in terms of  
 path-integral transformations.\foot{A natural idea   is to start 
 with the original superstring sigma model  path integral 
 in the conformal gauge  (i.e. with the delta-function insertions 
 $\delta(T_{++}) \delta(T_{--})$), fix the $\k$-symmetry gauge  and change variables from coset
 coordinates to $PSU(2,2|4)$ currents.  The  $\hat L_{gWZW} $ term  in the 
 path integral action may 
 then appear  due to this  change of variables. This procedure can work only if the original 
 path integral represents  a 
 2d  conformal theory:  in the reduction procedure  we used the residual conformal  symmetry.}

\bs  

Despite the fact that the 2d Lorentz invariance may appear to be  broken 
by  various gauge choices made  above and that 
$\Psi_{_L}$ and $\Psi_{_R}  $ originated from the 
2d vector components of the fermionic currents 
(cf. \rf{psa},\rf{psis}) 
it is remarkable that  it is still possible to 
assign  the fermions  the  $SO(1,1)$ Lorentz transformation rules 
of the  components of the left and right 
 2d  Majorana-Weyl spinors. Then  
  the Lagrangian \rf{L-tot}  becomes  invariant
under the standard  2d  Lorentz  symmetry 
\begin{equation} \la{lora}
\sigma^+\to \Lambda \sigma^+ \ , \ \ \  \ \
\sigma^-\to \Lambda^{-1} \sigma^- \ , \ \ \  \ \ \ \ \  \ \
\psl \to \Lambda^{1/2}  \psl\,,\qquad \psr \to \Lambda^{-1/2}\psr\,,
\end{equation}
with $g$  and $A_\pm$   having the usual   scalar and vector 
transformation laws.
Choosing  a  parametrisation for the matrix variables
$\PsiL$ and $\PsiR$  which satisfy 
 the ``parallel'' constraint in \rf{psis},\rf{mkm}\foot{The ``parallel''
subspace is formed by anti-diagonal matrices with fermionic $2\times2$ blocks.
}
one can put the  fermion kinetic terms in \rf{L-tot} into the familiar   form 
$\vpsiL \del_+ \vpsiL +   \vpsiR \del_- \vpsiR + ...$.

\bs 

As in the case of the bosonic reduced theory 
the  classical  conformal invariance 
of the original superstring sigma model in the conformal gauge is   broken 
by the $\km$-dependent interaction terms in \rf{L-tot}: the residual 
conformal diffeomorphism symmetry   was used (cf. \rf{ret}) 
to perform the reduction procedure. This   breaking is ``spontaneous''
being due to  the presence of the ``background field'' $T=T_+=T_-$. 
This is similar to what happened  in the light-cone gauge in the  plane-wave model
\ci{mt2}  where  the mass terms (proportional to the  light-cone momentum,
i.e.  appearing from the  $\del x^+$ terms) 
 were spontaneously breaking the classical  conformal invariance 
 of the original sigma model  action.

\bs \bs

Again as   in the bosonic case discussed in  section~\bref{sec:lag}, 
the  form of the reduced  Lagrangian expressed  in terms of only  ``physical''  
 bosonic and fermionic fields
   may be  found by imposing an 
$H$ gauge fixing condition 
on $g$  and then integrating out the $H$  gauge field components  $A_\pm$.
This   leads to a sigma-model   with  4+4 dimensional bosonic  part \rf{si} 
supplemented by the  fermionic terms, with  the following 
general structure 
(cf. \rf{si}) 
 \be \la{erm}
 \td L =
 G(x) \del_+ x \del_ - x - \km^2 U(x)   +  
  \vpsiL {\cal D}_+ \vpsiL +     \vpsiR {\cal D}_- \vpsiR +   
  F(x) \vpsiL \vpsiL \vpsiR  \vpsiR
 +   2 \km  H(x) \vpsiL \vpsiR   \ .
  \ee 
Here $x$ stands for 8 real  bosonic fields in \rf{si}  (i.e. for the  independent variables  in gauge-fixed 
 $g$
which parametrize $G/H$) 
and $\vpsiL,\vpsiR$  -- for    
 8+8 independent real Grassmann fields which are the components of the matrices 
  $\PsiL, \PsiR$.
 The quartic fermionic term originates from the $D_\pm$ terms  in \rf{L-tot} 
 upon integrating out $A_\pm$  ($ {\cal D}_\pm$  in \rf{erm} are the standard $x$-dependent 
 covariant derivatives). 
   As discussed below, the   structure  of 
 \rf{L-tot}  looks very similar to that of  the supersymmetric gWZW model 
 modified  by the bosonic  potential and the fermionic ``Yukawa'' terms, 
  and so 
 the presence of the  quartic fermionic terms in \rf{erm} may be interpreted as 
 reflecting    the curvature of the target space.

 \bs

 Let us now   discuss  the vacuum structure   and the corresponding 
 mass spectrum
 of  the reduced  model \rf{L-tot}. 
 %and  the corresponding  massive spectrum of small fluctuations near the vacuum.
 Since $[T, H]=0$ the  obvious  vacuum solution  of the equations of motion 
 \rf{main+psi},\rf{red-f} for  \rf{L-tot}  corresponds to $g$ being any  constant 
  element $h_0$ of $H$, i.e. 
 \be  \la{vaca}
 g_{\vac} = h_0= \const  \ , \ \   \ \ \  (A_+)_\vac=(A_-)_\vac=0
  \ ,\  \ \ \ \ \  (\Psi_{_L})_\vac=(\Psi_{_R})_\vac=0 \ , \ee
 i.e.   the  space of vacua is equivalent to $H=[SU(2)]^4$. By a global $H$ 
 transformation we can always set $h_0=1$, i.e. the mass spectrum should 
 not depend on $h_0$. 
 Expanding  the equations of motion  \rf{main+psi},\rf{red-f} near $g=\id$, i.e. 
  $g= \id + v + ...,$  and projecting to the algebra  of $H$ and its complement 
  in $\algg$  we find  a massive equation for 
  $v\in \algm\equiv \algf_0^\pp$ (i.e. $v= [[T,v],T]$, see \rf{mkm}) 
   as well as $F_{+-}=0$.\foot{Equivalently, expanding the action \rf{L-tot} 
   to quadratic order in fluctuations   the $A_+ A_-$ term will cancel while the 
   term linear in $A_+,A_-$ 
  will project $v$ to the coset  part $\algm$ of the algebra  $\algg$.
%$G$
}
  That all bosonic  coset directions get mass  $\mu$ 
  was mentioned already in section 5.3 and follows also  directly
  from the  equations of motion in the $A_+=A_-=0$ on-shell gauge
  in the parametrization used in  \rf{kkk},\rf{fluv}.
  The  linearized  bosonic and fermionic   equations are thus 
 \bea
 \la{fkk}
  \d_+\d_- v  + \mu^2 v &=&0\,, \\
\la{kyj}
  \commut{T}{\d_-\psr} +   \mu{\psl}=0\,, 
 \quad 
 \commut{T}{\d_+\psl  }+\mu  {\psr}   =0 \  \ \ &\to& \ \ \ 
 \d_+\d_-\Psi_{_{L,R}}   + \mu^2 \Psi_{_{L,R}} =0\ , 
\eea
where we used  that $[T,[T, \Psi_{_{L,R}}]]= -\Psi_{_{L,R}}  $  (see \rf{mkm},\rf{psis}).
 The 8+8 independent  real Grassmann components of the fermionic 
matrix fields  thus represent  8 massive  2d  Majorana  fermions 
having  the same mass $\mu$ as the bosonic modes.
The corresponding fermionic Lagrangian is then 
 $$ \vpsiL \del_+ \vpsiL +     \vpsiR \del_- \vpsiR    
  -   2 \km   \vpsiL \vpsiR + ...\ , $$
    where  the mass term originates from 
  the   last ``Yukawa''  term 
  in \rf{L-tot},\rf{erm}.\foot{This and other points discussed in this section can be 
  illustrated on the  $AdS_2 \times S^2$ example  discussed in  the next section
  (see, e.g., 
  \rf{aaac} 
   below 
  where  one is to expand near $\vp=\phi=0$).
  }

\bs 
The small-fluctuation spectrum we get is thus formally the same
as in the plane-wave limit   \ci{mt2}.
In contrast to the case of the original \adss  superstring expanded 
near the $S^5$  geodesic in the light-cone gauge where
one scatters ``magnons'' which are  small fluctuations  of the  
superstring  coordinates 
and 
 the remaining 
symmetry is  $[PSU(2|2)]^2$ \ci{bei,fpz}, here 
we scatter  the fluctuations of the 
current  components which are invariants of  the original supergroup  $PSU(2,2|4)$.
The manifest  global symmetry 
of the  S-matrix corresponding to \rf{L-tot}  in the vacuum \rf{vaca}
appears to  be just the bosonic $H= [SU(2)]^4$ one.\foot{If we start with the closed 
string picture with the  sigma model defined on a cylinder $R \times S^1$ 
we need to take the $\mu \to \infty$   limit (which ``decompactifies'' the spatial 
world sheet direction) 
 to  define the scattering 
matrix.
 An interesting question  then is how to generalize the {\it relativisic}
 (cf. \ci{bei}) 
  S-matrix for the CSG model \ci{doho} to 
 the  full reduced model for \adss. }

Indeed,   while the Lagrangian \rf{erm} obtained by integrating  out 
the $H$ gauge fields 
does not have manifest non-abelian global symmetry,  it is natural to expect 
that the tree-level S-matrix for scattering of the massive  excitations near the vacuum 
\rf{vaca} can be extracted  directly from the classical  equations 
 of motion \rf{main+psi},\rf{red-f}. The latter admit   larger on-shell $H \times H$  gauge
 symmetry   allowing us to  
 choose the $A_+=A_-=0$ gauge in which the global $H$-symmetry 
 of the remaining non-linear  equations  and thus of the resulting (gauge-independent)
 S-matrix becomes manifest. The same $H$ symmetry is expected also to be present in the 
 full quantum S-matrix.\foot{The S-matrix  should  also have higher hidden 
  symmetries presumably related to those of the S-matrix in \ci{bei};
 we  thank R.Roiban for a discussion of this point.}

 \bs 

Let us now   comment  on the meaning of the parameter 
   $\mu$ which  plays  a  crucial role in our reduction procedure 
   and sets the  mass scale.\foot{We thank S. Frolov for
   asking this question  and useful discussions.}
   $\mu$   entered first 
    through the conditions $P_+=\mu T, \ \ P_-=\mu  g^{-1} T g $ \rf{qc},\rf{ret} 
   on the $\pm $ components of the coset-space part of the current
   that solve the conformal gauge constraints. 
   In the vacuum \rf{vaca}  we thus have (cf. \rf{kkg},\rf{tkg})
\begin{equation} \la{jjd}
     (P_+)_\vac= (P_-)_\vac =\   \mu\  T   \ , \qquad\quad
   T= \frac{i}{2}{\rm diag}(1,1,-1,-1;1,1,-1,-1 )  \ . 
\end{equation}
Thus  $\mu$  determines the scale  while  $T$ -- the structure 
of the background values of the coset currents. The corresponding charges
(defined assuming the world sheet is a cylinder) 
thus have  both the $AdS_5$ and $S^5$  non-zero components.
Though  $P_\pm$ are invariants of $PSU(2,2|4)$ their non-zero  vacuum 
values appear to translate, in particular,  into the 
non-zero values of the quadratic Casimirs for $SO(2,4)$   and $SO(6)$ group.
This suggests  again a close relation to the BMN limit.\foot{In a certain sense, 
  our  reduction procedure  may then  be interpreted as 
 an ``invariant version''  of the expansion near the BMN vacuum.}

 In general, to relate the reduced or ``current''  formulation  of 
 the  theory to  the 
 original  \adss superstring model  \rf{lak}   (and thus to  gauge  theory 
 within the AdS/CFT  duality) one would  need to supplement the quantum 
 theory based on \rf{L-tot}  by a  list  of ``observables''  which are intrinsic to 
 the \adss  string in its original  coordinate-space  formulation.  
 This list  should include, in particular,  the components of the $PSU(2,2|4)$ 
 charges. They  cannot be computed directly without  supplementing the 
 reduced action with a  linear  problem for the associated Lax pair, 
 but according to   the above  remarks  about the vacuum   values of currents in \rf{jjd}
 we are  guaranteed  to have  at least some  components of the 
 $AdS_5$ and $S^5$   charges to be non-zero in the  vacuum 
 \rf{vaca}  of the reduced theory.

   % An interesting question  then is how to generalize the {\it relativisic}
  %(cf. \ci{bei})  S-matrix for the CSG model \ci{doho} to 
  %the  full reduced model for \adss. 

\bs \bs
%%%%%%%%%%%%%%%%%%%%%%%%%%%%%%%%%%%%%%%%%%%%%%%%%%%%%%%%%%%%%%

%\subsection{2d supersymmetry of the reduced Lagrangian}

%%%%%%%%%%%%%%%%%%%%%%%%%%%%%%%%%%%%%%%%%%%%%%%%%%%%%%%%%%%%%%%%

 Finally,  let us  discuss  possible  2d  supersymmetry
of the action corresponding to~\eqref{L-tot}. As was already mentioned above, 
the  number (8)  of independent  bosonic degrees of freedom 
 in the reduced Lagrangian \rf{erm}  matches 
that of  the fermionic ones (8+8), 
exactly as in a 2d supersymmetric model. 
Moreover, we  saw that the spectrum of small fluctuations near 
the vacuum state \rf{fkk},\rf{kyj}  is also  supersymmetric.

The structure  of \rf{L-tot}   is essentially  that of a 
supersymmetric gWZW model \ci{swz,swzw}, 
\be L_{\rm SgWZW}=
L_{\rm gWZW}  + \psiL D_+ \psiL + \psiR D_- \psiR \la{swn} \ , \ee  
 modified  by the $\km$-dependent  interaction terms. 
 If we first set $\km=0$, i.e. ignore the potential and Yukawa interaction terms 
 in \rf{L-tot}, then we should expect to find the same (1,1)   supersymmetry as found 
 in the component description of supersymmetric gWZW model \ci{swz,swzw}, i.e.
  \be \la{gws}
  \delta g  \sim  \epsL \psiR g + \epsR  g\psiL, \ \ \ \ 
  \delta \psiR \sim \epsL (g^{-1} D_+  g)_{_{G/H}}, \ \ \ \ 
  \delta \psiL \sim \epsR (g D_-  g^{-1})_{_{G/H}}, \ \ \ \ 
 \delta A_\pm =0 \ . \ee 
Here $\epsL$ and $\epsR$ are parameters of the (1,0) and (0,1) supersymmetries.

For this to work the fermions should transform under the $H$ gauge transformation
as elements of the coset part of $\algg$, i.e. 
$\algm=\alghf_0^\pp$,  considered as a representation of 
the gauge
 algebra $\algh=\alghf_0^\oo$.
It appears, however, that for the case of $psu(2,2|4)$ the fermions $\psr,\psl$ take values
in $\alghf_{1,2}^\pp$ which is,  in general, 
 a different representation of the gauge
 algebra $\algh$. More precisely, $\alghf_{1}^\pp$ and $\alghf_0^\pp$ 
 considered as representations of $\algh$ are
inequivalent representations related by an appropriate automorphism $\tau$ 
of the gauge algebra $\algh$.\foot{One can  see that $\alghf_{1}^\pp$ and $\alghf_0^\pp$
are inequivalent  by, e.g.,  
observing  that for a subalgebra $\algh_1$ represented 
by the upper-left block matrices there are no invariant vectors 
in $\alghf_{1,2}^\pp$ but all the elements from $\alghf_0^\pp$
 represented by lower-right block matrices are invariant.
The automorphism $\tau$ simply interchanges $su(2)$ factor in the upper left block
with the $su(2)$ factor in the lower-right block in the matrix representation of $\algh$.}
In the absence of   $\km$-dependent terms in \rf{L-tot}
one can of course modify the gauge transformation law of the fermions by
 replacing, e.g.,  $A_-$ with its image under that  automorphism   $\tau(A_-)$
in the kinetic term for $\psr$.
 This does not, however,   directly apply  for  $\km\neq 0$; for example, 
the gauge invariance of
the fermionic interaction 
 term $\km \str(g^{-1}\psl g\psr)$  in \rf{L-tot} 
 determines the gauge transformation
law of the fermions in terms of that of the field $g$.

We leave the question  whether the full \rf{L-tot}
in the   $psu(2,2|4)$ case 
does have a 2d supersymmetry, i.e. if it 
can  be  identified with 
a supersymmetric extension  of  the corresponding bosonic non-abelian  Toda theory,  
for a  future investigation.\foot{Supersymmetric extensions of generic 
non-abelian  Toda theories 
were not previously  discussed in the 
literature (apart from the complex sine-Gordon case \ci{naps,napsi,gas}).
 For some references on supersymmetric 
extensions of sigma models   with potentials   and, in particular, of  abelian  Toda 
models see \ci{spp,st}.}
Our conjecture is that the answer is yes and the supersymmetry should be 
the extended (2,2) one.\foot{The conditions for existence of 
the (2,2) supersymmetry  in the (1,1) supersymmetric $G/H$ gWZW model (i.e. 
in our  $\km=0$ case)  were discussed in ~\cite{swzw} (see also \ci{kaz,fig}).}

As we shall  show in the next section in a similar but 
simpler  case of the   $AdS_2 \times S^2$ superstring model 
 where $psu(2,2|4)$ is replaced by the $psu(1,1|2)$ superalgebra 
  (with trivial  $\algh$ so that 
 the  complication of extending the supersymmetry  from the ``free''
 to  $\km\not=0$ level is absent)
 the corresponding reduced Lagragian \rf{L-tot}  is indeed invariant under the
(2,2) supersymmetry.

\bs

An  interesting  question 
related to the existence of (2,2) supersymmetry
is  about  finiteness  property  of the quantum  theory defined by 
\rf{L-tot}. A  (supersymmetric) gWZW  model  corresponds to 
 a (super)conformal theory, 
but including   potential terms may in general introduce UV divergences. 
These divergences  should cancel out if  this  model  has  (2,2)
supersymmetry.
We conjecture that  this is indeed the case; then  
this  reduced model  has a chance to be useful for a  quantum description 
of the \adss superstring.

%\bs 

%\iffalse
%One may wonder if  there is
%more than  $\ns =1$  supersymmetry, i.e.  (2,2) one 
%also in the  higher-dimensional  reduced  Lagrangians. 
%For gWZW   models the general condition is that 
%$G/H$  should be Kaehler  \ci{kaz,swz,fig}.   
%An indication 
%Indeed, the bosonic  parts of the  reduced model in $AdS_2 \times S^2$  \rf{sinh}  and 
%$AdS_3 \times S^3$ \rf{hgi}   cases are  Kaehler, so we should expect $\ns=2$ (i.e. (2,2))
%supersymmetry. This is likely to be true also in 
%the $AdS_5 \times S^5$   case  assuming $G/H$ in this case is also Kaehler.
%The model \rf{L-tot} may  be  UV  finite at the quantum level
%if it does have $\ns=2$ supersymmetry.
%Indeed, the supersymmetric gWZW models define superconformal theories 
%while in the supersymmetric   case the potential and the associated Yukawa terms 
%should not be renormalized either. 
%\fi

%%%%%%%%%%%%%%%%%%%%%%%%%%%%%%%%%%%%%%%%%%%%%%%%%%%%%%%%%%%o the%%%%%%%%%%%%%%%%%

\section{Example: reduced model for  superstring in $AdS_2\times S^2$ \space{ }\\
as $\ns=2$ super sine-Gordon model }\label{sec:ads2s2}
%%%%%%%%%%%%%%%%%%%%%%%%%%%%%%%%%%%%%%%%%%%%%%%%%%%%%%%%%%%%%

Let us now  specialise the  construction of the previous section to  the  simplest 
 case of $AdS_2\times S^2$ superstring model \ci{oth,\berk}
  where $\alghf=psu(1,1|2)$. 
As we shall  see below,  here  the reduced Lagrangian \eqref{L-tot},\rf{erm} 
 is equivalent to that  of the 
 $\ns=2$ supersymmetric sine-Gordon theory. 
 This demonstrates the existence 
 of the (2,2) world-sheet supersymmetry in the reduced version of this 
  GS  superstring 
 model. Assuming  one  may  consider the reduced theory as a legitimate starting 
 point for the quantisation, this   also  implies the  UV finiteness 
  of the $AdS_2\times S^2$  superstring    and its quantum integrability.

%First,  we find an explicit form of the homogeneous $Z_4$ components. 

\subsection{Explicit  parametrisation of $psu(1,1|2)$}
%%%%%%%%%%%%%%%%%%%%%%%%%%%%%%%%%%%%%%%%%%%%%%%%%%%%%%%%%%%%%%%%

The bosonic subspaces 
$\alghf_0$ and $\alghf_2$ in \rf{decompos}
 here are represented by block-diagonal matrices of the form
\begin{equation}
 f=\left(
\begin{array}{cc}
A &0\\
0& B
\end{array}
\right) \ ,  \ \ \ \ \ \ \   \ \ \ \ \ \ \ \Sigma A^\dagger \Sigma =-A\,,\ \ \   \qquad B^\dagger=-B\,,
\end{equation}
with $A,B$ being traceless $2\times 2$ matrices and 
$\Sigma$ given by \eqref{Sigma-K-def}, i.e.  $A\in su(1,1)$
and $B\in su(2)$.
  The  subspace $\alghf_0$ is formed by matrices satisfying also 
\begin{equation}
 -KA_0^tK=A_0\,, \qquad\ \ \ \ \    -KB_0^tK=B_0\,,
\end{equation}
with $K=\Sigma$ in \eqref{Sigma-K-def}.
It is usefull to parametrise these matrices as
\begin{equation}
A_0=\left(
\begin{array}{cc}
0&\phi\\
\phi&0
\end{array}
\right)\,,\qquad
B_0=\left(
\begin{array}{cc}
0&\imath \varphi\\
\imath \varphi & 0
\end{array}
\right)\,,
\end{equation}
where $\phi,\varphi$ are real.
The elements of the subspace $\alghf_2$ are  determined by the additional conditions
\begin{equation}
 KA_2^tK=A_2^t\,,\qquad\ \ \ \ \ \   KB_2^tK=B_2^t\ ,  
\end{equation}
%along with the reality conditions. 
%It can be parametrized as 
\begin{equation}
A_2=\left(
\begin{array}{cc}
\imath b & \imath c\\
-\imath c & -\imath b
\end{array}
\right)\,,
\qquad
B_2=\left(
\begin{array}{cc}
\imath q & r\\
- r  & -\imath q
\end{array}
\right)\,,
\end{equation}
where $b,c,q,r$ are  real.
For the fermionic subspace  $\alghf_1$
%As for the fermions, 
the reality condition together with $M^\Omega=\imath M$  (see Appendix C)
%(for $\alghf_1$) 
imply
\begin{equation}
M=\left(
\begin{array}{cc}
0& X\\
Y  & 0
\end{array}
\right)\,, \ \ \ \ \ \ \ \ \ \ \
KY^tK=\imath X\,, \qquad \imath \Sigma Y^\dagger=X\,.\qquad 
\end{equation}
Since  $\Sigma=K$  gives  $Y^+=-Y^t K$, 
 $\alghf_1$ can be parametrized as
\begin{equation}
\label{1par}
Y_1=\left(
\begin{array}{cc}
\imath \alpha & \imath \beta \\
\gamma & \delta\ 
\end{array}
\right)\,,
\qquad\qquad
X_1=\left(
\begin{array}{cc}
\alpha & \imath \gamma \\
 - \beta & -\imath \delta
\end{array}
\right)\,.
\end{equation}
For  $\alghf_3$  we have  $KY^tK=-\imath X$ and 
$\imath\Sigma Y^\dagger=X$ giving 
$Y^\dagger =Y^t K$ and 
\begin{equation}\label{3par}
Y_3=\left(
\begin{array}{cc}
\lambda & \nu \\
\imath \rho & \imath \sigma\ 
\end{array}
\right)\,,
\qquad\qquad
X_3=\left(
\begin{array}{cc}
 \imath \lambda & \rho \\
 -\imath \nu & -\sigma
\end{array}
\right)\,.
\end{equation}
The fixed element $T=T^1+T^2$ in \rf{tk},\rf{jow} can be chosen in the form:
\begin{equation}\la{trw}
T=
\half\left(
\begin{array}{cccc}
\imath&0 &0 &0 \\
0 &-\imath &0 & 0\\
0&0& \imath &0 \\
0&0& 0&-\imath
\end{array}
\right)
\,.
\end{equation}
The subspaces $\alghf_1^\pp$ and $\alghf_3^\pp$ defined in \rf{mkm}
 are then 
 represented  by \eqref{1par} and \eqref{3par}
with
\be \la{spr}
\alpha=\delta=0\ , \ \ \ \ \ \ \ \ \ \ \lambda=\sigma=0 \ .\ee
The field $g\in G$   introduced in \rf{saf}
takes values in the direct product of two  one-dimensional subgroups
of $SU(1,1) \times SU(2)$  isomorphic to $SO(1,1)$ and $SO(2)$; it 
can be parametrized as
\begin{equation}
%\begin{multline}
g=
\exp\left(\begin{array}{cc}
A_0&0\\
0&B_0
\end{array}
\right)
=
\left(
\begin{array}{cccc}
\cosh \phi&\sinh\phi &0 &0 \\
\sinh \phi&\cosh\phi&0 & 0\\
0&0& \cos \varphi& \imath \sin\varphi \\
0&0& \imath \sin\varphi & \cos\varphi
\end{array}
\right)
\,.
%\end{multline}
\end{equation}

\subsection{Reduced Lagrangian}

Let us write down the explicit form of the reduced Lagrangian~\eqref{L-tot} using 
the parametrisation introduced above.
Here  the subgroup $H$ is trivial so  that  $A_+=A_-=0$. 
 The ``kinetic'' WZW term  is simply 
\begin{equation}
\half\mathrm{STr}(g^{-1}\d_+g g^{-1}\d_-g)=\d_+\phi\d_-\phi+\d_+\vp\d_-\vp\,.
\end{equation}
The potential term in  \rf{L-tot} is 
\begin{equation}
 \km^2\str(g^{-1}TgT)=-\frac{\km^2}{2}(\cosh 2\phi- \cos  2\vp  )\,.
\end{equation}
%Note that this term in the Lagrangian is explicitly real.
% confirming the general statement above.
The fermionic terms in   \eqref{L-tot}  are 
\begin{equation}
 \begin{aligned}
\half\str(\PsiR\commut{T}{\d_-\Psi_R})&=
\tr(\d_-Y_1\commut{T^1}{X_1})&=
-\tr(\d_-X_1\commut{T^2}{Y_1})&=
\beta\d_-\beta+\gamma\d_-\gamma\,, \\ 
\half\str(\PsiL\commut{T}{\d_+\PsiL})&=
\tr(\d_+Y_3\commut{T^2}{X_3})&=
-\tr(\d_+X_3\commut{T^2}{Y_3})&=
\nu\d_+\nu+\rho\d_+\rho\,, 
\end{aligned}
\end{equation}
\begin{equation}
\begin{aligned}
\label{ferm-large}
\km\str(g\PsiR g^{-1}\PsiL)=&
~\km\tr(g_1X_1g_2^{-1}Y_3)-\tr(g_2Y_1 g_1^{-1} X_3)\\
=&~-2\km[\cosh\phi\cos\vp\ (\beta\nu+\gamma\rho)+\sinh\phi\sin\vp\ 
(\beta\rho-\gamma\nu)]\,,
\end{aligned}
\end{equation}
where we have used the explicit form of the diagonal blocks 
$T^1=T^2=\frac{\imath}{2}\,{\rm diag}(1,-1)=\frac{\imath}{2}\Sigma$  in \rf{trw}.

Thus  the final expression of the corresponding 
 reduced  Lagrangian   \rf{L-tot} 
 in terms of the two bosonic $\phi, \varphi$ 
   and  the four fermionic  $\beta,\gamma,\nu,\rho$ field  
 variables  is given by (cf. \rf{sinh})\foot{As  expected, the Lagrangian is real 
(the fermionic fields  are real).}
\begin{multline}
\la{aaac}
L_{tot}=\d_+\vp\d_-\vp+\d_+\phi\d_-\phi  + {\km^2\ov 2 }(\cos 2\vp - \cosh 2\phi )
\\
\ \ \  +\beta\d_-\beta+\gamma\d_-\gamma+\nu\d_+\nu+\rho\d_+\rho\\
-2\km\left[\cosh\phi\ \cos\vp\ (\beta\nu+\gamma\rho)+\sinh\phi\ \sin\vp\ 
(\beta\rho-\gamma\nu)\right]\,.
\end{multline}
%As  expected, the Lagrangian is real 
%(the fermionic fields  are real).

\subsection{Equivalence to  $\ns=2$  supersymmetric sine-Gordon  model}
%%%%%%%%%%%%%%%%%%%%%%%%%%%%%%%%%%%%%%%%%%%%%%%%%

The bosonic  part of  the $AdS_2 \times S^2$   reduced Lagrangian in
\rf{sinh},\rf{aaac} happens to be exactly 
the same as the bosonic part of the $\ns=2$  supersymmetric 
sine-Gordon Lagrangian \ci{susy}.
Furthermore, the number of the fermionic fields in \rf{aaac} is the same 
as in the $\ns=2$ SG  theory. This suggests  
that the $AdS_2 \times S^2$   reduced model \rf{aaac} 
  may have a hidden $\ns=2$ world-sheet 
supersymmetry. 
\bs 

Indeed, \rf{aaac} is  equivalent to the  
 $\ns=2$ SG theory.
 A  generic $\ns=2$ (i.e. (2,2))  superfield Lagrangian is 
\begin{equation}
\begin{aligned}
L&= \int d^4 \vt \ \hat \Phi^* \hat \Phi + [ \int d^2 \vt \  W(\hat \Phi)  + h.c. ]\  , \cr
\hat \Phi&= \Phi + \vt_1 \psi_{_L} + \vt_2 \psi_{_R} + \vt_1\vt_2  {\cal D} \ , 
\la{sup}
   \end{aligned}
\end{equation}
where $ \hat \Phi$ is a chiral $\ns=2$ superfield, 
 $\Phi=\vp + i \phi $ is a complex scalar  and $ \psi_{_L},\psi_{_R}$ are complex fermions.
In components   
%\maxim{I have got of imaginary units: if I understand 
%correctly this corresponds to our choice of reality condition}
\begin{equation}
L= \d_+ \Phi \d_- \Phi^*  - | W'(\Phi)|^2  %+ i 
+\psi_{_L}^* \d_+ \psi_{_L} %+ i
+\psi_{_R}^* \d_-  \psi_{_R}
+ 
\ \big[
 W''(\Phi)\psi_{_L}\psi_{_R} +   W^*{}''(\Phi^*) \psi^*_{_L}\psi_{_R}^*\big]\ . 
\la{comp} 
 \end{equation}
The   sine-Gordon  choice is 
\be \la{poiy}   W(\Phi) = {\km  }  \cos \Phi \ , \ \ \ \ \ \ 
\ \ \ \ \
| W'(\Phi)|^2= {\km^2  \ov 2} ( \cosh 2 \phi - \cos 2 \vp) \ . \ee
Splitting $\psi_{_L}$, $\psi_{_R}$ into the real and imaginary parts 
\begin{equation}
\psi_{_L}=\nu+\imath \rho\,,\qquad \psi_{_R}=-\beta+\imath\gamma \ , 
\end{equation}
we indeed   find the agreement between \rf{comp} and \rf{aaac}.

%match the kinetic term in \rf{comp}  with the last line of
%\rf{aaac} and the interaction term with the second line \rf{aaac}.
%\foot{

%Indeed, its  target space is a product of two  2-dimensional Kaehler  spaces. It is not,
%however, hyperKaehler so cannot have $N=4$ supersymmetry} 

\bs

Let us note that  
it is possible  to write down the $\ns=2$ supersymmetry transformations
 of the fields in \rf{aaac}
in terms of the original 
matrix parametrisation used in \rf{L-tot}.
 Let  us consider separately the (2,0) and (0,2) 
supersymmetries.  To describe the (2,0) transformation 
 let us introduce a matrix fermionic
 parameter $\epsL$ taking values in $\alghf_1$ in \eqref{decompos} and satisfying
in addition $\commut{T}{\epsL}=0$.
 This  ensures that $\epsL$ contains 
  two independent fermionic parameters
($\alpha$ and $\delta$ in the  parametrisation \eqref{1par}).
 The (2,0) supersymmetry transformation of the  matrix fields in \rf{L-tot} 
 then reads as 
\begin{equation}
\label{susy-SG}
\delta_{\epsL} g=g\commut{T}{\commut{\PsiL}{\epsL}}\,,\qquad
\delta_{\epsL} \PsiL=\commut{g^{-1}\d_+g }{\epsL}\,,\qquad
\delta_{\epsL} \PsiR=\km\commut{T}{g\epsL g^{-1}}\,.
\end{equation}
%Here we do not distinguish $T_\pm$.
 In checking the invariance of the action we have to use
 (besides the $Z_4$
  grading and definition of $\epsL$)  that 
$
\commut{T}{\commut{T}{\psl}}=-\psl\,, 
\ \ 
\commut{\commut{T}{\commut{\psl}{\epsL}}}{\psl}=0\,,$ etc. 
%\end{equation}
%\maxim{To be checked as the notations have changed -- nearly for sure signs are wrong..}
The (0,2) transformation with parameter $\epsR$ looks similarly.

% \maxim{Up to the gauge field the SUSY should also work for the respective model obtained from $psu(2,2|4)$. The only thing I am not completely sure
% is how to make the second property in \eqref{prop-su} satisfied in this case. But probably that is what reduces the number of SUSY parameters
% (otherwise $\epsilon$ contains 8 real fermionic parameters)}.

\bs

 The  (2,0) 
  supersymmetry transformation law \eqref{susy-SG} can be formally 
  generalized
to the algebraically analogous models described by \eqref{L-tot} 
{\it provided}  $\alghf_1^{\oo}$ 
contains a nontrivial element commuting with the entire gauge algebra $\algh$. 
Indeed, suppose 
$\epsL$ belongs to $\alghf_1^\oo$ 
and is satisfying in addition $\commut{\epsilon}{h}=0$ for any $h\in\algh=\alghf_0^\oo$ 
(in other words, $\epsL$ should belong  to the centraliser of $\algh$ in $\algf_1^\oo$).
Then the supersymmetry  transformation reads 
\begin{equation}
\label{susy}
\begin{gathered}
 \delta_{\epsL} g= g \commut{T}{\commut{\psr}{\epsL}}\,, \qquad 
\delta_{\epsL} \psr
=\commut{(g^{-1}D_+g)^\pp}{\epsL}\,, \qquad
\delta_{\epsL} \psl
=\km\commut{T}{g\epsL g^{-1}}\,,\\
\delta_{\epsL} A_+=0\,,\qquad
\delta_{\epsL} A_-=\km\commut{(g^{-1}\psl g)^\oo}{\epsL}\,,
\end{gathered}
\end{equation}
where the superscript $\pp$ or $\oo$ denotes the projection to $\alghf^\pp$ or  
$\alghf^\oo$ respectively. Note  that 
for $\km\not=0$  the field $A_-$ 
starts  transforming  under the  supersymmetry.\foot{In checking the  invariance of the
 action one is to use  the  algebraic 
properties 
%\begin{equation}
$ \commut{\commut{T}{\psr}}{\psr}\in\alghf_0^\oo,\  \  
% \qquad 
% \qquad 
\commut
 {\commut{\epsL}{\psr}}{\psr}\in\alghf_0^\oo , $
%\end{equation}
which follow upon the application of  the projectors to $\alghf^{\pp,\oo}$
and the use of the identities \eqref{sa-ident}.}
 Since 
 the action is invariant under the exchange  $+\leftrightarrows -$, 
 $L \leftrightarrows R$, and $g \leftrightarrows g^{-1}$  one finds 
 also the ``right''  counterpart of the ``left''  supersymmetry  \rf{susy}
 with  $\epsL \to \epsR$   where 
 $\epsR$ is taking values in $\alghf_3^\oo$ and  is annihilated by $\algh$.
 
 In the case of $psu(1,1|2)$ 
 the subalgebra $\algh$ is empty and $\epsL$ 
%(as well   $\epsR$)
 is an 
arbitrary element of the two-dimensional space $\alghf_1^\oo$ 
(and similarly $\epsR \in \alghf_3^\oo$)  so that \eqref{susy}
defines  a  consistent (2,0) (and also  (0,2)) supersymmetry  transformation.
However, in   the case of $psu(2,2|4)$,  none of the  elements in $\alghf_{1,2}$ commute
 with the entire
$\algh$  so  that \rf{susy} does not  directly apply
%appear define the   supersymmetry
(cf.  the discussion  at the end of  section~\eqref{Lorentz}).
The existence of 2d supersymmetry of \rf{L-tot}  in the \adss   case thus
 remains an interesting 
 open question.\foot{Among  other interesting questions let us mention 
  also the construction of reduced models  for  non-critical $AdS_n $ 
superstrings \ci{polk,oz}  and their possible  world-sheet  supersymmetry.}

\bs 
Let us   finally
mention   that the  complex sine-Gordon  model \rf{csgi} 
%(and thus also its ``double''  \rf{hgi})
also admits an  $\ns=2$   supersymmetric  version  \ci{naps,napsi}.
The same  applies to its  ``double''  in  \rf{hgi}
which has 2+2 dimensional  target space which is a direct sum of the two K\"ahler spaces.
We expect that the corresponding  $\ns=2$ model  should   be  equivalent to  
the reduced model for  the superstring on 
 $AdS_3 \times S^3$  \ci{adst}  with  \rf{hgi} as its  bosonic part.

\bs

\section*{Acknowledgements }

%%%%%%%%%%%%%%%%%%%%%%%%%%%
We are grateful to G. Arutyunov, S. Frolov,  
 A. Mikhailov   and R. Roiban for many 
useful discussions,  explanations and questions. 
We also thank  to I. Bakas, G. Papadopoulous  and K. Sfetsos 
for useful  remarks on related subjects. 

MG  acknowledges the support of Dynasty foundation, RFBR Grant 05-01-00996, and
the Grant LSS-4401.2006.2.
He would like also to thank the organizers of the workshop
``Poisson sigma models, Lie algebroids, deformations and higher
analogues'' at the Erwin Schr\"odinger International Institute for Mathematical Physics in  Vienna, Austria for the hospitality while this work was in progress.

AAT acknowledges the support of the  EU-RTN network grant MRTN-CT-2004-005104, 
 the  INTAS 03-51-6346 grant    and the RS Wolfson award.
 Part of this work was done while AAT was a participant of the 
programme ``Strong Fields, Integrability and Strings'' at the  Newton's 
Institute in Cambridge, U.K.

AAT is most grateful to R.Roiban for an initial collaboration on 
   part of the material 
  discussed in  sections 6 and 7.
  Some preliminary results of this paper (in particular, 
  possible existence of 2d supersymmetry in Pohlmeyer  reduction of \adss string) 
    were mentioned  in  \ci{atc}.
AAT  also thanks  to A. Mikhailov for  sending him a draft of an  unpublished 
 work \ci{msa} which also attempted to uncover world-sheet supersymmetry 
 in the  formulation of \adss string model in terms of currents.
  
 While this paper was in preparation we were informed 
 by  A. Mikhailov and S. Sch\"afer-Nameki  about  their   closely related 
 forthcoming paper  \ci{MS} in which an equivalent 
  reduced action for \adss  superstring is  found.

\appendix
\addcontentsline{toc}{section}{Appendices}
%\addcontentsline{toc}{section}{Appendices}
\subsection*{Appendix A:  Proof of gauge equivalence in section \bref{sec:gWZW}}
\refstepcounter{section}
\label{sec:l2wzw}
\addcontentsline{toc}{subsection}{Appendix A:  Proof of gauge equivalence in section \textbf{3.2}}
\def\theequation{A.\arabic{equation}}
\setcounter{equation}{0}

%Equivalence of the Lax representation 
%and the Lagrangian gWZW 
%equations}

Here we provide some details of the argument in section \bref{sec:gWZW}.
Let us introduce the following combinations 
\begin{equation}\la{comb}
 \hat A_+=g^{-1}\d_+g + g^{-1}A_+ g\,,\qquad \hat A_-=g\d_-g^{-1}+gA_-g^{-1}
\end{equation}
Under the gauge transformations \eqref{gs}  $\hat A_\pm$ transform as follows:
\begin{equation}
\label{c-transf}
 {\hat A}_{+}\to {\bar h}^{-1}{\hat A}_{+}{\bar h}+{\bar h}^{-1}\d_+ \bar h\,,
 \qquad
{\hat A}_{-}\to h^{-1}{\hat A}_+h +h^{-1}\d_- h\,.
\end{equation}
It follows from the commutation relations $\commut{\algh}{\algm}\subset\algm$ 
and $\commut{\algh}{\algh}\subset \algh$ that their $\algh$ projections also transform in 
the same way.
Then  the constraints \eqref{lwc} take the form
\begin{equation}
 A_+=(\hat A_+)_\algh\,, \qquad A_-=(\hat A_-)_\algh\,.
\end{equation}
They are not invariant under the transformations \eqref{c-transf} unless $h=\bar h$.
Using~\eqref{gs} one can then set
\be
%\label{rrr}
 (\hat A_+)_\algh = A_+=   (g^{-1}\d_+g+g^{-1}A_+g)_\algh \,.
\ee
This condition can be satisfied by applying the transformation \rf{gs}
 with $h=\id$.
Under this transformation $A_+$ is unchanged while $(\hat 
A_+)_\algh=(g^{-1}\d_+g+g^{-1}A_+g)_\algh$ transforms  as an 
 $H$ connection, so it is possible
to find $\bar h$ so that transformed value of $(\hat A_+)_\algh$ is equal to $A_+$.

Next, once $(\hat A_+)_\algh=A_+$,   eq.~\rf{weom} implies
that $A_+,A_-$  are components of a flat 2d connection, i.e.
 satisfy \eqref{fla}.\foot{Note that 
contrary to the discussion before \eqref{fla} now we do not assume 
that both constraints \eqref{lwc} 
are satisfied.}
 This, together with the equation on $g$ contained in \rf{weom} and
  the remaining  part of gauge 
 invariance \rf{gs} allows one to show that the second relation in  \rf{lwc} 
 can also be satisfied. 
 
 Indeed, let us show that  one can find such $h_0$ that the 
 transformation \eqref{gs} with $h=h_0$ 
 and $\bar h=\id$
preserves $A_+=(\hat A_+)_\algh$ and transforms $A_-$ and $g$ so
 that $A_-=(\hat A_-)_\algh$ (note that $\hat A_-$ is unchanged under such transformation).
It is enough to find  $h_0$ in any admissible gauge that can be reached by the
 gauge transformation 
with $h=\bar h$ (both conditions $(\hat A_+)_\algh=A_+$ and $(\hat A_-)_\algh=A_-$ are invariant
under such gauge transformations). Without loss of generality we can choose this  gauge  to be 
 $A_+=A_-=0$ (this gauge can always be reached   by a gauge transformation with $h=\bar h$). 
 In this gauge the equation
\eqref{weom} and the constraint $(\hat A_+)_\algh=A_+$ take the form \eqref{ol} and
 the first equation 
in \eqref{olk} respectively. Equation \eqref{ol} can  be written equivalently as
\begin{equation}\la{jl}
\d_+(g\d_-g^{-1})=\km^2 \commut{T_-}{gT_+g^{-1}}\,,
% \qquad \Pi_\algh(g^{-1}\d_+g)=0\,,
\end{equation}
implying $\d_+(g\d_-g^{-1})_\algh=0$. This means that
 $(g\d_-g^{-1})_\algh$ is a function of $\sigma^-$ only
and therefore can be represented as $(g\d_-g^{-1})_\algh=h_0\d_-h_0^{-1}$ for some $H$-valued function
$h_0(\sigma^-)$. By performing the gauge transformation with $\bar h=\id$ and $h=h_0$ one then 
arrives at $(\hat A_-)_\algh=(g\d_-g^{-1})_\algh=0$ while 
still  satisfying $A_\pm=0$ and $(\hat A_+)_\algh=0$.

\subsection*{Appendix B: Vanishing of the antisymmetric tensor  coupling \\ in the reduced 
Lagrangian in section \bref{sec:gen-struct}}
\refstepcounter{section}
\label{sec:a-vanish}
\addcontentsline{toc}{subsection}{Appendix B: Vanishing of the antisymmetric tensor
coupling in the reduced Lagrangian in section \textbf{5.1}}
\def\theequation{B.\arabic{equation}}
\setcounter{equation}{0}

Here we provide details of the argument  mentioned at the end of section \bref{sec:gen-struct}
 that the reduced Lagrangian 
\rf{si} does not contain  a WZ-type  term. Indeed, 
all possible 
 antisymmetric tensor 
 contributions that may result from integrating out the gauge field of the gWZW model 
 vanish.
 
 Let us  consider the following automorphism 
of the orthogonal matrix group
and its Lie  algebra:
\begin{equation}
 \tilde M ^i_{j}=M^i_j (-1)^{i+j}\,, \qquad \widetilde{MN}=\tilde M \tilde N\,.
\end{equation}
It is easy to check that
\begin{equation}
\tr\tilde M=\tr  M \,,\qquad \det \tilde M=\det M \,, \quad  \tilde M ^{-1}=\tilde{ M^{-1}}\,, \quad
\tilde M^T=\tilde{M^T}\,.
\end{equation}
%In particular, it is the automorphism of the orthogonal group (as well as its Lie algebra)
%in the standard matrix
%representation. Indeed, their defining conditions are invariant under the automorphism.
If $g$ has the  gauge-fixed form \eqref{gah}
 then $\tilde g=g^{-1}$: 
 this is obviously correct
for any $g_k=e^{\theta_k R_k}$ because $\tilde R_k=-R_k$ while
$g^{-1}$
has the same form with all $g_k$ replaced
with $g_k^{-1}$.  

The integrand of the WZ term in \rf{pi},\rf{gaui} 
 then satisfies
\begin{multline}
 \tr(g^{-1}dgg^{-1}dgg^{-1}dg)=
\tr({\widetilde{(g^{-1}dgg^{-1}dgg^{-1}dg)}})\\=~
\tr(gdg^{-1}gdg^{-1}gdg^{-1})=
-\tr(g^{-1}dgg^{-1}dgg^{-1}dg)\,,
\end{multline}
and thus should vanish.

Another possible contribution may originate from the gauge field 
dependent  
term in the gWZW  Lagrangian \rf{gaui} 
\begin{equation}
L_{A} =\Tr \big(   A_+
 \d_- g  g\inv - A_- \,g\inv\d_+ g   - g\inv  A_+ g   A_-  + A_+  A_- \big)\,,
\end{equation}
where $ A_\pm $ should be replaced by the  solutions of their
equations of motion 
\begin{equation}
\label{a-const}
  A_+ = (g\inv \d_+ g + g\inv  A_+ g)_\algh\,, \qquad 
 A_- = (g\d_-g\inv + gA_- g\inv)_\algh \ . 
\end{equation}
%Using the equations in the Lagrangian one obtains the following representations:
This gives
\begin{equation}
L_{A} =\Tr \big(   A_+\d_- g  g\inv)= -\Tr(A_- \,g\inv\d_+ g)\,.
\end{equation}
It follows from the explicit form of Eqs. \eqref{a-const} that there exists a 
function $\mathbf{A}(g,\d g)$ such that
\begin{equation}
A_+(g,\d_+g)=\mathbf{A}(g,\d_+g)\,, \qquad A_-(g,\d_-g)=\mathbf{A}(g^{-1},\d_-g\inv)\,.
\end{equation}
Moreover, assuming the  analyticity in $g$ one finds
\begin{equation}
 \widetilde{\mathbf{A}(g,\d_\pm g)}=\mathbf{A}(g^{-1},\d_\pm g^{-1})\,,
\end{equation}
provided $\tilde g =g\inv$. In particular, this holds in the gauge \eqref{gah}).

Since  $A_\pm$ are linear in $\d_\pm g$ the 
vanishing of the antisymmetric part of 
the metric is equivalent to
$L_{A}(g,\d_+g,\d_-g)=L_{A}(g,\d_-g,\d_+g)$. 
Assuming $\tilde g =g\inv$ one gets 
\begin{multline}
L_{A}(g,\d_-g,\d_+g)=\Tr(\mathbf{A}(g,\d_-g) 
\d_+g g\inv)=\Tr(\widetilde{\mathbf{A}
(g,\d_-g\inv) \d_+g g\inv})\\=
\Tr({\mathbf{A}(g\inv,\d_-g\inv) \d_+ g\inv  
g})=-\Tr(A_- g\inv \d_+g)=L_{A}(g,\d_+g,\d_-g)\,.
\end{multline}
This shows that the antisymmetric tensor contribution to the 
reduced Lagrangian  indeed vanishes in the gauge \eqref{gah}.

\subsection*{Appendix C: Matrix superalgebras: definitions and  notations}
\refstepcounter{section}
\label{sec:smatrix}
\addcontentsline{toc}{subsection}{Appendix C: Matrix superalgebras: definitions and  notations}
\def\theequation{C.\arabic{equation}}
\setcounter{equation}{0}

Here we  summarize some basic definitions and notation used in sections~\bref{sec:spohl} and \bref{sec:ads2s2}.

Let $\Lambda$ be a Grassmann algebra. The 
algebra $Mat(n,l;\Lambda)$ 
is that 
of $(n+l)\times (n+l)$ matrices 
over  $\Lambda$ whose diagonal block entries are even elements of $\Lambda$ while off-diagonal block
 entries are odd.\footnote{This corresponds to considering
even matrices. In general one can also allow for both even and odd ones; this would 
lead to additional sign factors in the equations   below.}
The super-transposition $^{st}$ is defined as follows:
\begin{equation}
\left(
\begin{array}{cc}
A&X\\
Y&B
\end{array}
\right)^{st}=
\left(
\begin{array}{cc}
A^t &-Y^t\\
X^t &B^t
\end{array}
\right)\,, \qquad
(MN)^{st}=N^{st}M^{st}\,.
\end{equation}
Note that in general $(M^{st})^{st}\neq M$. More precisely,
 $(M^{st})^{st}=WMW$ where 
$W$ is the parity automorphism
given by 
\be\la{wer}
 W={\rm diag}(1,\ldots,1,-1,\ldots,-1) \ . \ee
A real form of a complex matrix Lie (super)algebra 
can be described in terms of an 
antilinear anti-automorphism
$*$
satisfying 
\begin{equation}
(MN)^*=M^*N^*  \,, \qquad (M^*)^*=M\,, \qquad (aM)^*=\bar a M^*\,,\,\,\, a \in\fC\,.
\end{equation}
The real subspace of elements satisfying $M^*=-M$ is then a real Lie superalgebra.

We are interested in the case of $n=l$, i.e.  $Mat(n|n,\Lambda)$.
Suppose first that the corresponding  $*$ operation 
is defined on $\Lambda$ so that $(a^*)^*=a$ and 
$(ab)^*=a^*b^*=(-1)^{|a||b|}b^*a^*$ where $|a|$ denotes the Grassmann parity of $a$.
Let us extend $*$ to arbitrary supermatrices according to
\begin{equation}
\left(
\begin{array}{cc}
A&X\\
Y&B
\end{array}
\right)^*=
\left(
\begin{array}{cc}
\Sigma^{-1} A^\dagger \Sigma & - \imath \Sigma^{-1} Y^\dagger \\
-\imath X^\dagger\Sigma &B^\dagger
\end{array}
\right)\,,
\end{equation}
where $\dagger$ applied to the block denotes standard hermitian conjugation, i.e.
 transposition combined with the
$*$-conjugation of entries. It is useful to represent it as 
\begin{equation} M^*=\mathbf{\Sigma}^{-1}
 M^\dagger \mathbf{\Sigma} \ , \ \ \ \ \ \ \ \ \ 
\mathbf{\Sigma}=\left(
\begin{array}{cc}
\Sigma & 0\\
0 & \id
\end{array}
\right)\,,
\qquad
\left(
\begin{array}{cc}
A&X\\
Y&B
\end{array}
\right)^\dagger=
\left(
\begin{array}{cc}
 A^\dagger &-\imath Y^\dagger\\
-\imath X^\dagger  & B^\dagger
\end{array}
\right)\,.
\end{equation}
It is easy to see that $^*$ is involutive provided $\Sigma^2=\id$ and $\Sigma^\dagger=\Sigma$.
 Note that $(MN)^{\dagger}=N^\dagger M^\dagger$ and $(M^\dagger)^\dagger=M$. Note also 
 that $(M^\dagger)^{st}=W(M^{st})^\dagger W$
where $W$ is the parity automorphism introduced above. Let us also note that the $*$
 conjugation induces the real form of the respective Lie group. Namely, the condition 
 $g^*=g^{-1}$ selects the real subgroup of the complex group. It is obviously compatible 
 with the conjugation for the Lie algebra due to the representation $g=e^{M}$ and $M^*=-M$.

To define   $Z_4$ anti-automorphism let us first  consider the following automorphism
\begin{equation}
\left(
\begin{array}{cc}
A&X\\
Y&B
\end{array}
\right)^{\Omega}=
-\left(
\begin{array}{cc}
K^{-1}A^t K&- K^{-1}Y^tK\\
K^{-1}X^tK &K^{-1}B^t K
\end{array}
\right)\,,
\end{equation}
where $K$ is some matrix required to satisfy $K^2=\pm \id$ and $K^t=\pm K^{-1}$.
It is useful to represent $~^\Omega$ as follows
\begin{equation}
M^\Omega=-\mathbf{K}^{-1}M^{st}\mathbf{K}\,,
\qquad
\mathbf{K}=\left(
\begin{array}{cc}
K&0\\
0 & K
\end{array}
\right)\,,
\end{equation}
so that we have the property
\begin{equation}
(MN)^\Omega=-N^\Omega M^\Omega \ . 
 \end{equation}
%holds transparently. 
 A  Lie superalgebra $\algf^\fC$  admits a $Z_4$ automorphism 
if  it can be 
decomposed into a direct sum  of eigenspaces of $\Omega$-anti-automorphism
\begin{equation}
\algf^\fC=\algf^\fC_0\oplus \algf_1^\fC\oplus\algf_2^\fC\oplus\algf_3^\fC\,,
\end{equation}
where $\algf^\fC_l$ denotes the eigenspace with eigenvalue $\imath^l$, i.e.
\begin{equation}
 M^\Omega= \imath^m M\,, \quad
 (\commut{M}{N})^\Omega=\imath^{m+n}\commut{M}{N}\,,\qquad M\in \algf^\fC_m,
  \quad N\in \algf^\fC_n\,.
\end{equation}
To see under which conditions $\Omega$ is compatible with the 
reality condition we note that 
\begin{multline}
-\mathbf{K}^{-1} (\mathbf{\Sigma}^{-1} M^\dagger \mathbf{\Sigma})^{st} \mathbf{K}
=
-((\mathbf{K}^{-1} \mathbf{\Sigma}^{-1} M \mathbf{\Sigma} 
 \mathbf{K})^\dagger)^{st}\\
=~
-W( \mathbf{\Sigma}^{-1} \mathbf{K}^{-1} M^{st} \mathbf{K} \mathbf{\Sigma})^\dagger W=
(-\imath)^m W\mathbf{\Sigma}^{-1} M^\dagger \mathbf{\Sigma}W\,,
\end{multline}
where we  used 
\begin{equation}
 \mathbf{K}^{st}=\pm \mathbf{K}^{-1}\,, \qquad 
 \mathbf{\Sigma}^\dagger=\mathbf{\Sigma}^{-1}=\mathbf{\Sigma}\ , 
\end{equation}
and also  assumed that 
\begin{equation}
\commut{\Sigma}{K}=0\,, \qquad \mathbf{K}^\dagger=\pm \mathbf{K}^{-1}\,,\qquad
\mathbf{\Sigma}^{st}=\mathbf{\Sigma}\ . 
\end{equation}
If in addition the eigenvectors with odd $m$ belong 
to the off-diagonal blocks (which 
is the case for $psl(2m|2m)$ superalgebra)
one finds
\begin{equation}
 (-\imath)^{m} W \mathbf{\Sigma^{-1}} M^\dagger \mathbf{\Sigma} W=
\imath^m \mathbf{\Sigma^{-1}} M^\dagger\mathbf{\Sigma}\,,
\end{equation}
so that $(M^*)^\Omega=\imath^m M^*$ provided $M^\Omega=\imath^m M$.
This proves that $Z_4$ grading restricts to the real form implying its decomposition~
\eqref{decompos}.

The explicit form of $\Sigma$ and $K$ in the case
of  $psu(2,2|4)$ is\footnote{Here we follow 
the notation of  \ci{arutu,ald}.}
%\maxim{Here notation are taken from Arutyunov's 
%lectures. This also seem to agree with those of Mikhailov
%provided one changes the basis appropriately}.}
\begin{equation}
 \Sigma=\left(
\begin{array}{cccc}
1&0&0&0\\
0&1&0&0\\
0&0&-1&0\\
0&0&0&-1
\end{array}
\right)\,,\qquad
K=\left(
\begin{array}{cccc}
0&-1&0&0\\
1&0&0&0\\
0&0&0&-1\\
0&0&1&0
\end{array}
\right)\,.
\end{equation}
%It is easy to see that all the conditions on $\Sigma$ and $K$ are fulfilled.
In the case of  $psu(1,1|2)$ we  take\footnote{
%\maxim{Note that in BBHZZ they use different
 %definition which doesn't satisfy our present requirements. To be understood.}
 This choice is  different from the one used in \ci{\berk}.}
\begin{equation}
\label{Sigma-K-def}
 \Sigma=\left(
\begin{array}{cc}
1 &0\\
0&-1
\end{array}
\right)\,,\qquad
K=\left(
\begin{array}{cc}
1&0\\
0&-1
\end{array}
\right)\,,
\end{equation}
which satisfy all the conditions above.

%%%%%%%%%%%%%%%%%%%%%%%%%%%%%%%%%%%%%%%%%%%%%%%%%%%%%%%%%%%%%%%%%%%%%%%%%%%%%%%
\subsection*{Appendix D: $\kappa$-symmetry transformations and gauge fixing in section \bref{sec:spohl}}
\refstepcounter{section}
\label{sec:kappa}
\addcontentsline{toc}{subsection}{Appendix D: $\kappa$-symmetry transformations and gauge fixing in section \textbf{6}}
\def\theequation{D.\arabic{equation}}
\setcounter{equation}{0} 
%%%%%%%%%%%%%%%%%%%%%%%%%%%%%%%%%%%%%%%%%%%%%%%%%%%%%%%%%%%%%%%%%%%%%%%%%%%%%%%%%

To prove that the gauge condition $Q_{1-}=Q_{2+}=0$  \rf{kg} is reachable it is useful 
to introduce the tangent frame field $e^a_\alpha$ so that the 2d metric is expressed
as $g^{ab}=e^a_\alpha e^b_\beta \eta^{\alpha\beta}$ where $\eta^{\alpha\beta}$
 is the tangent-space  metric.
We shall use the standard local frame where in the $\pm$ basis 
%$\eta^{\alpha\beta}$ is a fixed fibre metric which we assume
% to have a standard conformal form 
$\eta^{+-}=\eta^{-+}=1$ and $\eta^{++}=\eta^{--}=0$.
The frame 
% \maxim{I would say fiber components}
 components of the currents are defined in
the standard way as $J_\alpha=e_\alpha^a J_a$. 

In terms of this parametrization the Lagrangian density for the superstring sigma-model
 can be written as (cf. \rf{lak}) 
\begin{equation}
 L_{\rm GS}=\str \big[P_+P_-+\half(Q_{1+}Q_{2-}-Q_{1-}Q_{2+}) \big]\ e^+\wedge e^-  \,.
\end{equation}
Recall that the $\pm$ components of the currents are defined as $J_\pm=f^{-1}e_\pm^a\d_a f$. 
\foot{Note that here we use $\pm$ for the light-cone frame components contrary to the 
genuine light-cone components in the conformal gauge 
in  the main text. They of course coincide if one chooses the adapted frame and
$\sigma^\pm$ coordinates.}
 The WZ term can be  written also as $Q_1\wedge Q_2$ and
does not of course depend on the frame field.
% (similarly to the metric formulation
%where it is metric-independent). 
Using  $e_a^\alpha$ instead of $\gamma^{ab}$ introduces a 
local 2d Lorentz invariance (with the corresponding 
 the new gauge degree of freedom entering 
through $e_a^\alpha$).
The analog of the Virasoro constraints in this formulation
 are the equations of motion obtained by
varying the action with respect to the frame field. 
Note the following useful relations:
\begin{equation}
\dl{e^a_-}{L_{\rm GS}}= e^+_a \str(P_+P_+)\  e^+\wedge e^- 
\,, \qquad  \dl{e^a_-}{L_{\rm GS}}=  e^-_a \str{}(P_-P_-) \ e^+ \wedge e^- \,,
\end{equation}
where  $e^+\wedge e^-=d\sigma^1\wedge d \sigma^2 (\det\, e^a_\alpha )^{-1}$.

The variation of the Lagrangian under the $\kappa$-transformation of the currents 
$\delta_\kappa J_a=\d_a\epsilon+\commut{J_a}{\epsilon}$ with $\epsilon=\epsilon_1+\epsilon_2=
\scommut{P_+}{\imath k_{1-}}+\scommut{P_-}{\imath k_{2+}}$ is given by:
\begin{multline}
 \delta^{J}_{\kappa}L_{\rm GS}=2 \, \str\big(\commut{P_+}{Q_{1-}}\scommut{P_+}{\imath k_{1-}}
+\commut{P_-}{Q_{2+}}\scommut{P_-}{\imath k_{2+}}\big)\ e^+\wedge e^-  \\
=2\str\big( P_+P_+\commut{Q_{1-}}{\imath k_{1-}}+P_-P_- \commut{Q_{2+}}{\imath k_{2+}}\big)\ 
e^+\wedge e^- \,.
\end{multline}
The last expression can be rewritten as
\begin{multline}
 \delta^{J}_{\kappa}L_{\rm GS}
=\frac{1}{2m} \bigg( \str(P_+P_+)\str(W\commut{Q_{1-}}{\imath k_{1-}})+
\str(P_-P_-)\str(W \commut{Q_{2+}}{\imath k_{2+}})\bigg)\ e^+\wedge e^-  \,,
\end{multline}
where $m$ is the integer in the definition of  $psu(m,m)|2m)$. 

To show thus  (e.g. for the first term)  it is convenient to use
the gauge \rf{ec} where $P_+=p_1T^1+p_2T^2$. The matrices $T^1,T^2\in\alghf_2$ are defined in
\eqref{kkg},\eqref{tkg} for $m=1,2$ (and can be obviously generalized to other $m$).
In this gauge 
$P_+P_+=-\frac{1}{4}(p_1^2 \id_1+p_2^2 \id_2)$ where $\id_1$ and $\id_2$
are matrices with unit upper-left and lower-right blocks respectively so
that one finds
\begin{equation}
 \str\left( P_+P_+\commut{Q_{1-}}{\imath k_{1-}}\right)=
\frac{1}{4m}\str(P_+P_+)\str(W\commut{Q_{1-}}{\imath k_{1-}})
\end{equation}
where $W$ is the parity automorphism \rf{wer}
and we used that 
$\str(\commut{Q_{1-}}{\imath k_{1-}})=0$ and $p_1^2-p_2^2=-\frac{2}{m}\str(P_+P_+)$. 

The variation $\delta^{J}_{\kappa}L_{\rm GS}$ can be compensated by the 
following variation of the frame field
\begin{equation}
\delta_\kappa e^a_-=-\frac{1}{2m}e^a_+\str(W\commut{Q_{1-}}{\imath k_{1-}})\,,\qquad
\delta_\kappa e^a_+=-\frac{1}{2m}e^a_-\str(W\commut{Q_{2+}}{\imath k_{2+}})\,.
\end{equation}
In particular, for the variation of the metric $g^{ab}=e^a_\alpha e_\beta^b\eta^{\alpha\beta} 
=e^a_+e^b_-+e^a_-e^b_+$ one finds
\begin{equation}
 \delta_\kappa g^{ab}=\frac{1}{m}\big[e^a_+e^b_+
\str(W\commut{\imath k_{1-}}{Q_{1-}})+e^a_-e^b_-\str(W\commut{\imath k_{2+}}{Q_{2+}})\big]\,.
\end{equation}
This can be rewritten in terms of the tangent components as 
\begin{equation}
 \delta_\kappa g^{ab}=\frac{1}{m\sqrt{-g}}\big[
\str(W\commut{\imath k^b_{1(-)}}{Q^a_{1(-)}})+\str(W\commut{\imath k_{2(+)}^b}{Q_{2(+)}^a})\big]\,,
\end{equation}
where we have used that (cf. \rf{vev})
$V^a_{(\pm)}=\sqrt{-g}e^a_\mp V_\pm=(\det e)^{-1}e^a_\mp V_\pm$. 
 Taking into account  the fact
that $\delta_\kappa \sqrt{-g}=0$ one indeed finds that this variation determines the variation of
$\gamma^{ab}=\sqrt{-g}g^{ab}$ given in \eqref{kappa}.

\bs\bs

Let us now turn to the question of $\kappa$-symmetry  gauge fixing 
 in terms of the  current components.
The $\kappa$-variation of the frame components of the current is 
\begin{equation}
 \delta J_\alpha=(\delta_\kappa e_\alpha^a)J_a+ e_\alpha^a (\d_a \epsilon+\commut{J_a}{\epsilon})=
(\delta_\kappa e)_\alpha^a e_a^\beta J_\beta + e_\alpha^a \d_a \epsilon+\commut{J_\alpha}{\epsilon}\,.
\end{equation}
The fermionic equations of motion  written in terms of the  frame   components $\pm$ 
of the currents 
take exactly
the same form as in the usual ``light-cone'' coordinates (cf. last line in \eqref{eom})
\begin{equation}
 \commut{P_+}{Q_{1-}}=0\,, \qquad \commut{P_-}{Q_{2+}}=0\,.
\end{equation}
As we have seen above the same applies to  the Virasoro constraints expressed in terms of the frame components:
\begin{equation}
 \str(P_+P_+)=0\,, \qquad \str(P_-P_-)=0\,.
\end{equation}
Under the gauge transformation with  $G$-valued gauge parameter the components $P_\pm$ transform as
$P_\pm\to g_{0}^{-1}P_\pm g_0$. Using the Virasoro constraints and applying exactly the same
 argument as in the discussion of the reduction gauge in terms of the original
  light-cone components in section 6.2 one can assume that $P_+=p_+ T$
and $P_-=p_-g^{-1}Tg$ where $p_\pm$ are some real functions and $g$ is a $G$-valued function.

In this gauge the $\kappa$-transformation of the component $Q_{1-}$   becomes 
\begin{equation}
\label{q1-k}
\delta_\kappa {Q_{1-}}=(\delta_\kappa e)^a_{-} e_a^\alpha Q_{1\alpha}+e_-^a\d_a \epsilon+
\commut{\A_-}{\epsilon_1}+\commut{P_-}{\epsilon_2}+\commut{Q_{1-}}{h}\,,
\end{equation}
where $h=h(J,\epsilon_1,\epsilon_2)$ is the $\alghf_0$-valued parameter
 of the compensating gauge transformation
needed to maintain the gauge condition $P_+=p_+T$. 
In fact,
 in this gauge $\commut{P_-}{\epsilon_2}=0$ because
$\epsilon_2=\imath\scommut{P_-}{k_{2+}}$ and $\commut{T}{\scommut{T}{M}}=0$ 
vanishes for any matrix $M$. The term with the $\kappa$-symmetry transformation of the 
frame field is given explicitly by
\begin{equation} \la{jpl}
 (\delta_\kappa e)^a_{-} e_a^\alpha Q_{1\alpha}=f_-^+Q_{1+}\,, \qquad
f_-^+=\frac{1}{2m}\str(W\commut{\imath k_{1-}}{Q_{1-}})\,.
\end{equation}
The transformation \eqref{q1-k} then takes  the form (cf. \rf{kappa})
\begin{equation}
\label{1-var}
\delta Q_{1-}=e_-^a\d_a \epsilon_1+\commut{\A_-}{\epsilon_1}+Q_{1+}f^+_-+\commut{Q_{1-}}{h}\,.
\end{equation}
Applying the decomposition
 $\alghf=\alghf^\oo\oplus\alghf^\pp$ to the 
 $\kappa$-symmetry transformation of 
 $Q_{1-}$ in the reduction gauge where $P_+=p_+T$
one observes that $\epsilon_1$ takes values in $\alghf_1^\oo$ (cf. \rf{kappa})
and  at the same time the equation $\commut{P_+}{Q_{1-}}=0$ implies
that $Q_{1-}$ is also $\alghf_1^\oo$-valued. 
Because \eqref{1-var} is the
 symmetry of the equation $\commut{P_+}{Q_{1-}}=0$ preserving the structure 
 of $P_+$, the variation $\delta Q_{1-}$
also belongs to $\alghf_1^\oo$. One then 
concludes that $Q_{1-}$ can be put to zero by an appropriate choice 
of $\alghf_1^\oo$-valued $\epsilon_1$.
This in turn implies that such $\epsilon_1$ can be represented as $\imath\scommut{P_+}{k_{1-}}$.

Note that once $Q_{1-}$ is set to zero,  any transformation with 
an arbitrary $\epsilon_2=\imath\scommut{P_-}{k_{1+}}$ and 
$\epsilon_1=\imath\scommut{P_+}{k_{1-}}$ satisfying
$e_-^a\d_a\epsilon_1+\commut{\A_-}{\epsilon_1}=0$ preserves $Q_{1-}=0$ because
$f_-^+$ in \rf{jpl} also vanishes when $Q_{1-}=0$. This statement is invariant under the 
$\alghf_0$-gauge transformations
and therefore holds in any $\alghf_0$-gauge. Analogous considerations for $Q_{2+}$ in the
gauge where $P_-=p_-T$ show that one can also set $Q_{2+}=0$.
Finally, using a local Lorentz transformation and choosing the
 appropriate coordinates
$\sigma^\pm$ one can bring $e_a^\alpha$ to the standard
 form where the only nonvanishing components
are $e_+^+=e_-^-=1$. We then  arriving at the gauge  choice \rf{kg} for the two 
components of the fermionic currents.

%%%%%%%%%%%%%%%%%%%%%%%%%%%%%%%%%%%%%%%%%%%%
\subsection*{Appendix E:  Details  of gauge fixing in section \bref{Lorentz}}
\refstepcounter{section}
\label{sec:ils}
\addcontentsline{toc}{subsection}{Appendix E:  Details  of gauge fixing in section \textbf{6.4}}
%%%%%%%%%%
\def\theequation{E.\arabic{equation}}
\setcounter{equation}{0}

In order to show that the reduced model of section \bref{sec:red-ferm} is indeed described by ~\eqref{L-tot}
one   is to demonstrate that  the constraint equations that arise from 
 varying this  action with respect to
$A_{\pm}$ represent  an  admissible gauge condition for the equations of motion 
\eqref{main+Psi},\eqref{ferm+Psi}.
To see this let us introduce the following quantities (cf. \rf{comb})
\begin{align}
\hat A_+ &= g^{-1}\d_+g+g^{-1}A_+g 
-\frac{\km}{2}\commut{\commut{T}{\psr}}{\psr}\,,\\ 
\hat A_- &= g\d_-g^{-1}+gA_-g^{-1} 
-\frac{\km}{2}\commut{\commut{T}{\psl}}{\psl}\,.
\end{align}
Under the gauge transformation~\eqref{gs+f}, \eqref{gs-opsi} they 
transform as follows
\begin{equation}
 \hat A_{+}\to {\bar h}^{-1}{\hat A}_{+}{\bar h}+{\bar h}^{-1}\d_+ \bar h\,,
 \qquad
{\hat A}_{-}\to h^{-1}{\hat A}_+h +h^{-1}\d_- h\,.
\end{equation}
Their $\algh$ projections $({\hat A}_\pm)_\algh$ obviously have the same 
transformations properties.
%In terms of $\hat A_\pm$ 
The variation of the action \rf{L-tot}  with respect to  $A_\pm$ gives 
\begin{equation}
 A_+=(\hat A_+)_\algh\,, \qquad A_-=(\hat A_-)_\algh\,.
\end{equation}
The first equation 
in~\eqref{main+Psi} can be written (upon using the other two equations) as 
\begin{equation}
\label{hat-form}
\d_-\hat A_+-\d_+A_-+\commut{A_-}{\hat A_+} +\km^2\commut{g^{-1}Tg}{T}-\frac{\km}{2}
\commut{T}{\commut{D_-\psr}{\psr}}=0\,,
\end{equation}
or,  equivalently, as
\begin{equation}
\label{hat-form-2}
\d_+\hat A_- -\d_-A_+ +\commut{A_+}{\hat A_-} +\km^2\commut{gT g^{-1}}{T}-\frac{\km}{2}
\commut{T}{\commut{D_-\psl}{\psl}}=0\,.
\end{equation}
%This equivalent representation can be obtained by multiplying \eqref{hat-form} by $g$ and $g^{-1}$
%from the left and the right respectively.
Since  $(\commut{T}{u})_\algh=0$  (note that  $\commut{T}{u}\in\alghf^\pp$ while
$\algh=\alghf_0^\oo$) and projecting this 
equation on $\algh$ one finds that $A_-$ and $(\hat A_+)_\algh$ are the two 
components of a flat 
connection. Repeating  the argument used in  the bosonic case one then concludes that
one can set $A_+=(\hat A_+)_\algh$
 by an appropriate
gauge transformation with $h=\id$.  In this gauge $A_-$ 
 and $A_+$ are then components of a flat connection and can be put to zero by a
  gauge transformation with $h=\bar h$.

In the gauge $A_+=A_-=0$ the equation \eqref{hat-form-2} implies:
\begin{equation}
\d_+(\hat A_-)_\algh=0\,,
\end{equation}
where we again  made  use of the fact that $(\commut{T}{u})_\algh=0$ for
 any $u\in\alghf_0\oplus \alghf_2$. Then $(\hat A_-)_\algh$ is a function 
 of $\sigma^-$ only and therefore can be set to zero
  by a gauge transformation with $\bar h=\id$ and $h=h(\sigma^-)$. As 
   in the bosonic case such a gauge transformation
does not spoil the conditions $A_+=A_-=(\hat A_+)_\algh=0$.

%\addcontentsline{toc}{section}{\refname}

\end{document}